\documentclass[a4paper,fleqn,usenatbib,useAMS]{mnras}

\usepackage[T1]{fontenc}
\usepackage{ae,aecompl}

\usepackage{graphicx}	
\usepackage{amsmath}	
\usepackage{amssymb}	
\usepackage{bm}		
\usepackage{pdflscape}	

\usepackage{ulem}
\usepackage{times}
\usepackage{txfonts}
\usepackage{color}

\newcommand{\fracb}[2]{\left(\frac{#1}{#2}\right)}

\newcommand{\mean}[1]{\langle{#1}\rangle}
\newcommand{\lamep}{\fracb{\lambda}{\epsilon}}

\usepackage[dvipsnames]{xcolor}
\definecolor{blazeorange}{rgb}{1.0, 0.4, 0.0}
\definecolor{seagreen}{rgb}{0.18, 0.55, 0.34}
\definecolor{rufous}{rgb}{0.66, 0.11, 0.03}
\definecolor{royalfuchsia}{rgb}{0.79, 0.17, 0.57}
\definecolor{scarlet}{rgb}{1.0, 0.13, 0.0}
\definecolor{royalpurple}{rgb}{0.47, 0.32, 0.66}

\title[GRB Spectrum from Gradual Dissipation in a Magnetized Outflow]
{GRB Spectrum from Gradual Dissipation in a Magnetized Outflow}

\author[Gill, Granot, \& Beniamini]{
Ramandeep Gill$^{1,2}$\thanks{E-mail: rsgill.rg@gmail.com},
Jonathan Granot$^{2,1}$,
Paz Beniamini$^{3}$
\\
$^{1}$Department of Physics, The George Washington University, Washington, DC 20052, USA\\
$^{2}$Department of Natural Sciences, The Open University of Israel, P.O Box 808, Ra'anana 43537, Israel\\
$^{3}$Theoretical Astrophysics, California Institute of Technology, Mail Code 350-17, Pasadena, CA 91125, USA \\
}

\date{Accepted XXX. Received YYY; in original form ZZZ}

\pubyear{2020}

\begin{document}
\label{firstpage}
\pagerange{\pageref{firstpage}--\pageref{lastpage}}
\maketitle

\begin{abstract}
Modeling of many GRB prompt emission spectra sometimes requires a (quasi) thermal spectral component in addition to the Band function 
that sometimes leads to a double-hump spectrum, the origin of which remains unclear. In photospheric emission models, 
a prominent thermal component broadened by sub-photospheric dissipation is expected to be 
released at the photospheric radius, $r_{\rm ph}\sim10^{12}\,$cm. We consider an ultra-relativistic strongly magnetized steady outflow 
with a striped-wind magnetic-field structure undergoing gradual and continuous magnetic energy dissipation at $r<r_s$ that 
heats and accelerates the flow to a bulk Lorentz factor $\Gamma(r)=\Gamma_\infty\min[1,(r/r_s)^{1/3}]$, where typically 
$r_{\rm ph}<r_s$. Similar dynamics and energy dissipation rates are also expected in highly-variable magnetized outflows without 
stripes/field-reversals. Two modes of particle energy injection are considered: (a) power-law electrons, 
e.g. accelerated by magnetic reconnection, and (b) distributed heating of all electrons (and $e^\pm$-pairs), e.g. due to magneto-hydrodynamic instabilities. Steady-state spectra are obtained using a numerical code that evolves coupled kinetic equations for a photon-electron-positron plasma. We find that
(i) the thermal component consistently peaks at $(1+z)E_{\rm pk}\sim0.2-1\,$MeV, for a source at redshift $z$, and becomes subdominant 
if the total injected energy density exceeds the thermal one, (ii) power-law electrons cool mainly by 
synchrotron emission whereas mildly relativistic and almost monoenergetic electrons in the distributed heating scenario cool by Comptonization on 
thermal peak photons, (iii) both scenarios can yield a low-energy break, and (iv) the $\sim0.5(1+z)^{-1}\,$keV X-ray 
emission is suppressed in scenario (a) whereas it is expected in scenario (b). Energy-dependent linear polarization can differentiate between the 
two particle heating scenarios.
\end{abstract}


\section{Introduction}

The emission mechanism that powers the prompt gamma-ray emission in both short-hard 
($T_{90}\lesssim2\,$s; \citealt{Kouveliotou+93}) and long-soft ($T_{90}\gtrsim2\,$s) gamma-ray bursts 
(GRBs; see, e.g., \citealt{Kumar-Zhang-15,Piran-04} for reviews) is still a matter of debate. The typical 
prompt emission spectrum is non-thermal and is 
traditionally described by the empirical ``Band-function'' \citep{Band+93}, 
representing a smoothly broken power law. The  break photon energy where the $E L_E$ spectrum peaks is on average 
measured to be around $\mean{E_{\rm pk}}\simeq250\,$keV and the mean power-law photon indices 
below and above $E_{\rm pk}$ are $\mean{\alpha}\simeq-1$ and $\mean{\beta}\simeq-2.3$, respectively 
\citep[e.g.,][]{Preece+00,Kaneko+06}. One of the main difficulties in understanding the emission mechanism has been 
our ignorance of the jet composition, i.e. whether it is kinetic energy dominated \citep{Rees-Meszaros-94} or 
Poynting-flux dominated \citep{Thompson-94,Lyutikov-Blandford-03}, which dictates the mode of energy dissipation 
in the outflow, e.g., internal shocks or magnetic reconnection, respectively. When the outflow is loaded with 
protons and neutrons, nuclear collisions between the two particle species can also dissipate energy \citep[e.g.][]{Beloborodov-10}.

The localization of the spectral peak to $100\,{\rm keV}\!\,\!\lesssim\!\,\!E_{\rm pk}\!\,\!\lesssim\!\,\!1\,$MeV finds a natural explanation 
in photospheric emission models with sub-photospheric dissipation \citep[see, e.g.,][for a review]{Beloborodov-Meszaros-17}. 
In this scenario, the spectral peak and the low-energy part of the spectrum at photon energies $E<E_{\rm pk}$ are formed by quasi-thermal 
Comptonization of soft seed photons up to the thermal peak by mildly relativistic electrons when the flow is optically thick 
with Thomson optical depth $1\lesssim\tau_T\lesssim100$ 
\cite[e.g.,][]{Eichler-Levinson-00,Peer-Waxman-04,Rees-Meszaros-05,Giannios-Spruit-07,Vurm+13,Beloborodov-13,Thompson-Gill-14,Bhattacharya-Kumar-20}. 
Continued dissipation as the flow becomes optically thin ($\tau_T<1$) then gives rise to the high-energy part of the spectrum at photon energies $E>E_{\rm pk}$ 
\citep[e.g.,][]{Giannios-06,Peer+06,Giannios-08,Gill-Thompson-14,Vurm-Beloborodov-16}. The end result is a broadened spectrum 
that resembles the typical non-thermal Band-function as compared to a narrow thermal one.

Many GRBs, however, show deviations from the typical single-component Band spectrum by having multiple spectral components \citep[e.g.,][]{Guiriec+16b}, 
namely a double-hump spectral profile \citep{Guiriec+11,Guiriec+17} with, sometimes, an additional underlying power-law component 
that in some cases features a high-energy cutoff \citep{Ryde-04,Ryde-05,Guiriec+10,Guiriec+15a,Guiriec+15b,Guiriec+16a}. When two 
spectral humps are present, one of them is modeled as being thermal and the other non-thermal, where the latter is generally interpreted 
as fast-cooling synchrotron emission from electrons with a power-law energy distribution. 
The emergence of the two-component spectrum in the internal-shock model \citep[e.g.,][]{Meszaros-Rees-00} as well as in a magnetized 
outflow \citep[e.g.,][]{Beniamini-Giannios-17} has been demonstrated analytically. 
Better and consistent time-resolved spectral fits have been obtained in the observational works mentioned above when using a two-component 
\textit{thermal} + \textit{non-thermal} model over the traditional single-component Band-function. Typically, the thermal component is 
sub-dominant. However, in some (albeit rare) cases the entire pulse is dominated by thermal (or quasi-thermal) emission \citep[e.g.,][]{Ryde-04}. 
The presence of the thermal component in the spectra of many bursts gave the initial motivation to consider photospheric emission 
models.  Now, increasing incidence for such components in GRB spectra, attributed to the wider energy range of {\it Fermi}/Gamma-ray Burst Monitor 
as well as the use of multi-component spectral fits, gives further credence to this idea. 

In addition, many GRBs have been shown to feature a low-energy spectral break at $E_{\rm br}\sim0.03E_{\rm pk}$, 
with photon indices $\alpha_1 = -0.66\pm0.35$ for $E<E_{\rm br}$ and $\alpha_2 = -1.46\pm0.31$ for $E_{\rm br}<E<E_{\rm pk}$ 
\citep{Oganesyan+17,Ravasio+19}. 
Such a break would be naturally produced in models that considered optically-thin synchrotron emission from fast-cooling electrons, 
where the break would represent the cooling break due to synchrotron emission from electrons cooling at the dynamical time, with 
photon indices $\alpha_1 = -2/3$ and $\alpha_2 = -3/2$ \citep{Katz-94,Rees-Meszaros-94,Tavani-96,Sari+98,Granot-Sari-02,KM2008,Beniamini2013}.
Under this interpretation, the relative proximity of the two break energies suggests that the particle injection Lorentz factor (LF) 
is close to the cooling LF (recall that $\gamma_m \propto E^{1/2}$) and therefore that synchrotron emission is produced in the 
marginally fast cooling regime \citep{Daigne2011,BBG2018}. Photospheric emission models generally lack such a break at low energies apart 
from that produced by synchrotron self-absorption of the soft seed photon source, which typically features a much harder Rayleigh-Jeans 
spectrum below $E_{\rm br}$ with $\alpha_1 = 1$.

To investigate spectral formation, detailed numerical simulations have been performed for models featuring 
energy dissipation in internal shocks with power-law electrons emitting synchrotron photons \citep{Peer-Waxman-04,Peer-Waxman-05}, 
neutron-proton collisional heating with monoenergetic $e^\pm$-pair injection \citep{Vurm+11,Vurm-Beloborodov-16}, and distributed heating 
with quasi-thermal Comptonization of soft photospheric component \citep{Peer+06,Gill-Thompson-14} or self-absorbed cyclo-synchrotron emission 
\citep{Stern-Poutanen-04,Vurm+13,Thompson-Gill-14} as the main emission mechanisms. These works self-consistently 
included the effects of $e^\pm$-pair cascades and conducted a thorough parameter space study. Many works considered a single-collision model, 
in which the final spectrum was derived from dissipation occurring over a single dynamical time. Some only considered sub-photospheric 
dissipation in a flow coasting at its terminal bulk LF, $\Gamma_\infty$. Earlier numerical works that explored dissipation in a Poynting-flux 
dominated outflow, using Monte Carlo simulations \citep{Giannios-Spruit-05,Giannios-06,Giannios-Spruit-07,Giannios-08}, considered a thermal 
distribution of electrons at (comoving) temperature $T_e'$ set by the balance between volumetric heating of all particles due to magnetic 
energy dissipation and their cooling due, mostly to Comptonization, as well as synchrotron emission. These works did not include the effects 
of $e^\pm$-pairs on the final spectrum. 

The appearance of the double-hump spectrum and low-energy break offers additional clues for understanding the prompt GRB emission 
mechanism. To this end, we consider a photospheric emission model with sub-photospheric dissipation, occurring continuously from 
$r_{\tau0}\ll\; r_{\rm ph}<r_s$ until the saturation radius $r_s$, from magnetic reconnection or magneto-hydrodynamic (MHD) instabilities in a striped-wind Poynting-flux dominated ultra-relativistic outflow. \citet{Beniamini-Giannios-17} carried out analytic modeling of this scenario that yielded 
two-component spectra for a range of values of the model parameters. Alternatively, in a highly time-variable and magnetized relativistic 
outflow impulsive magnetic acceleration takes place even without any magnetic-field reversals (i.e. striped-wind) \citep[e.g.,][]{Granot+11} 
that leads to a similar dynamical evolution and energy dissipation per unit radius, where dissipation occurs through internal shocks rather 
than magnetic reconnection \citep{Granot+11,Granot12b,Komissarov12}. 
Therefore, the results of this work are generally applicable to a wider class of Poynting-flux dominated models. 

Here we consider two energy dissipation scenarios that accelerate/heat the electrons (and created $e^\pm$-pairs) differently. 
(i) Magnetic reconnection in the striped wind is assumed to accelerate a fraction $\xi$ of the total baryonic electrons in the 
emission region, whose initial Thomson optical depth is $\tau_{T0}$, into a power-law energy distribution. The remaining fraction, 
$(1-\xi)$ of the total, forms a cold Maxwellian distribution which is initially in thermal equilibrium with the entrained thermal 
radiation field for $\tau_T>\tau_{T0}$. 
(ii) MHD instabilities, e.g. the Kruskal-Schwarzchild instability \citep{Lyubarsky-10,Gill+18}, lead to distributed heating of 
all the particles that form a narrowly peaked distribution at a critical energy defined by the balance between heating and cooling. 

In both scenarios energy dissipation commences at a given 
optical depth $\tau_{T0}$ and coupled kinetic equations for both particles and photons are self-consistently evolved using a one-zone time-dependent 
kinetic code that includes all relevant radiation processes and interactions between both distributions. Most importantly, we include the 
effect of $e^\pm$-pair cascades that was ignored in some earlier works due to its highly non-linear nature.

The main model is presented in \S\ref{sec:model}, where we describe the flow dynamics of an ultra-relativistic steady spherical flow 
(\S\ref{sec:dynamics}) followed by energy dissipation and particle acceleration (\S\ref{sec:dissipation}) and details 
of the thermal radiation (\S\ref{sec:radiation}). A brief description of the one-zone code is provided in \S\ref{sec:numerical}. 
The two particle heating scenarios are discussed in \S\ref{sec:particle-heating} 
and the results of the simulations including radial evolution of the spectrum, particle distribution, and flow parameters for scenario (i) 
are presented in \S\ref{sec:scenario-I}. Likewise, results for the distributed heating scenario (ii) are presented in \S\ref{sec:scenario-II}. 
In \S\ref{sec:Params-Study}, we carry out a parameter space exploration and present spectra for different outflow parameters. Low-energy 
spectral breaks are discussed in \S\ref{sec:spectral-break} followed by a summary of this work and discussion in \S\ref{sec:discussion}.

\section{Gradual Energy Dissipation in a Relativistic Spherical Flow}
\label{sec:model}
\subsection{Flow Dynamics}\label{sec:dynamics}
We consider a steady Poynting-flux dominated relativistic (locally) spherical flow with a striped wind magnetic field structure 
\citep[e.g.,][]{Lyubarsky-Kirk-01,Begue+17}, where we follow the treatment in \citet{Beniamini-Giannios-17} and present the salient points below. The characteristic 
length scale ($\lambda$) over which the magnetic field lines reverse polarity is set by the size of the light cylinder ($r_L$), 
such that $\lambda\sim\pi r_L=\pi c/\Omega = cP/2 = 1.5\times10^7P_{-3}\,$cm, where $\Omega=2\pi/P$ is the central engine's rotational 
angular frequency, $P=10^{-3}P_{-3}\,$s is the spin period, and $c$ is the speed of light. 
While this description of a striped wind flow is relevant for a millisecond magnetar central engine 
\citep[e.g.,][]{Metzger+11}, more generally a magnetized outflow from an accreting black hole arguably features stochastic flips in 
magnetic field polarity over length scales $\lambda\gtrsim r_L$ \citep[e.g.,][]{Mckinney-Uzdensky-12,Parfrey+15}. 
It is worth pointing out that a broadly similar scenario may take place even without magnetic field flips or reversals, for a time-variable 
Poynting-flux dominated outflow. In this case impulsive magnetic acceleration leads to a very similar global flow dynamics \citep{Granot+11} 
in terms of $\Gamma(r)$ and the fraction of the total energy that is dissipated up to a radius $r$, $f_{\rm dis}(r)$. While there is no 
magnetic reconnection in this picture, energy dissipation is driven by internal shocks within the outflow \citep{Granot+11,Granot12b,Komissarov12} 
including multiple weak shocks at $r\ll r_s$ where $\sigma\gg1$ that gradually become more efficient and become strongest and most efficient 
when $\sigma\lesssim1$ is reached at $r\gtrsim r_s$. In this scenario the effective shell (rather than stripe) width is 
$\lambda\sim ct_\varv$ where $t_\varv = 5\times10^{-4}P_{-3}\,$s is the central engine's variability time, which is reflected in the 
observed variability timescales of the prompt GRB emission (up to cosmological time dilation). The observed variability time is 
typically $\Delta t_\varv = (1+z)t_\varv\sim1\,$s, and so $\lambda\lesssim10^{10}\,$cm.

Magnetic energy is dissipated in the flow when field lines of opposite polarity are brought together 
and undergo reconnection. The rate of reconnection is set by the inflow plasma velocity, $\varv_{\rm in}=\epsilon \varv_A$, which is 
a fraction $\epsilon\sim0.1$ of the Alfv\'{e}n speed. For a strongly magnetized flow, the initial magnetization (ratio of 
magnetic to particle energy flux ratio) at the jet launching radius $r_0$ is
\begin{equation}
\label{eq:sigma}
    \sigma_0 =  \frac{L_{B,\Omega,0}}{L_{k,\Omega,0}} 
    = \frac{\beta_0c(B_0r_0)^2}{4\pi\Gamma_0\dot M_\Omega c^2} 
    = \frac{{B'_0}^2}{4\pi {n'_0}m_pc^2}\gg1\,.
\end{equation}
Here, $L_{B,\Omega,0} = \beta_0c(\Gamma_0B_0'r_0)^2/4\pi$ and $L_{k,\Omega,0} = \Gamma_0\dot M_\Omega c^2=\beta_0(r_0\Gamma_0)^2n_0'm_pc^3$ are the initial 
power per unit solid angle carried by the magnetic field, with comoving strength $B_0'=B_0/\Gamma_0$, and kinetic power per unit solid 
angle carried by the cold baryons, with comoving number density $n_0'$. The flow is assumed to achieve magnetization $\sigma_A = \sigma_0^{2/3}$ 
at the Alfv\'{e}n radius $r_A\sim{\rm few}\times r_L$ \citep{Drenkhahn-02},
at which point its proper velocity is $u_A=(\Gamma_A^2-1)^{1/2}=\sqrt{\sigma_A}$ and 
$\beta_A=u_A/\Gamma_A=(1-\Gamma_A^{-2})^{1/2}=\varv_A/c=\sigma_A/(1+\sigma_A)\approx1$, and therefore $\varv_{\rm in}=\epsilon c$. 

Under the assumption that a reasonable fraction of the dissipated energy in the flow goes towards its acceleration, the condition 
$\Gamma(r)\sigma(r)=\Gamma_0\sigma_0$ always holds a long as $\sigma\gg1$ (more generally $\Gamma(r)(1+\sigma(r))=\Gamma_0(1+\sigma_0$) 
from conservation of the total specific energy, i.e. neglecting radiative losses etc., where $\sigma=B'^2/4\pi w$ and $w$ is the proper enthalpy density), 
which eventually leads to $\Gamma(r>r_s)\approx\Gamma_\infty\approx\Gamma_0\sigma_0=\sigma_0=\sigma_A^{3/2}$ where $\sigma(r>r_s)<1$ 
\citep[see, e.g.,][]{Granot+11}. 
At this point, the flow becomes kinetic energy dominated and starts to coast at its terminal LF $\Gamma_\infty$ until it is decelerated by its 
interaction with the external medium -- interstellar medium (ISM) for short-hard GRBs and stellar wind of the massive star progenitor of 
long-soft GRBs. Beyond the Alfv\'{e}n radius the outflow's bulk LF grows as a power law in radius
\begin{equation}
    \Gamma(r) = \Gamma_\infty\fracb{r}{r_s}^{1/3}\,,\quad\quad r_A<r<r_s\ ,
\end{equation}
until the saturation radius\footnote{Throughout this work, the notation $Q_x$ denotes the value of the quantity $Q$ in units of $10^x$ times its (cgs) units}, 
\begin{equation}\label{eq:r_s}
    r_s = \frac{\Gamma_\infty^2\lambda}{6\epsilon} = 1.7\times10^{13}\Gamma_{\infty,3}^2\fracb{\lambda}{\epsilon}_8~{\rm cm}\,,
\end{equation}
at which point all of the magnetic energy in the flow has been dissipated with nothing left 
for further acceleration. However, further dissipation can still occur due to internal shocks which become efficient when $\sigma<1$ for 
$r>r_s$, as argued above.

\begin{figure}
    \centering
    \includegraphics[width=0.42\textwidth]{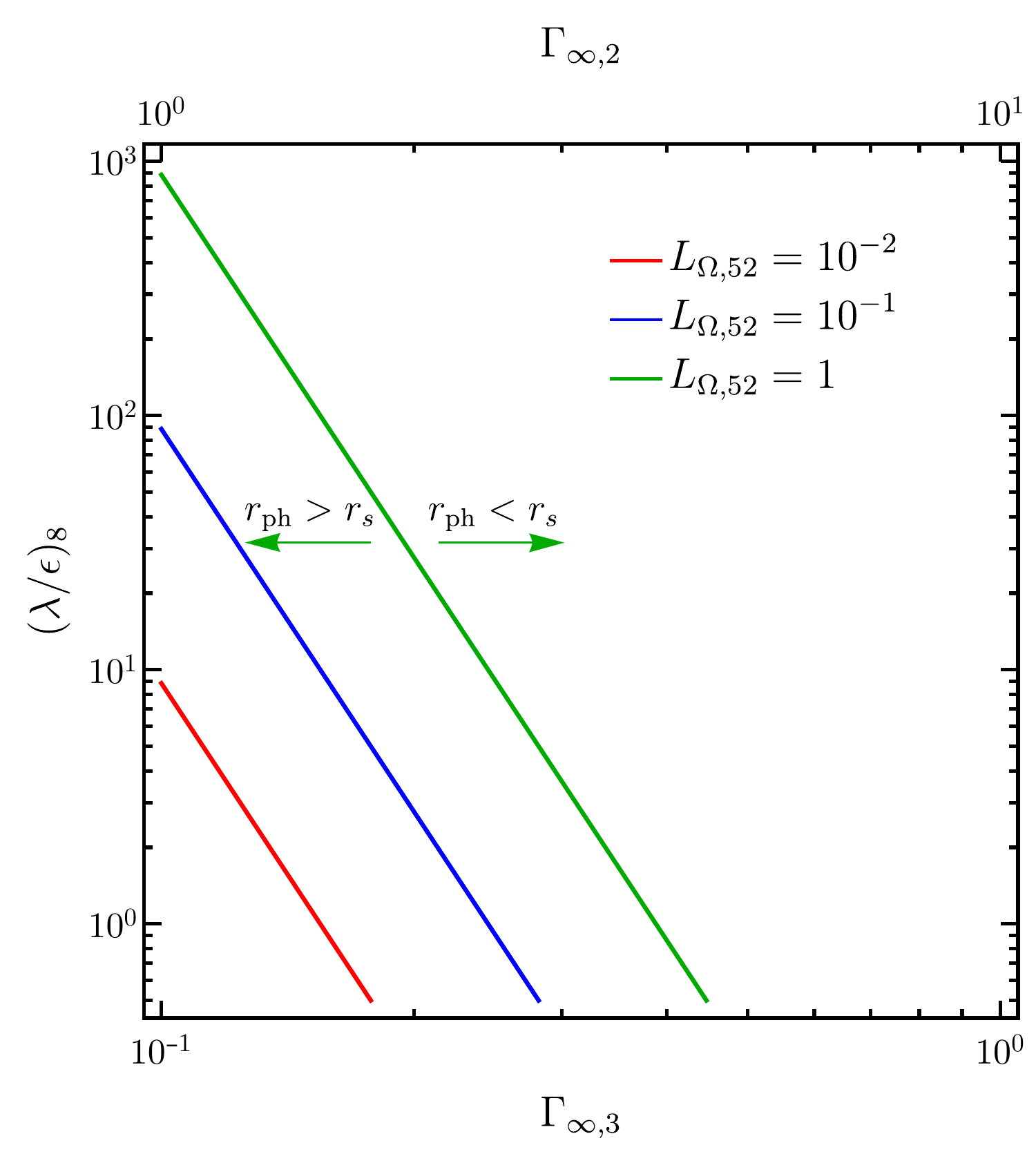}
    \caption{Parameter space for which the photospheric radius $r_{\rm ph}$, due to baryonic electrons, is equal to the saturation 
    radius $r_s$, shown as a function of $\Gamma_{\infty}$ and $(\lambda/\epsilon)$ for fixed jet power per unit solid angle $L_{\Omega}$ 
    (note that the outflow's total isotropic equivalent power is $4\pi L_\Omega = 1.26\times10^{53}L_{\Omega,52}\;{\rm erg~s}^{-1}$). 
    We only consider the regime where $r_{\rm ph} < r_s$ (to the right of the lines) when the flow is heated continuously from the 
    optically thick to thin regime.}
    \label{fig:Rph-Rs}
\end{figure}{}

The flow is launched Poynting-flux dominated and the total power per unit solid angle crossing radius $r$ is given by 
$L_\Omega = L_{B,\Omega}+L_{k,\Omega}+L_{\gamma,\Omega}$, where the last term represents the emitted radiation.
In the absence of any dissipation $L_{\gamma,\Omega}=0$, and the power carried by the Poynting flux can be expressed in terms of the 
total jet power, $L_{B,\Omega}=L_\Omega(1-\Gamma/\Gamma_\infty)\approx L_\Omega$ for $r_A<r\ll r_s$ (where $1<\Gamma\ll\Gamma_\infty$ 
and $\beta\approx1$), which yields an estimate of the comoving magnetic field 
\begin{equation}
    B' \approx \fracb{4\pi L_\Omega}{\Gamma^2r^2c}^{1/2}
    = 4.1\times10^6\,\frac{L_{\Omega,52}^{1/2}\lamep_8^{1/3}}{r_{12}^{4/3}\Gamma_{\infty,3}^{1/3}}\,{\rm G}\;.
\end{equation}
The comoving number density of the baryonic electrons in the flow is given by 
$n' = L_\Omega/r^2\Gamma\Gamma_\infty m_pc^3\propto(r^2\Gamma)^{-1}$, 
which contributes a characteristic Thomson optical depth of
\begin{equation}
    \tau_T = \frac{n'\sigma_Tr}{\Gamma} = \frac{\sigma_TL_\Omega}{r\Gamma^2\Gamma_\infty m_pc^3}
    = 1\,\frac{L_{\Omega,52}\lamep_8^{2/3}}{\Gamma_{\infty,3}^{5/3}r_{12}^{5/3}}\,,
\end{equation}
where $\sigma_T$ is the Thomson cross-section. 
For $r<r_s$, $\Gamma\propto r^{1/3}$ and therefore $\tau_T\propto r^{-5/3}$. However, when the flow starts to coast at $\Gamma=\Gamma_\infty$ 
the Thomson optical depth drops more slowly with radius, $\tau_T\propto r^{-1}$. 
At $\tau_T=1$ matter and radiation decouple, allowing the radiation to stream freely, which defines the photospheric radius,
\begin{equation}
    \label{eq:r_ph}
    r_{\rm ph} \approx 10^{12}\,\frac{L_{\Omega,52}^{3/5}\lamep_8^{2/5}}{\Gamma_{\infty,3}}~{\rm cm}\;.
\end{equation}
Here we have only considered the Thomson optical depth of baryonic electrons. In Fig.~\ref{fig:Rph-Rs}, we show the different 
parameters for which $r_{\rm ph} = r_s$. The vertical axis shows the typical range expected for $\lamep_8$, where the lower end 
is relevant for a millisecond magnetar central engine and the higher end reflects the typical values based on the observed variability 
timescale of prompt GRB emission. The solid lines show the model parameter space for fixed values of the jet power per unit solid angle, 
$L_\Omega\lesssim10^{52}\,{\rm erg~s}^{-1}{\rm sr}^{-1}$, the fiducial value adopted in this work. We consider the regime with 
$r_{\rm ph}<r_s$ when the flow is heated continuously as it transitions from the optically thick to thin regimes. We will show below that 
copious pair-production ensues when energy dissipation leads to particle acceleration into a power-law energy distribution that emits 
energetic synchrotron radiation. The created pairs extend the photospheric radius by factors of a few.

\subsection{Energy Dissipation \& Particle Acceleration}
\label{sec:dissipation}
Energy is dissipated gradually in the flow, for $r<r_s$, as magnetic field lines of opposite polarity come into contact and undergo 
magnetic reconnection. The rate of energy dissipation at any given radius can be obtained from the Poynting-flux power, such that \citep{Giannios-Spruit-05}
\begin{equation}
    \frac{dL_{\rm diss,\Omega}}{dr} 
    = -\frac{dL_{B,\Omega}}{dr} 
    = -\frac{d}{dr}\left[L_\Omega\left(1-\frac{\Gamma}{\Gamma_\infty}\right)\right]
    = \frac{1}{3}\frac{L_\Omega}{\Gamma_\infty}\frac{\Gamma}{r}\propto r^{-2/3}\,.
\end{equation}
This implies a differential dissipation $dL_{\rm diss,\Omega}\propto r^{-2/3}dr$ or a cumulative dissipation 
$L_{\rm diss,\Omega}(<\!r)\propto r^{1/3}$ at $r_0<r<r_s$. At $r=r_s$, when $\Gamma = \Gamma_\infty$, magnetic 
energy dissipation peaks and stops, so that $f_{\rm dis}(r) = L_{\rm diss,\Omega}(<\!r)/L_\Omega = \min[1,(r/r_s)^{1/3}]$. 
Next, we relate the dissipated power to the comoving dissipated energy density, $dL_{\rm diss,\Omega}=r^2\Gamma^2c\,dU'_{\rm diss}$, 
and express $dr=\Gamma\beta cdt'\approx\Gamma c dt'$ for $\Gamma\gg1$ and $\beta\approx1$, which yields
\begin{equation}
    \label{eq:dUdt}
    \frac{dU'_{\rm diss}}{dt'} = \frac{1}{3}\frac{L_\Omega}{\Gamma_\infty r^3}~.
\end{equation}
Magnetic reconnection leads to the acceleration of electrons into a non-thermal power-law energy distribution, 
with $dn'\propto\gamma_e^{-p}d\gamma_e$ for $\gamma_m<\gamma_e<\gamma_M$, for which the mean energy per unit rest mass is 
$\langle\gamma_e\rangle_{\rm nth}=[(p-1)/(p-2)]\gamma_m$ when $p>2$.
The power-law index $p$ has been shown to depend sensitively on the value of $\sigma$ \citep[e.g.,][]{Sironi-Spitkovsky-14,Guo+15,Kagan+15,Werner+16}, 
where it can be approximated to follow the scaling \citep{Beniamini-Giannios-17}
\begin{equation}\label{eq:sigma}
 p=4\sigma^{-0.3}\,.
\end{equation}
In models featuring internal shocks, $2\lesssim p\lesssim3$ is left to vary as one of the model parameters, whereas the $\sigma$ 
dependence of $p$, as employed here, reduces the total number of model parameter by one.

It is assumed here for simplicity that half of the dissipated energy $E'_{\rm diss}$ goes directly into the flow's kinetic energy 
\citep[see, e.g.,][]{Drenkhahn-Spruit-02}, while the other half goes towards particle acceleration and is divided between electrons 
($\epsilon_e E'_{\rm diss}/2$) and protons ($(1-\epsilon_e)E'_{\rm diss}/2$), where most of the latter energy is also typically quickly 
converted into kinetic energy. In scenario $(i)$ we further assume that 
only a fraction $\xi<1$ of electrons are actually accelerated during magnetic reconnection, and the remaining fraction $(1-\xi)$ form a thermal distribution. 
The mean energy per baryon is limited to $\sigma m_pc^2$, as this is the total dissipated energy per baryon-electron for complete 
magnetic dissipation, however some particle may in principle exceed the mean energy. Therefore, the mean energy per accelerated electron, 
for a total of $N_e$ electrons, is given as $\xi\langle\gamma_e\rangle m_ec^2 = (\epsilon_e/2) E'_{\rm diss}/N_e = \epsilon_e \sigma m_pc^2/2$, 
which yields an estimate of the mean energy per rest mass energy of the non-thermal electrons \citep{Beniamini-Giannios-17}
\begin{equation}\label{eq:gamma_nth}
    \langle\gamma_e\rangle_{\rm nth} = \frac{\epsilon_e}{2\xi}\sigma\frac{m_p}{m_e} 
    = 2.3\times10^3\fracb{\epsilon_e}{\xi}\frac{\lamep_8^{1/3}\Gamma_{\infty,3}^{2/3}}{r_{12}^{1/3}}\,.
\end{equation}
For a given set of flow parameters, the ratio of the parameters $\epsilon_e$ and $\xi$ controls the mean energy of the power-law accelerated electrons. Since 
$\epsilon_e$ also controls the amount of energy put into the power-law electrons, it also sets the normalization of the non-thermal synchrotron 
emission component with respect to thermal component.


\subsection{Thermal Radiation}
\label{sec:radiation}
The magnetic energy in the flow is dissipated over a range of radii ($r_0<r<r_s$) and as the flow expands to larger radii 
its Thomson optical depth drops. Therefore, for a given set of model parameters it is possible that energy dissipation 
proceeds continuously from the optically thick to thin regions. Where most of the energy is dissipated has consequences for the 
emergent radiation field spectrum. If most of the dissipation occurs at smaller radii, when the flow is optically thick ($\tau_T\gg1$), 
Compton interactions between the electrons (or pairs) and the radiation field ensure that the flow maintains (quasi-)thermal 
equilibrium. In this case, the flow expands adiabatically and since it is radiation-dominated, the scaling of comoving energy density 
with comoving volume follows $U'_{\rm th}\propto V'^{-4/3}$. The energy density of the thermal radiation field can be related to 
its comoving temperature, $U'_{\rm th}= (4\sigma_{\rm SB}/c)T_{\rm th}'^4$, where $\sigma_{\rm SB}$ is 
the Stefan-Boltzmann constant, which yields $T'_{\rm th}\propto V'^{-1/3}$. 
For a steady relativistic spherical flow expanding radially, the continuity equation yields, 
$r^2\Gamma(r) n'\varv=\,$constant, so that $V'\propto r^2\Gamma(r)$.
This finally implies that $T'_{\rm th}(r)\propto r^{-7/9}$ when $\Gamma\propto r^{1/3}$. The scaling of the thermal luminosity with radius 
can now be expressed as $L_{\rm th,\Omega}=(4/3)r^2\Gamma^2cU'_{\rm th}\propto r^{-4/9}$. If an amount $dL_{\rm diss,\Omega}$ of power 
is dissipated at radius $r_{\rm diss}$, then the thermal luminosity surviving till any radius $r>r_{\rm diss}$ is given by 
$dL_{\rm th,\Omega}(r) = (1/2)dL_{\rm diss,\Omega}(r_{\rm diss})(r/r_{\rm diss})^{-4/9}$, 
such that the integrated thermal luminosity is, $\int_0^r dL_{\rm th,\Omega}(r)\propto r^{1/3}$, for $r<r_s$, and its value at the photosphere is
\begin{equation}
    \label{eq:L_th}
    L_{\rm th,\Omega}(r_{\rm ph}) \approx \frac{3}{14}L_\Omega\fracb{r_{\rm ph}}{r_s}^{1/3} 
    = 8.3\times10^{50}\,\frac{L_{\Omega,52}^{6/5}}{\Gamma_{\infty,3}\lamep^{1/5}_8}\,{\rm erg~s^{-1}~sr^{-1}}\,,
\end{equation}
and its comoving temperature at the photosphere is
\begin{equation}
    \label{eq:T_th}
    k_BT'_{\rm th}(r_{\rm ph}) = k_B\fracb{3L_{\rm th,ph,\Omega}}{16r_{\rm ph}^2\Gamma_{\rm ph}^2\sigma_{\rm SB}}^{1/4}
    \simeq 0.2\,\frac{\Gamma_{\infty,3}^{1/4}}{L_{\Omega,52}^{1/10}\lamep_8^{3/20}}\,{\rm keV}\,,
\end{equation}
and the corresponding observed energy of the Wien peak is
\begin{equation}
    E_{\rm pk,th}(r_{\rm ph}) = \frac{\Gamma_{\rm ph}}{1+z}3k_BT'_{\rm th}(r_{\rm ph})
    = \frac{210}{1+z}\,\frac{L_{\Omega,52}^{1/10}\Gamma_{\infty,3}^{1/4}}{\lamep_8^{7/20}}\,{\rm keV}\,,
\end{equation}
where $k_B$ is the Boltzmann constant. The above peak energy estimate corresponds to that for the spectral luminosity $L_E$. 
The $EL_E$ Wien spectrum peak energy occurs at $4\Gamma k_BT'_{\rm th}/(1+z)$ instead. Since $T'_{\rm th}(r)\propto r^{-7/12}$, 
the peak energy scaling with radius is $E_{\rm th,pk}\propto r^{-1/4}\propto \tau_{T0}^{3/20}$. In deriving the estimates 
above, we have made the assumption that deeper in the flow, at very large optical depths, the energy imparted to particles is 
readily thermalized and the efficiency of thermalization is high.

Several works have studied the importance of the various radiative processes that shape the (quasi-)thermal spectrum at 
different optical depths \citep[e.g.,][]{Beloborodov-13,Vurm+13,Thompson-Gill-14,Begue-Peer-15,Vurm-Beloborodov-16} and its radiative 
efficiency in a Poynting flux dominated flow \citep{Peer-17}. The radiation field is able to 
maintain a blackbody spectrum only at extremely high optical depths ($\tau_T\gg10^2$), where softer seed photons are provided 
by double Compton scattering and/or bremmstrahlung (in a weakly magnetized flow, $\sigma\ll1$) or cyclo-synchrotron emission 
(in a strongly magnetized flow, $\sigma>1$). At larger radii, the efficiency of completely thermalizing the flow drops and a Wien 
spectrum emerges instead at $\tau_T\gtrsim10^2$. Further dissipation at lower optical depths, but still below the photosphere, 
acts to broaden the Wien spectrum, producing a softer spectral slope below the spectral peak energy and a harder one above it.

\section{Numerical Treatment}\label{sec:numerical}
We model the emission region using a one-zone kinetic code \citep[see][for code details]{Gill-Thompson-14}, where we include all relevant 
high-energy radiation processes in a relativistic photon-$e^\pm$-pair plasma, including Compton scattering, cyclo-synchtrotron emission and 
self-absorption, pair production and annihilation, and Coulomb interactions among the pairs.

The escape of radiation from an optically thin ($\tau_T<1$) 
region of comoving causal size $r/\Gamma(r)$ is implemented using a simple `leaky-box' geometrical prescription \citep[see, e.g.,][]{Lightman-Zdziarski-87}. 
When the flow is optically thick ($\tau_T>1$), radiation is assumed to remain within the dissipation region with no leakage. To obtain the steady-state spectrum in 
the observer frame, we integrate over the comoving spectral emissivity \citep[see, e.g.,][]{Granot+99} from the photospheric radius, $r_{\rm ph}(\tilde\theta)$, 
which depends on the polar angle $\tilde\theta$ measured from the line-of-sight \citep{Abramowicz+91,Peer-08,Beloborodov-11}, to a large radius 
$\gg\max(r_s,r_{\rm ph})$ where $\tau_T\ll1$ and the emission and absorption become negligible. 

Since we employ a one-zone code, which lacks any spatial and angular information of the flow and the radiation field, the emission is approximated to arise 
from essentially a blob of comoving causal size $r/\Gamma$ that is radially localized at $r$ and moving with bulk LF $\Gamma(r)$. In addition, the leaky-box 
prescription is not particularly well suited to describe the optically thin parts of the flow when radiation is expected to stream freely. Instead, under the 
current prescription radiation leaks out over a (comoving) dynamical time, $t_{\rm dyn}'=r/\Gamma c$, at the rate of $dn_\gamma'/dt'=-n_\gamma'/t_{\rm dyn}'$ 
where $n_\gamma'$ is the comoving number density of photons. Then, for a coasting flow, for which $t'\propto r$, this would mean that the remaining photon 
number density, $n_\gamma'(r) = n_{\gamma,0}'(r_0/r)$, is still half at $r = r_0+\Delta r = 2r_0$ of that emitted a dynamical time (radius doubling time) 
ago at $r=r_0$. As a result, the radiation field accumulates in the emission region over multiple dynamical times, which is unphysical and may produce some artefacts. 
For example, this would cause a larger suppression of the high-energy part of the spectrum due to $\gamma\gamma$-annihilation for which a test photon with energy 
$E>\Gamma m_ec^2/(1+z)$ `sees' a larger optical depth $\tau_{\gamma\gamma}$ due to larger number density of annihilating low-energy target photons at energy 
$\sim (\Gamma m_ec^2)^2/E(1+z)^2$. This also leads to the emergence of a power-law spectral break at high-energies instead of an exponential one 
\citep[e.g.,][]{Granot+08}. Therefore, a more accurate radiation transfer treatment, which is outside the scope of this work, is needed to avoid such artefacts 
and include the angular dependence of the radiation field \citep[see, e.g.,][]{Vurm-Beloborodov-16}.

\section{Two Different Particle Heating Scenarios}\label{sec:particle-heating}

Magnetic energy dissipation due to either magnetic reconnection or MHD instabilities commences when the flow is highly optically 
thick. It continues to inject energy in the form of either power-law (baryonic) electrons or via distributed heating of all particles, 
respectively. The details of how particle injection/heating is implemented in the simulation are presented in Appendix (\ref{sec:part-inject}).

Our starting point is an optically thick flow with initial Thomson optical depth $\tau_{T0}=100$. At this point, the comoving radiation field 
spectrum resembles a Wien-like thermal spectrum,
\begin{equation}
    \label{eq:Wien}
    \frac{dn_\gamma'}{d\ln E'} = \frac{U_0'}{6(k_BT_{\rm th})^4}E'^3\exp\left(-\frac{E'}{k_BT_{\rm th}'}\right)
\end{equation}
characterized by its temperature $T'_{\rm th}$ from Eq.~(\ref{eq:T_th}) and normalization given by 
$U_0'=L_{\rm th,\Omega}/(4/3)r^2\Gamma^2c$ with 
$L_{\rm th,\Omega}=L_{\rm th,ph,\Omega}(r/r_{\rm ph})^{1/3}=L_{\rm th,ph,\Omega}\tau_T^{-1/5}$ in Eq.~(\ref{eq:L_th}) for 
$r<r_s$.

\begin{figure*}
    \centering
    \includegraphics[width=0.45\textwidth]{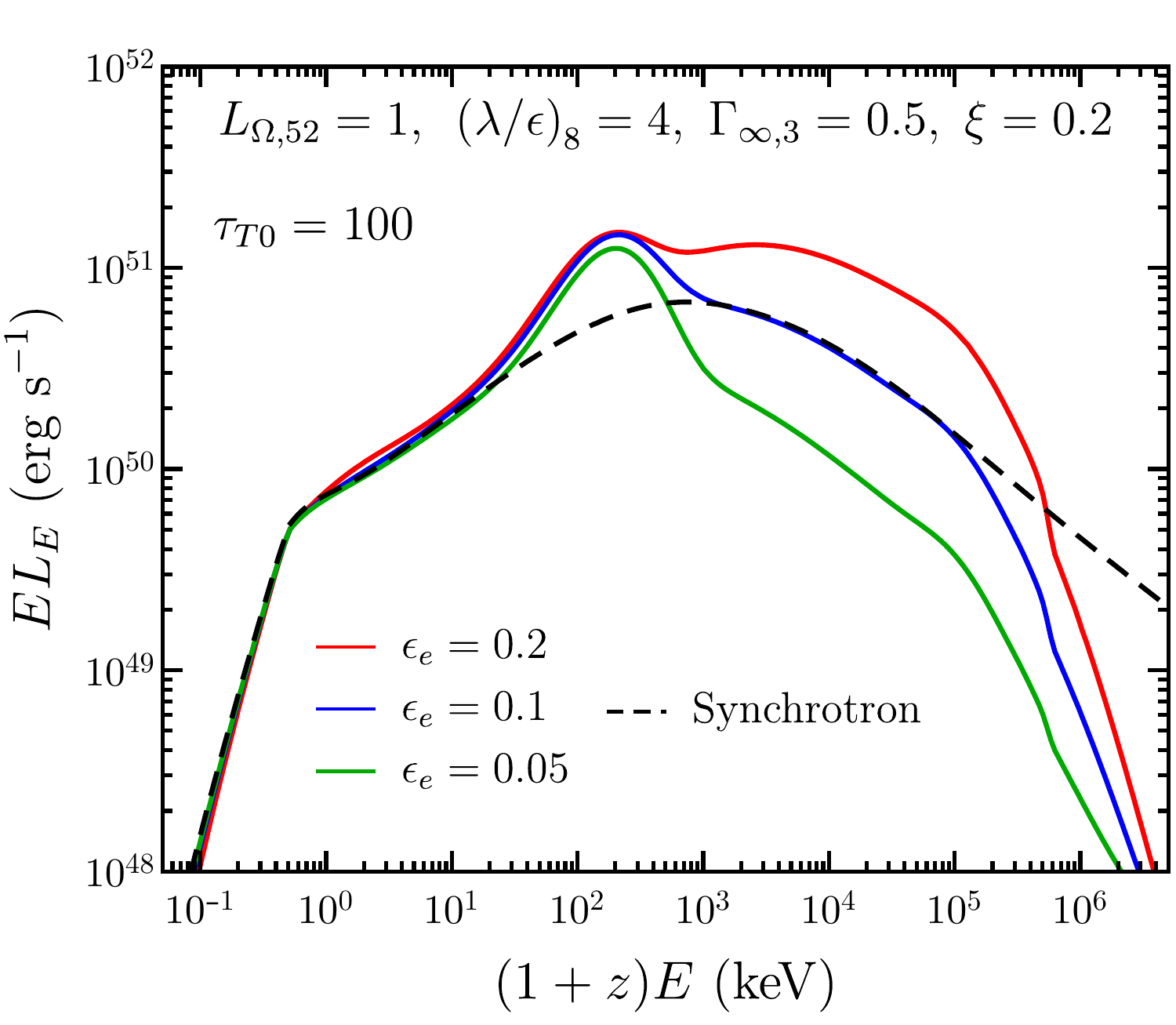}\quad\quad\quad
    \includegraphics[width=0.45\textwidth]{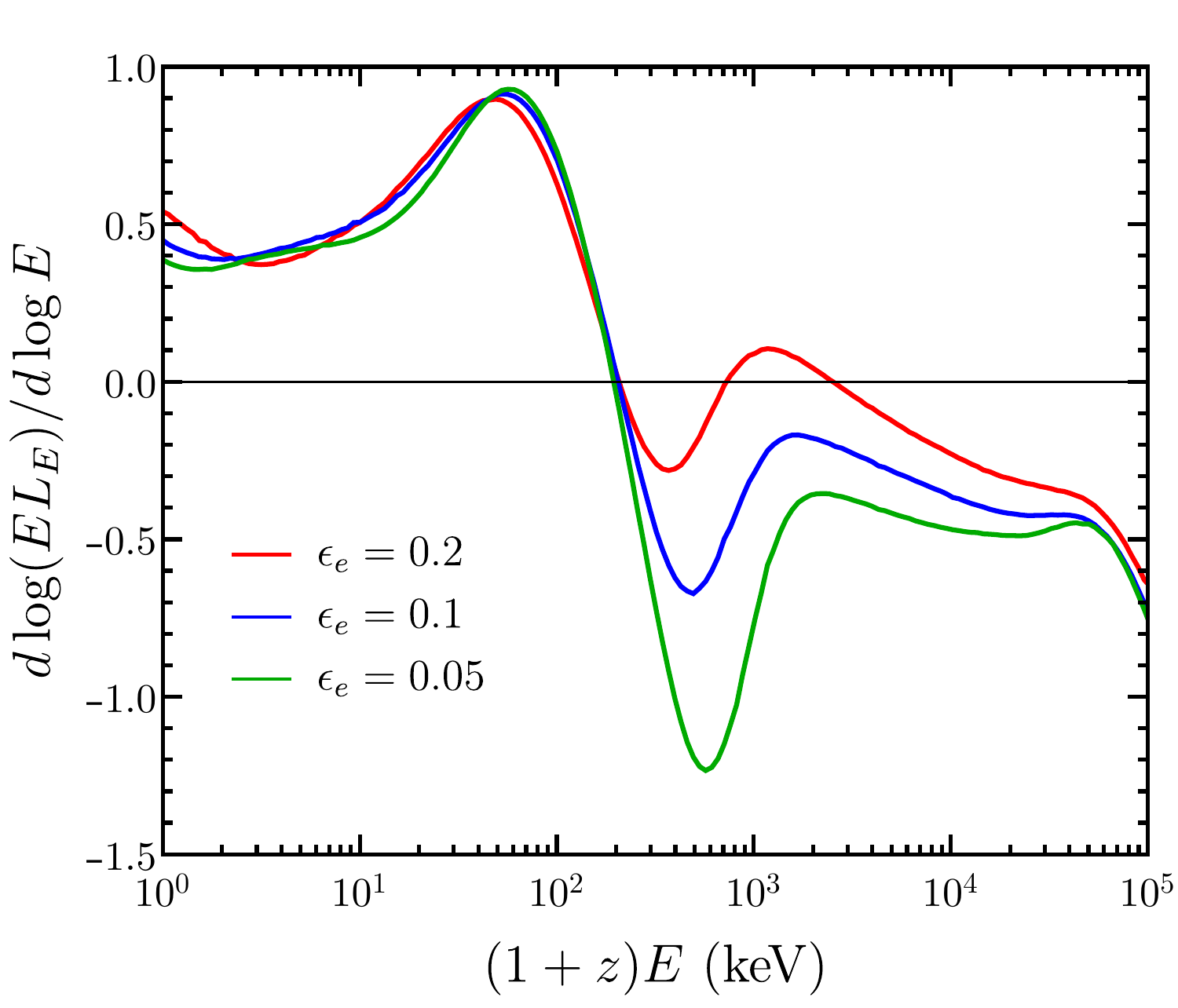}
    \includegraphics[width=0.45\textwidth]{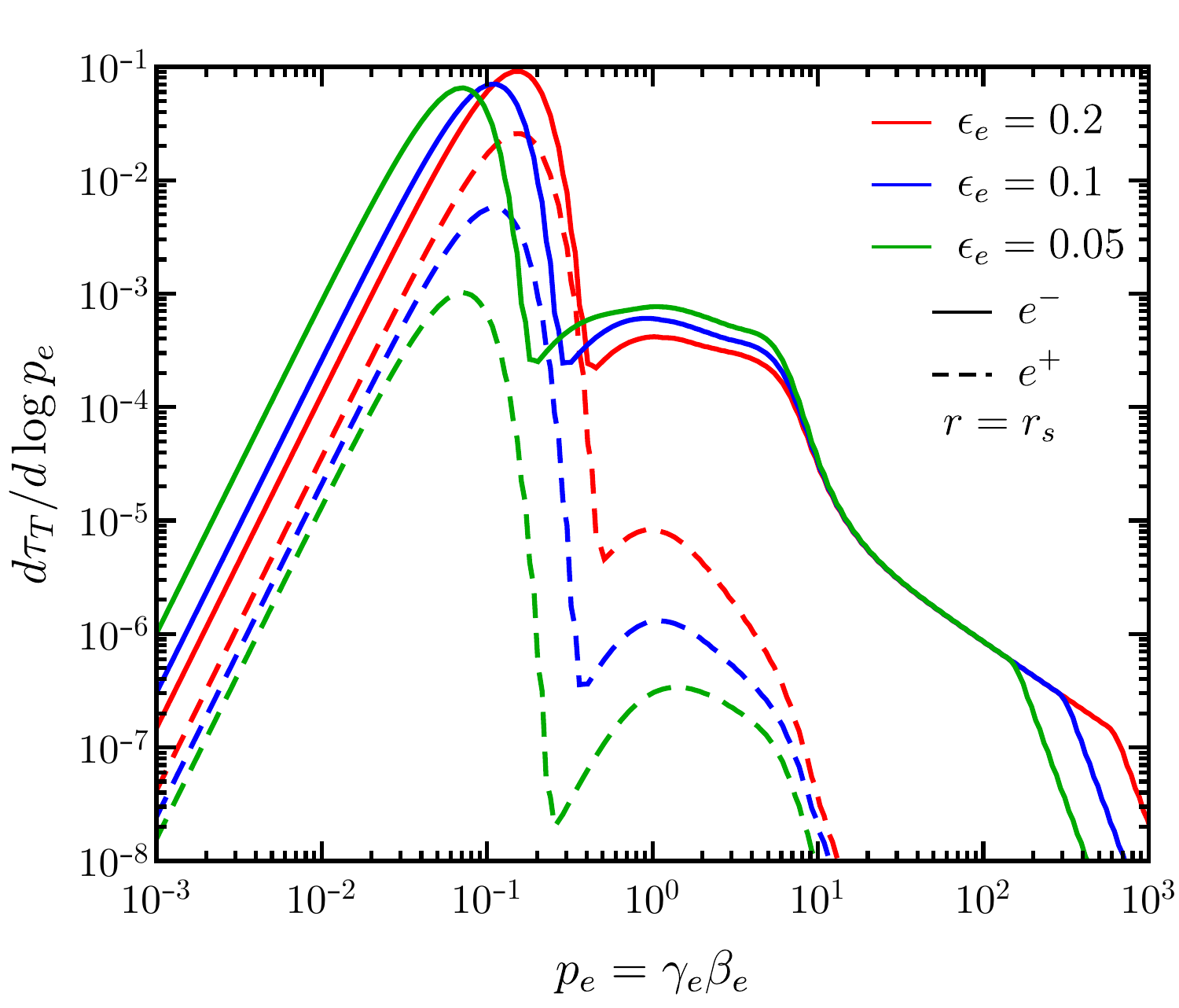}\quad\quad\quad
    \includegraphics[width=0.45\textwidth]{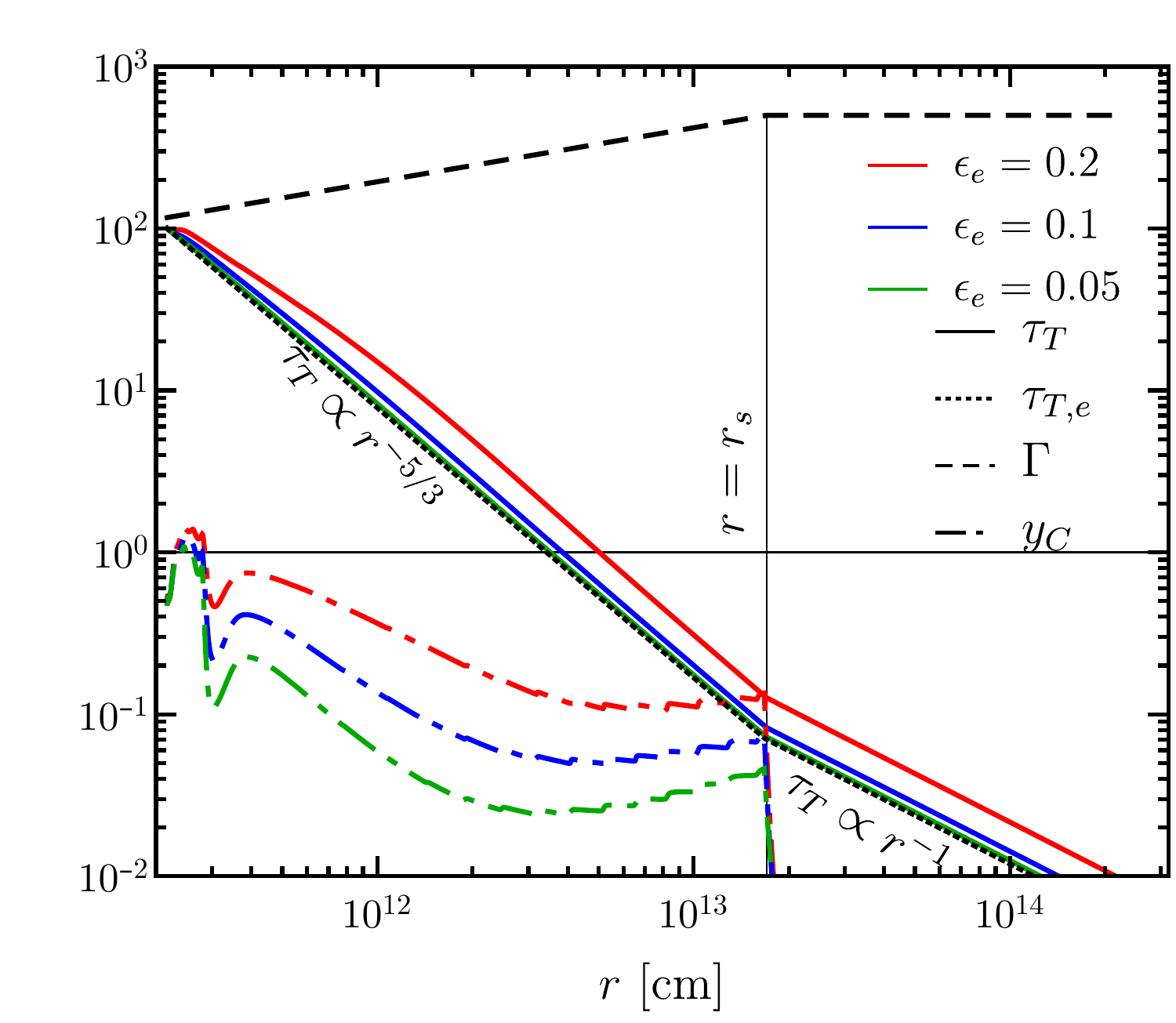}
    \caption{{\bf Top-Left}: Observed steady-state spectrum with injection of power-law electrons commencing at $\tau_{T0}=100$. 
    The final spectrum is obtained when the flow becomes optically thin with $\tau_T\ll1$. The black dashed line shows 
    the synchrotron emission from power-law electrons (without the effects of $\gamma\gamma$-annihilation at the highest 
    energies). 
    {\bf Top-Right:} Spectral slopes where the photon index $\alpha=-2+d\log(EL_E)/d\log E$. 
    {\bf Bottom-Left:} Electron and positron momentum distributions at $r=r_s$ shown using the optical depth. 
    {\bf Bottom-Right}: Radial evolution of the bulk LF $\Gamma$, total optical depth of $\tau_T=\tau_{T,e}+\tau_{T,\pm}$ including 
    that due to produced $e^\pm$-pairs ($\tau_{T,\pm}$), optical depth of baryonic electrons only ($\tau_{T,e}$) if no pairs were 
    produced, and Compton-y parameter of pairs ($y_C$). Magnetic energy dissipation and acceleration of the flow halts at the saturation 
    radius $r=r_s$, beyond which the flow coasts at constant $\Gamma=\Gamma_\infty$. The photospheric radius due to baryonic electrons 
    ($\tau_{T,e}=1$) is extended due to production of $e^\pm$-pairs ($\tau_T=1$).
    }
    \label{fig:diff-epse-spectrum}
\end{figure*}

\begin{figure*}
    \centering
    \includegraphics[width=0.45\textwidth]{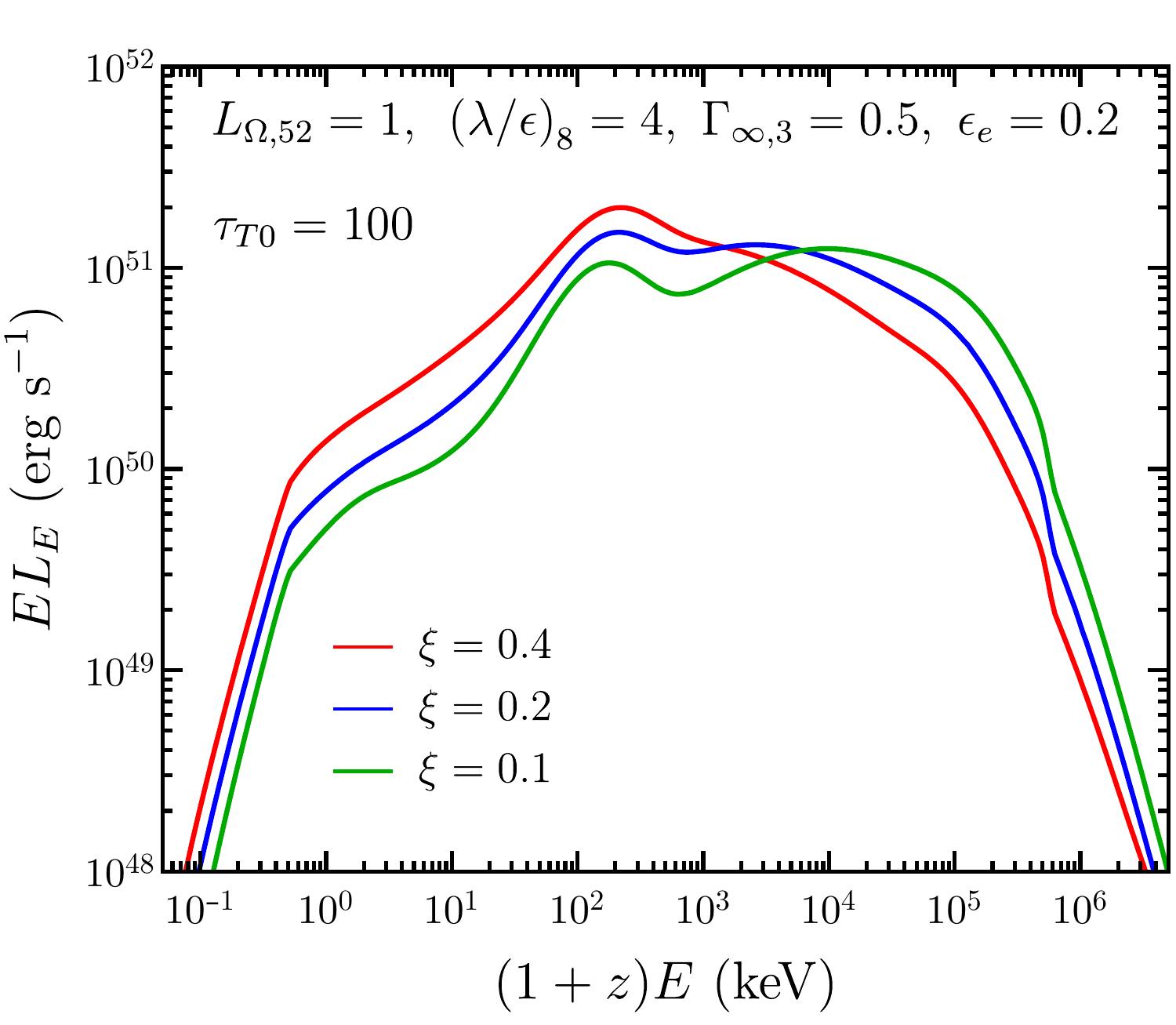}\quad\quad\quad
    \includegraphics[width=0.45\textwidth]{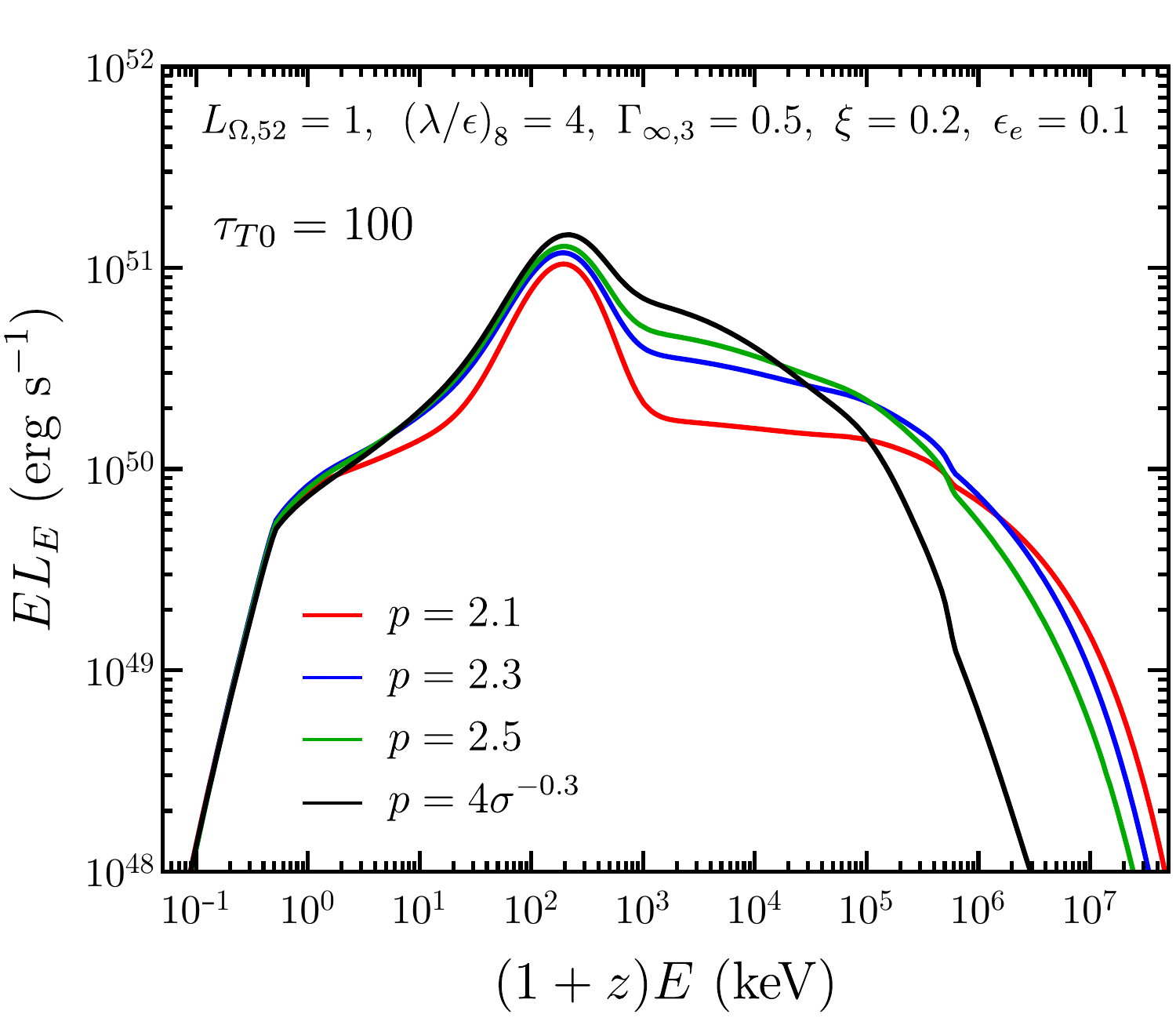}
    \caption{\textbf{Left:} Observed steady-state spectrum from the injection of power-law electrons for different fraction 
    $\xi$ of total incoming electrons accelerated in magnetic reconnection layers.
    \textbf{Right:} Spectrum for different power-law index $p$ of the injected electrons with energy distribution $n'(\gamma_e)\propto\gamma_e^{-p}$.
    }
    \label{fig:diff-xi-diff-p}
\end{figure*}

\subsection{Injection of Power-Law Electrons}\label{sec:scenario-I}

Power law electrons injected with $\gamma_e>\gamma_m = [(p-2)/(p-1)]\langle\gamma_e\rangle_{\rm nth}$, where the last equality is valid for $p>2$ 
which is obtained for $\sigma<10$, emit synchrotron radiation for which the peak of the $\nu F_\nu$ or $EL_E$ synchrotron spectrum occurs at the 
characteristic energy (with $p=4$ when $\sigma=1$ at $r=r_s$ according to our parameterization in Eq.~(\ref{eq:sigma}))
\begin{eqnarray}
    E_m &=& \frac{\Gamma}{1+z}h\nu'_m = \frac{\Gamma}{1+z}\gamma_m^2\fracb{\hbar eB'}{m_ec} \\
    &\approx& \frac{1.3\times10^5}{1+z}\fracb{p-2}{p-1}^2\fracb{\epsilon_e}{\xi}^2
    \frac{L_{\Omega,52}^{1/2}\Gamma_{\infty,3}^{4/3}\fracb{\lambda}{\epsilon}_8^{2/3}}{r_{12}^{5/3}}\,{\rm keV} \nonumber \\
    &\approx& \frac{530}{1+z}\fracb{\epsilon_e}{\xi}^2
    \frac{L_{\Omega,52}^{1/2}}{\Gamma_{\infty,3}^2\fracb{\lambda}{\epsilon}_8}\,{\rm keV}\quad(p=4) \nonumber
\end{eqnarray}
when particles are fast-cooling, i.e. when the characteristic cooling break energy $E_c < E_m$, where
\begin{equation}
    \label{eq:Ec}
    E_c = \frac{36\pi^2}{1+z}\frac{\hbar em_ec^3}{\sigma_T^2}\frac{\Gamma^3}{B'^3r^2} 
    \approx \frac{2.6\times10^{-9}}{1+z}\frac{\Gamma_{\infty,3}^2r_{12}^3}{L_{\Omega,52}^{3/2}\fracb{\lambda}{\epsilon}_8^2}\,{\rm keV}\,,
\end{equation}
and $h=2\pi\hbar$ is the Planck's constant. Another characteristic break in the synchrotron spectrum appears when the emission 
becomes self-absorbed by the emitting electrons. A simple estimate of the self-absorption break energy can be obtained by noticing that 
at $E=E_{\rm sa}$ the synchrotron specific intensity cannot exceed that of a blackbody. We approximate the latter using the Rayleigh-Jeans 
specific intensity, $I_{E'}'^{\rm RJ}(E_{\rm sa}') = (2E_{\rm sa}'^2/h^3c^2)\gamma_e(E'_{\rm sa})m_ec^2$, 
where $\gamma_e(E'_{\rm sa})=(E'_{\rm sa}m_ec/e\hbar B')^{1/2}$ is the LF of electrons radiating at the self-absorption energy. 
The synchrotron specific intensity can be obtained from 
$I_{E'}' \sim (P_{E'}'/4\pi)\xi n'(R/\Gamma)$, where $R/\Gamma$ is the comoving size of the emission region and $\xi n'$ is the number 
density of baryonic electrons that were accelerated into a power-law distribution. The synchrotron spectral power at $E'=E'_{\rm sa}$ is given by
$P_{E'}'(E'_{\rm sa})=P_{E',\rm max}'(E_{\rm sa}'/E_c')^{-1/2}$ for $E'_c<E_{\rm sa}'<E_m'$. The peak spectral power at $E'=E_c'$ can be approximated 
using the total power emitted by a single electron, $P_{\rm syn}'=(4/3)\sigma_Tc\gamma_e^2(B'^2/8\pi)$, at the characteristic synchrotron energy, 
$E'_{\rm syn}=\gamma_e^2e\hbar B'/m_ec$, such that $P_{E',\rm max}'\sim P_{\rm syn}'/E_{\rm syn}' = \sigma_TB'm_ec^2/3eh$. From 
$I_{E'}'^{RJ} = I_{E'}'$, we find the synchrotron self-absorption energy
\begin{equation}\label{eq:E_sa}
    E_{\rm sa} \sim \frac{\Gamma}{1+z}\left(\frac{h^3}{8\pi m_p}\frac{\xi L_\Omega}{\Gamma_\infty}\frac{1}{r^2\Gamma}\right)^{1/3} 
    \approx \frac{1.4}{1+z}\frac{\xi^{1/3}L_{\Omega,52}^{1/3}}{\Gamma_{\infty,3}^{1/9}\lamep_8^{2/9} r_{12}^{4/9}}\,{\rm keV}\,.
\end{equation}
This estimate is only valid when $E_c<E_{\rm sa}<E_m$. 
In addition, it only accounts for the number of baryonic electrons and not the total number of particles that includes the $e^\pm$-pairs, 
and therefore, the true value is slightly higher by a factor $(\tau_T/\tau_{T,e})^{1/3}$, where $\tau_T=\tau_{T,e}+\tau_{T\pm}$ is the total 
optical depth and $\tau_{T,e}$ is the optical depth due to baryonic electrons.

At $E<E_{\rm sa}$ a photon index of $\alpha=1$ is usually assumed. This indeed holds for a uniform emission region, as is assumed in this 
work, and is physically expected in our scenario $(ii)$ for volumetric heating. However, when the particles are heated at a moving front, be it 
a shock or magnetic reconnection front as may be relevant in our scenario $(i)$, then the time they had to cool is proportional to their distance 
from that front, so that beyond a thin cooling layer where the minimal $\gamma_m$ electrons start cooling the electrons become locally essentially 
mono-energetic with an energy inversely proportional to their distance from the front. Once the emission becomes optically thick at $E<E_{\rm sa}$ 
the location of an optical depth of unity from which the photons reach the observer gets closer to the front as $E$ decreases, corresponding to a 
higher temperature $T'\propto E'^{-5/8}$ so that altogether the observed spectral slope becomes $I_{E'}^{\prime\rm RJ}\propto E'^2T'\propto E'^{11/8}$ 
or $\alpha=3/8$ \citep{Granot+00,Granot-Sari-02}. Once the location of optical depth of unity reaches the thin cooling layer where 
$k_BT'\sim\gamma_m m_ec^2 = {\rm const}$, the usual $\alpha =1$ photon index is recovered (corresponding to a second break energy $E_{\rm ac}$, so that 
$\alpha = 1$ at $E<E_{\rm ac}$ while $\alpha=3/8$ at $E_{\rm ac}<E<E_{\rm sa}$.)

In the top-left panel of Fig.~\ref{fig:diff-epse-spectrum}, we show the spectrum in the cosmological rest-frame of the central engine 
for different values of $\epsilon_e$. The spectrum shows a distinct peak at $(1+z)E\approx200\,$keV, which represents the adiabatically cooled 
thermal component. The spectrum below and above this peak energy is shaped by fast-cooling synchrotron emission from power-law electrons, 
as shown by the black dashed line, which peaks at $E=E_m\approx1(1+z)^{-1}\,$MeV for the $\epsilon_e=0.1$ case. For smaller values of $\epsilon_e$ 
the synchrotron peak moves to smaller energies and the normalization of the non-thermal component with respect to the thermal one declines 
while producing a distinct thermal bump. On the other hand, larger values of $\epsilon_e$ result in a two-hump spectrum until the non-thermal 
synchrotron component starts to dominate the spectrum completely. 

The spectrum drops off sharply at two characteristic energies. At low energies near $E=E_{\rm sa}\approx0.5(1+z)^{-1}\,$keV, the synchrotron 
spectrum becomes self-absorbed resulting in a sharp break. At high energies near $(1+z)E=\Gamma m_ec^2\approx0.2\,$GeV, the emission is suppressed 
due to $\gamma\gamma$-annihilation. The position of the high-energy spectral break is affected by the leaky-box prescription adopted in this work, as 
argued in Sec.~\ref{sec:numerical}, and therefore the actual break is expected to occur at a larger energy.

In the top-right panel of Fig.~\ref{fig:diff-epse-spectrum}, we show the spectral slopes by plotting $d\log EL_E/d\log E$, 
where the peak (or local minima/maxima) of the spectra occurs when the different curves cross zero. At energies just above $E_{\rm sa}$, 
the spectrum is dominated by fast-cooling synchrotron emission, and therefore has the expected slope with $L_E\propto E^{-1/2}$. Closer to the 
$EL_E$-peak, the spectrum deviates from this trend and becomes harder below the peak and softer above it. This is due to the predominance of 
the thermal component. However, the peak is not as hard as expected for a Wien spectrum ($EL_E\propto E^4$), the initial condition here. 
Instead, the spectral slope just below the peak is much softer and remains below unity which is observed for a large fraction of GRBs 
\cite[e.g.,][]{Kaneko+06}. At larger energies above the peak, the synchrotron component again tends to 
dominate for which $L_E\propto E^{-p/2}$ when $E>E_m$. In our model, the value of $p$ depends on the magnetization $\sigma$ according 
to Eq.~(\ref{eq:sigma}) and evolves over time, approaching $p=4$ near the end of dissipation at $r=r_s$. 

The particle distribution for both electrons and positrons at $r=r_s$, just before the injection of power-law electron ceases, is shown in the 
bottom-left panel of Fig.~\ref{fig:diff-epse-spectrum} as a function of the dimensionless momentum $p_e=\gamma_e\beta_e$. Since $\xi=0.2$ here, 
the colder baryonic electrons dominate the Thomson optical depth of the flow. However, for larger values of $\epsilon_e$, the fraction of produced 
$e^\pm$-pairs increases and starts to dominate the optical depth. Starting at high momentum, for $p_e>\gamma_m$, with $\gamma_m>100$, the curves reflect the 
distribution of the injected power law electrons that cools via synchrotron emission. The distribution of cooled electrons at $10\lesssim p_e<\gamma_m$ 
reflects their steady-state distributed due to cooling, where the differential number of particles at a given $\gamma_e$ reflects the cooling time 
at that $\gamma_e$, such that $dn=\gamma_edn/d\gamma_e=\gamma_en_e(\gamma_e)\propto t_c(\gamma_e)\propto\gamma_e^{-1}$ which yields 
$n_e(\gamma_e)\propto \gamma_e^{-2}$. In momentum space, $n_e(p_e)=(d\gamma_e/dp_e)n_e(\gamma_e)=(p_e/\gamma_e)n_e(\gamma_e)$, and 
therefore $n_e(p_e)\propto p_e/\gamma_e^3$. For $p_e\gg1$, $p_e\approx\gamma_e$ and so $n_e(p_e)\propto\gamma_e^{-2}$ and 
$dn_e/d\log p_e\propto d\tau_T/d\log p_e\propto p_e^{-1}$. At low $p_e<1$, the particle distribution is a Maxwellian that represents the initial colder baryonic electrons as well as the cooled injected power-law electrons and the produced $e^\pm$-pairs. 
Energy exchange between the cooler baryonic electrons and the injected power-law electrons and produced $e^\pm$-pairs occurs via Coulomb scattering, 
which is included in the numerical code. For larger values of $\epsilon_e$, the mean energy of incoming power-law electrons is also larger, which results 
in the respective Maxwellian distribution having a larger temperature.

For smaller values of $\epsilon_e<0.1$, the total optical depth is dominated by the baryonic electrons as shown in the bottom-right panel of 
Fig.~\ref{fig:diff-epse-spectrum}. When $\epsilon_e$ is increased, more energy is put into the non-thermal component that results in increasing 
the number of produced $e^\pm$-pairs, as evident for the $\epsilon_e=0.2$ case. Due to pair production the photospheric radius is extended 
to slightly larger radii by a factor $(1+\tau_{T,\pm}/\tau_{T,e})^{3/5}$ over the baryonic one given in Eq.~(\ref{eq:r_ph}). For example, 
$\tau_{T,\pm}\approx\tau_{T,e}$ for $\epsilon_e=0.2$ which yields an enhancement in the photospheric radius by a factor $\sim2^{3/5}\approx1.52$. This is 
demonstrated in the figure where the dotted black line shows the radial evolution of the optical depth in the absence of pair-production and the 
solid lines show the total optical depth including $e^\pm$-pairs. After a surge in $\tau_T$ due to the produced pairs, the solid lines display 
similar radial evolution as compared to the black dotted line that follows $\tau_T\propto r^{-5/3}$ for $r<r_s$ and $\tau_T\propto r^{-1}$ for $r>r_s$. 
For all the cases, the Compton-$y$ parameter, $y_C = (4/3)(\langle\gamma_e^2\rangle-1)\tau_T$, which measures the importance of Compton scattering, 
remains smaller than unity since the mean energy of the particles is dominated by the cooler baryonic electrons. As we discuss below, particles 
in this scenario mainly cool via synchrotron emission and Compton scattering is not important.

In the left panel of Fig.~\ref{fig:diff-xi-diff-p}, we show the spectrum for different values of $\xi$, which sets the fraction of the injected electrons 
accelerated into a power law. As a result, $\xi$ affects the mean energy of power-law electrons and consequently $\gamma_m$, where both are 
inversely proportional to $\xi$. The effect of decreasing $\xi$ is similar to that of increasing $\epsilon_e$. Since the number of electrons 
injected into the emission region remains fixed, increasing the mean energy of the distribution also increases the contribution of the non-thermal 
synchrotron component. Consequently, the optical depth due to pair production also increases with increasing $\xi$. 

The right panel of Fig.~\ref{fig:diff-xi-diff-p} shows the effect on the spectrum when the power-law index $p$ of incoming electrons is fixed 
rather than left to vary with the magnetization, as assumed in the model here in Eq.~(\ref{eq:sigma}). As the value of $p$ is lowered, the 
synchrotron spectrum at $E>E_m$ becomes harder since $L_E\propto E^{-p/2}$. By using 2D and 3D PIC simulations \citet{Sironi-Spitkovsky-14} find 
that $p\gtrsim1.5$ for $\sigma\lesssim50$, which means that the synchrotron spectrum can become even harder than shown in the figure if $\sigma$ 
is larger in the emission region. Indeed, this type of spectrum with a quasi-thermal peak and a hard power law component has been observed in, e.g, 
GRB$\,$090902B \citep{Abdo+09}. This type of scenario can also explain the observation of a $31.5[(1+z)/2.82]\,$GeV photon in the central-engine 
frame in this GRB during the prompt emission since the hard synchrotron spectrum extends to GeV energies.

\begin{figure}
    \centering
    \includegraphics[width=0.4\textwidth]{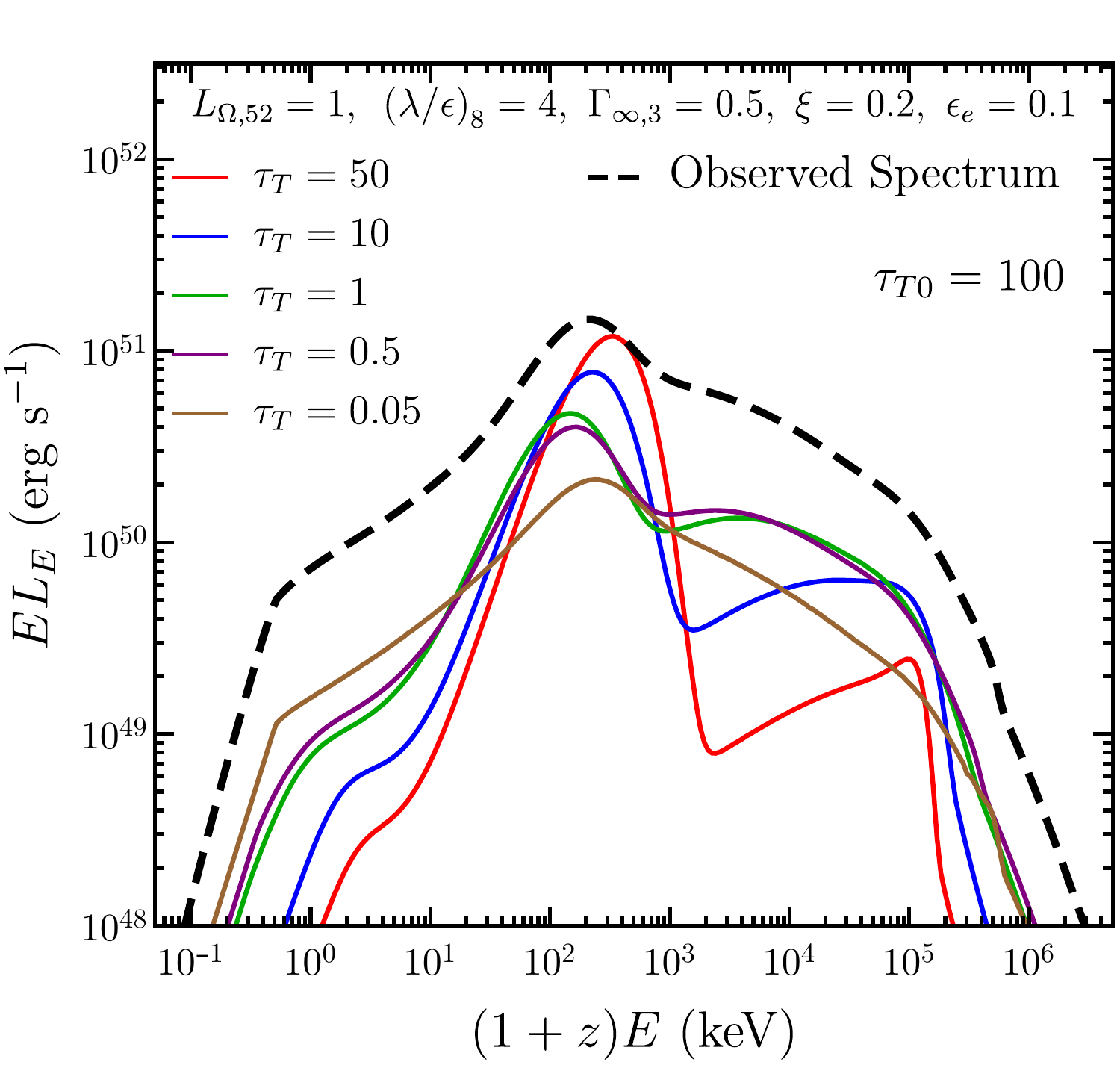}
    \includegraphics[width=0.4\textwidth]{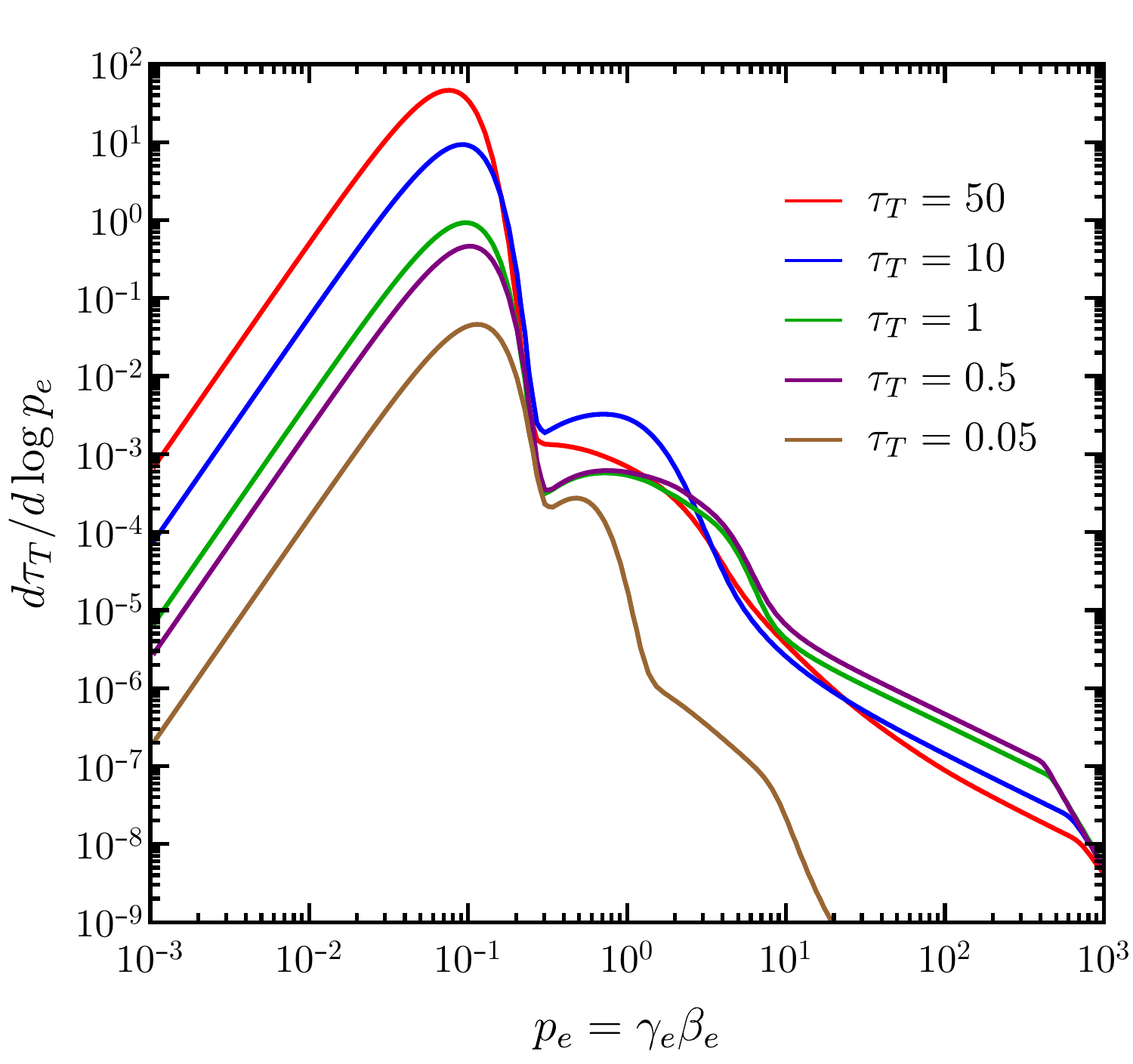}
    \includegraphics[width=0.4\textwidth]{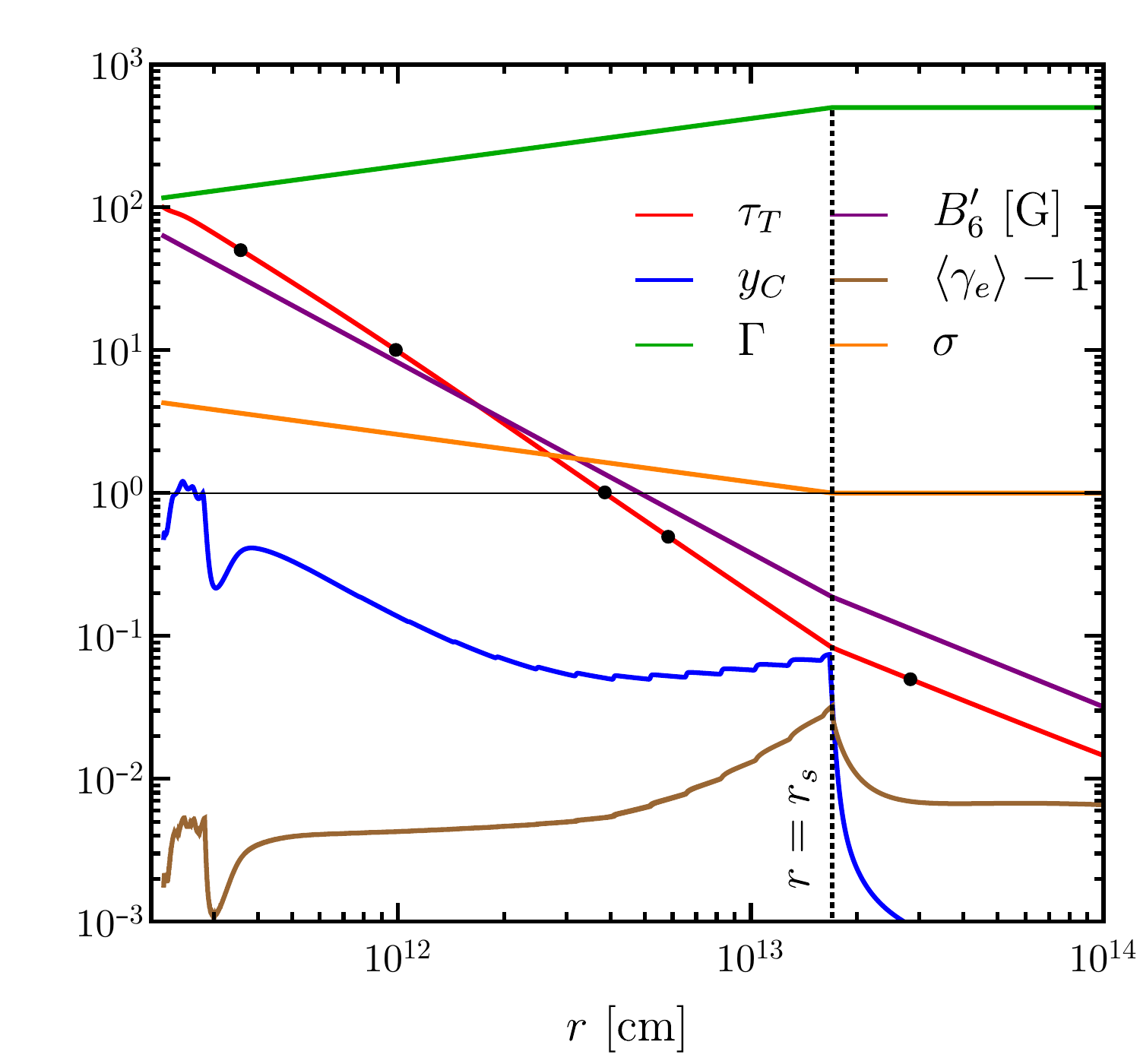}
    \caption{{\bf Top}: Evolution of the spectrum sampled at different total optical depth $\tau_T$ that was emitted over half a dynamical time 
    ($\Delta r/r=1/2$) centered at the radius corresponding to $\tau_T$. 
    The observer only sees the final spectrum, shown using a dashed black line, which is effectively a sum over the optically thin spectra with 
    emission arising from $r>r_{\rm ph}(\tilde\theta)$.
    {\bf Middle}: Evolution of particle distribution that remains dominated by the initial thermal component since $\xi=0.2$ in this case. 
    {\bf Bottom}: Radial evolution of flow parameters with black dots marking the optical depth $\tau_T$ for which the spectra is shown in the 
    top panel.}
    \label{fig:time-res-spec}
\end{figure}

\subsubsection{Radial Evolution of the Spectrum and Particle Distribution}

We present the radial evolution of the spectrum, the corresponding particle distribution, and flow parameters for the case 
with $\epsilon_e=0.1$ in Fig.~\ref{fig:time-res-spec}. The spectrum is obtained for different optical depths, as shown by the 
black dots on the red curve in the bottom-panel of Fig.~\ref{fig:time-res-spec}, and correspondingly different radii where we 
integrate the comoving emissivity over radial extent $\Delta r/r = 1/2$ centered on the radius corresponding to the chosen $\tau_T$. 
The observed steady-state spectrum, shown by the black dashed line, is effectively a sum over the optically thin spectra where the 
radial integration of the comoving emissivity is performed for $r>r_{\rm ph}(\tilde\theta)$. 
At early times, the spectrum is dominated by the initial condition given by the Wein-like 
spectrum from Eq.~(\ref{eq:Wien}). Injection of power law electrons gives rise to the fast-cooling synchrotron spectrum, which 
builds up over time while the thermal peak cools and dilutes due to adiabatic expansion of the outflow. After the flow becomes 
optically thin ($\tau_T<1$), the thermal peak starts to shift to higher energies since the radiation field is no longer adiabatically 
cooled and the thermal peak is simply blue-shifted to higher energies by the increasing $\Gamma$ from its value attained in the 
comoving frame at $\tau_T=1$. High-energy spectrum at energies $E>\Gamma m_ec^2/(1+z)$ is suppressed due to ($\gamma\gamma\to e^\pm$) 
pair-production. The produced $e^\pm$-pairs annihilate and yield a sharply peaked spectral component at $E=\Gamma m_ec^2/(1+z)$ at 
very early times.

\begin{figure*}
    \centering
    \includegraphics[width=0.45\textwidth]{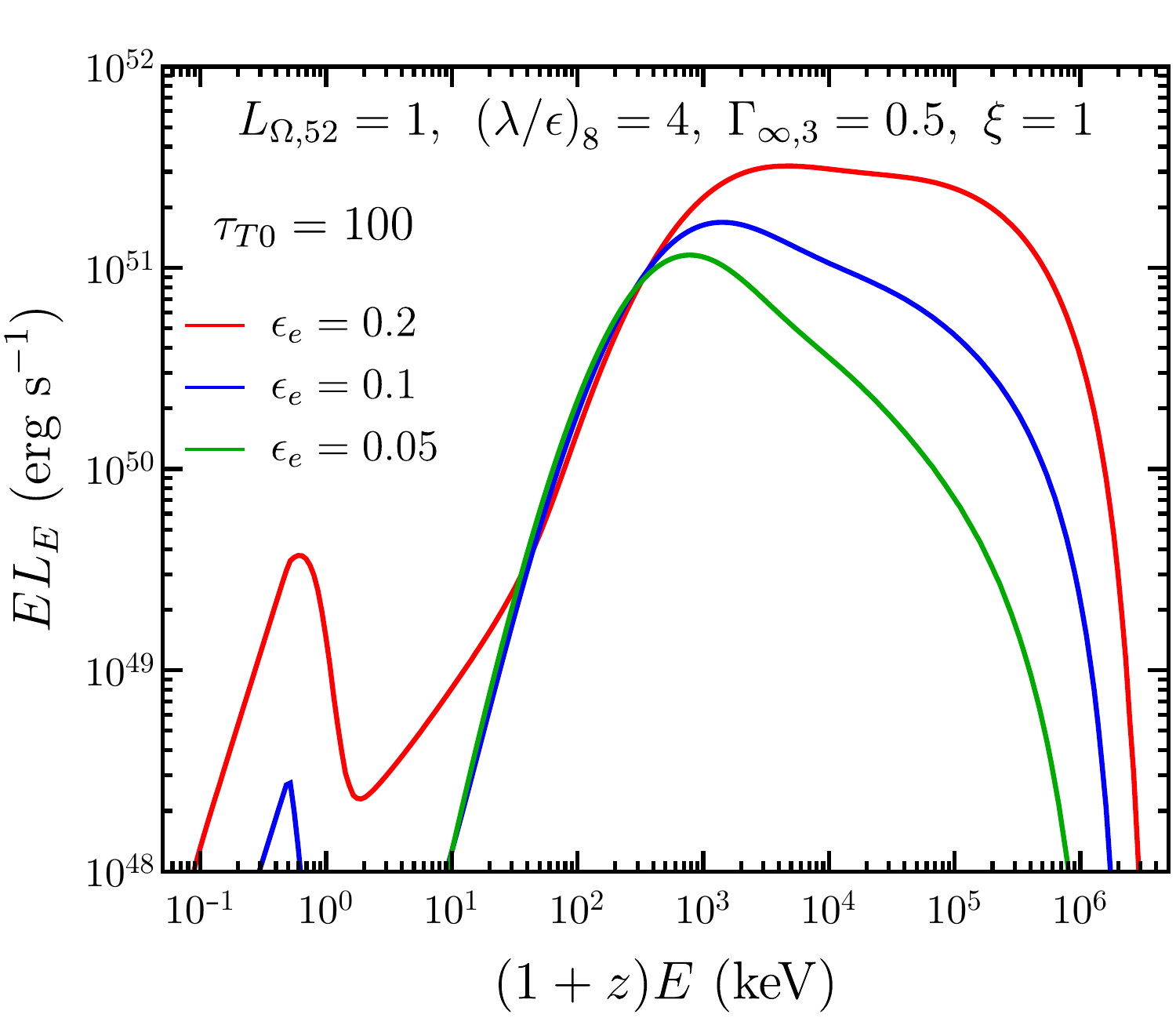}\quad\quad
    \includegraphics[width=0.45\textwidth]{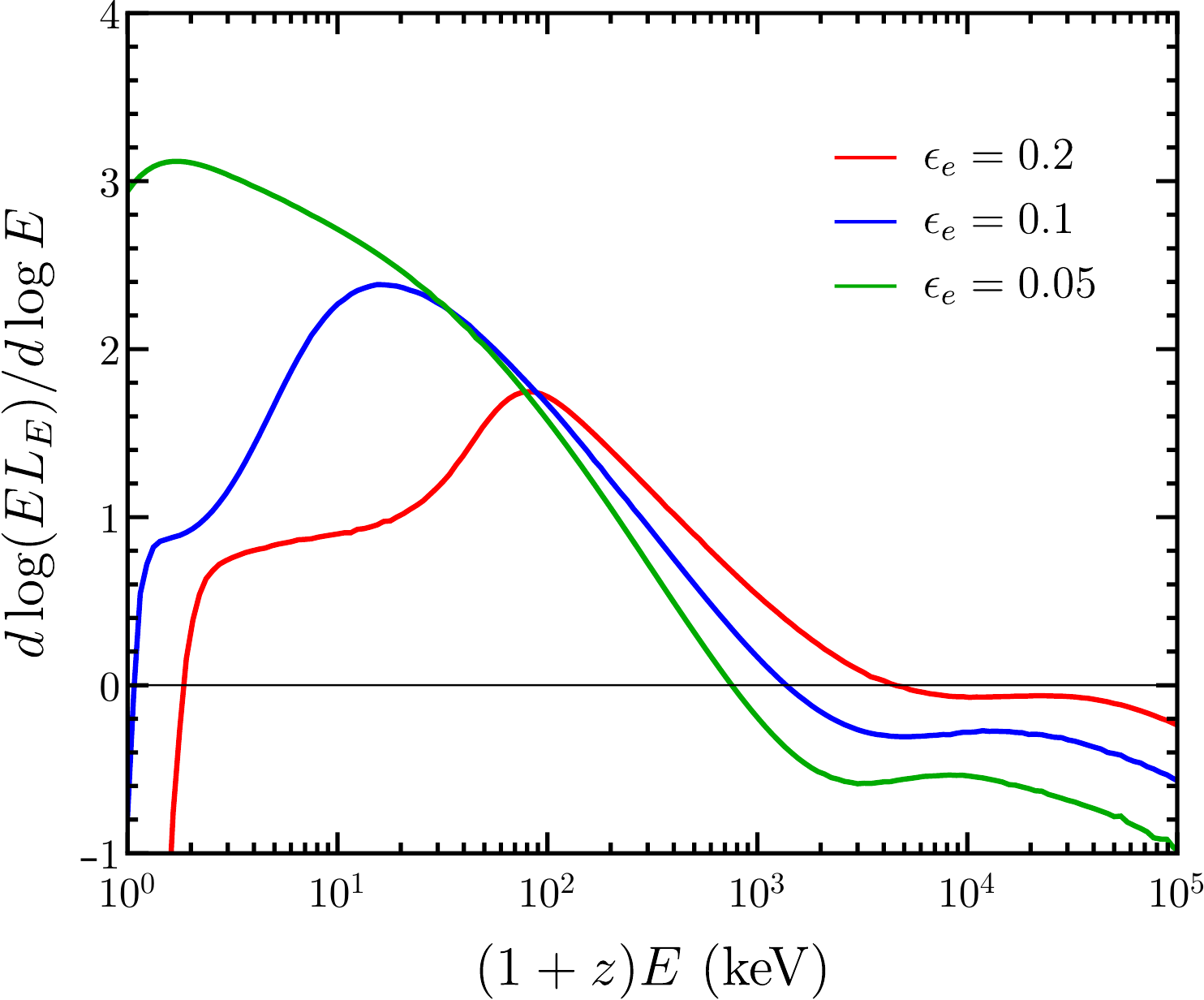}
    \includegraphics[width=0.45\textwidth]{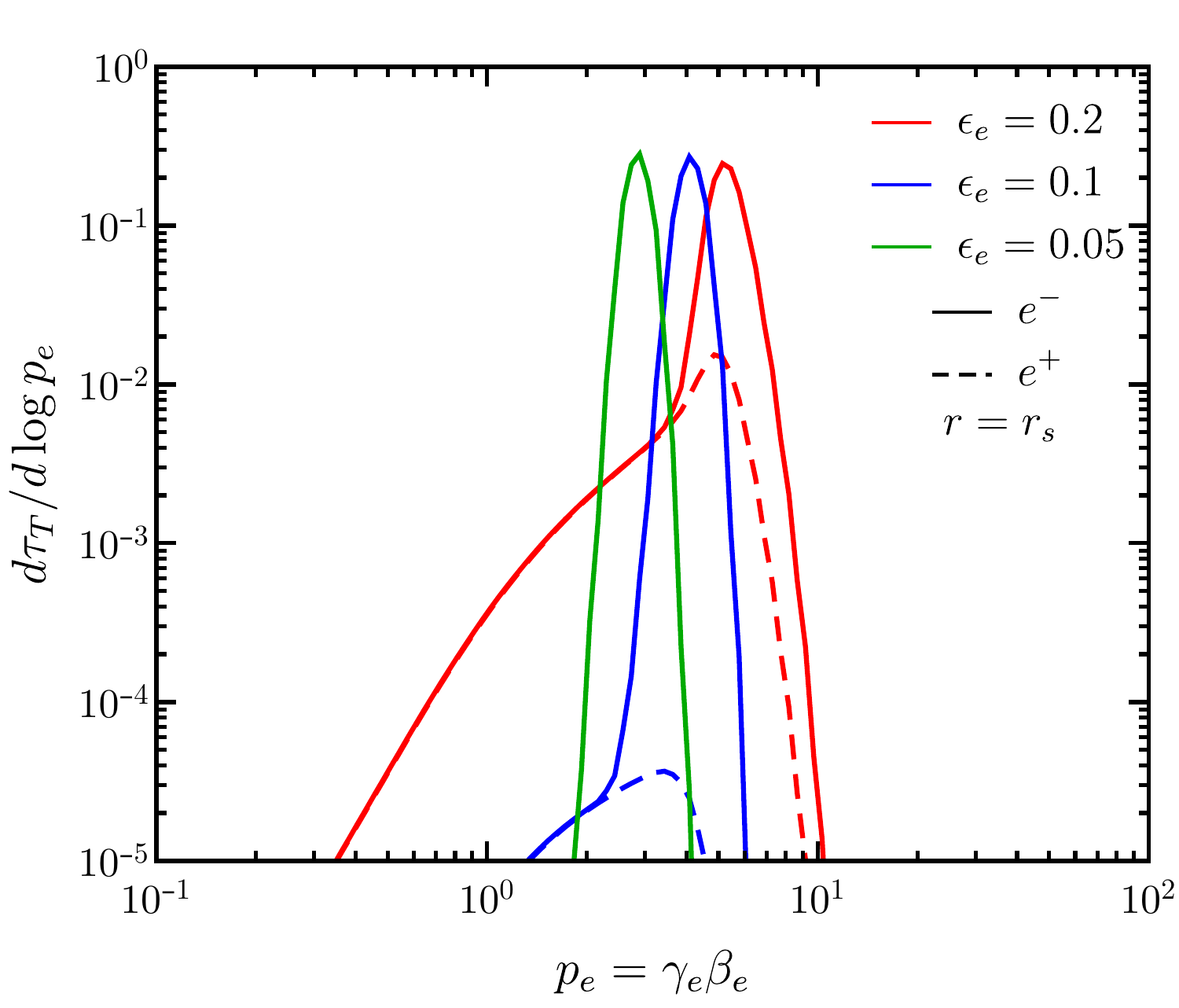}\quad\quad
    \includegraphics[width=0.45\textwidth]{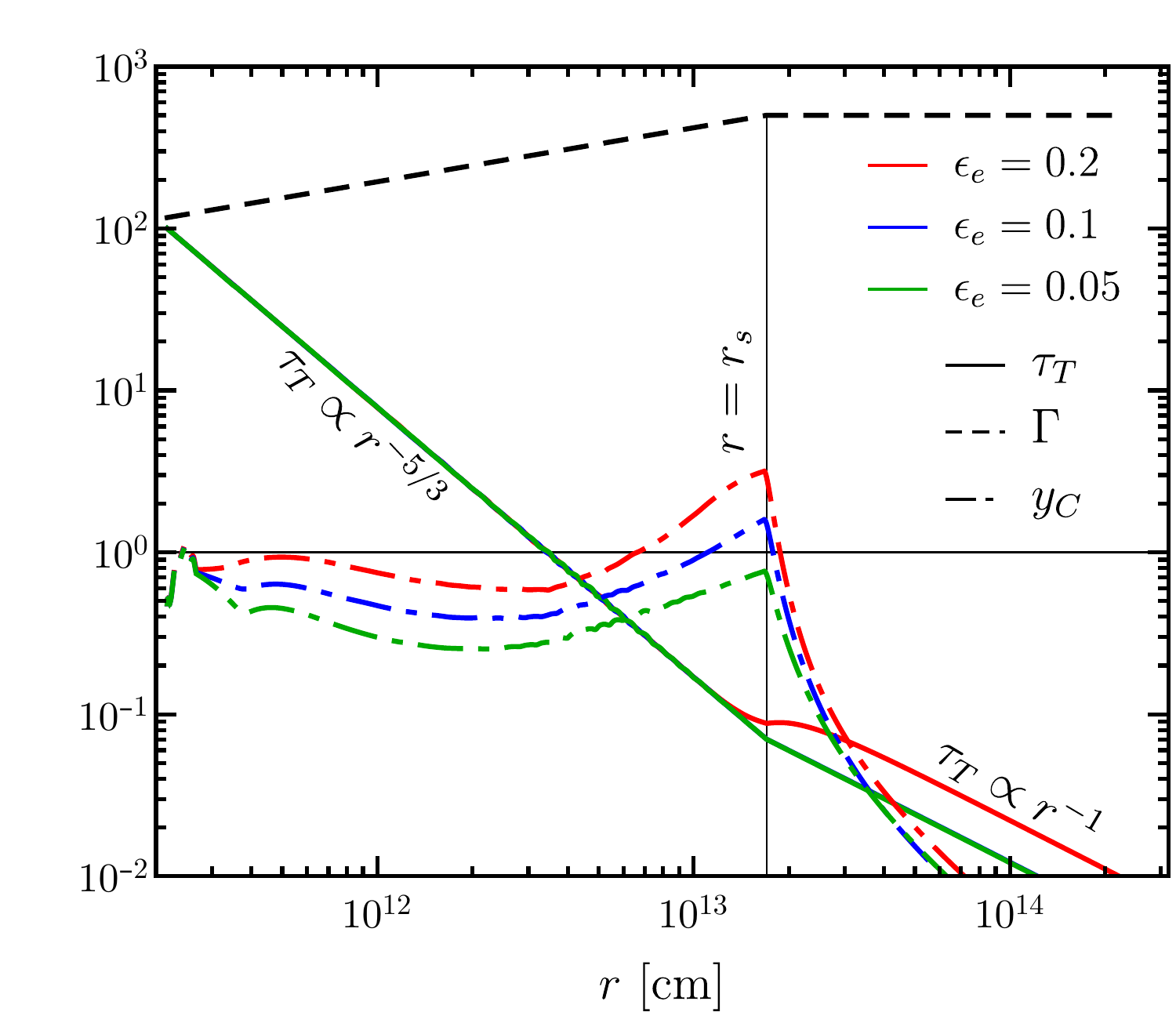}
    \caption{{\bf Top-left:} Observed steady-state spectrum with distributed heating of particles commencing at $\tau_{T0}=100$. 
    The final spectrum is obtained at $\tau_T\ll1$ when the flow is optically thin. For the chosen fiducial 
    parameters the equilibrium optical depth is $\tau_{\rm eq}\approx32$. 
    {\bf Top-right:} Spectral slopes with the photon index given by $\alpha=-2+d\log(EL_E)/d\log E$. 
    {\bf Bottom-left:} Electron and positron particle distribution at $r=r_s$, the radius at which the mono-energetic distribution 
    are expected to be the hottest.  
    {\bf Bottom-right:} Evolution of flow parameters with radius.
    }
    \label{fig:heated-spectrum-diff-epse}
\end{figure*}

In this case $\xi=0.2$, and therefore the initial optical depth is dominated by thermal baryonic electrons. However, copious pair production 
ensues after the injection of power law electrons and $e^\pm$-pairs start to become comparable to the baryonic electrons in optical depth. 
The injection of power law electrons also raises the mean energy per particle $\langle\gamma_e\rangle$, as can be seen from the rightward shift 
of the peak of the thermal particle distribution in the middle panel of Fig.~\ref{fig:time-res-spec} as well as from the radial evolution of 
$\langle\gamma_e\rangle-1$ shown in the bottom panel. However, the Compton-$y$ parameter remains below unity as the rate of heating is insufficient 
to make Comptonization important. 
The cooled power law electrons as well as the produced pairs ultimately join the thermal distribution.

The power-law electrons cool primarily due to synchrotron emission. This can be understood by comparing the magnetic field 
energy density to that of the thermal radiation field. For $r>r_{\tau0}$, where $r_{\tau0}$ is the radius corresponding to 
$\tau_{T0}$ when injection of power-law electrons commences, the comoving energy density of the thermal component is 
$U_{\rm th}'(r)=U_0'(r/r_{\tau0})^{-28/9}=U_0'\tau_{T0}^{-28/15}(r/r_{\rm ph})^{-28/9}$ since the injected energy is no longer 
completely thermalized. Therefore, the thermal component simply adiabatically cools for $r>r_{\tau0}$. The initial 
energy density is given by $U_0'\approx(4\sigma_{\rm SB}/c)[T_{\rm ph}'(r_{\tau0}/r_{\rm ph})^{-7/12}]^4=(4\sigma_{\rm SB}/c)T_{\rm ph}'^4\tau_{T0}^{7/5}$.
The energy density of the magnetic field is given by $U_B'=B'^2/8\pi$, which then yields
\begin{equation}
    \label{eq:UB_Uth}
    \frac{U_B'}{U_{\rm th}'} = 69 \frac{\Gamma_{\infty,3}\lamep_8^{1/5}}{L_{\Omega,52}^{1/5}\tau_{T,e}^{4/15}}\,,
\end{equation}
indicating that power-law electrons mainly cool by synchrotron emission. 
In addition, Compton cooling of injected electrons is suppressed as it occurs in the Klien-Nishina regime for photons with energy above
\begin{equation}
    E = \frac{\Gamma}{(1+z)\gamma_m} m_ec^2 
    = 171\left(\frac{\xi}{\epsilon_e}\right)\frac{r_{12}^{2/3}}{\Gamma_{\infty,3}^{1/3}\lamep_8^{2/3}}\,{\rm keV}\,,
\end{equation}
which suggests that the non-thermal synchrotron component above the thermal peak cannot cool the power-law electrons by inverse Compton 
scattering. 

The injected energy density at a given radius $\tilde r$ is reduced as the flow expands adiabatically, such that 
$dU_{\rm inj}'(r)=dU_{\rm inj}'(\tilde r)(\tilde r/r)^{-28/9}$, 
where the injected energy density between $\tilde r$ and $\tilde r+d\tilde r$ is  
$dU_{\rm inj}'(\tilde r)=(\epsilon_e/2)[dU_{\rm diss}'(\tilde r)/dt']d\tilde r/\Gamma c$. The total injected energy density surviving 
at $r\gg r_{\rm inj}$, where $r_{\rm inj}$ is the radius where energy injection commences, is obtained by integrating over $\tilde r$ that yields
\begin{equation}
    U_{\rm inj}'(r) = \frac{3}{7}\frac{\epsilon_eL_\Omega}{c\Gamma_\infty}\frac{1}{r^2\Gamma} 
    = \frac{9}{7}\epsilon_e\frac{dU_{\rm diss}'}{dt'}\frac{r}{\Gamma c}\,.
\end{equation}
The non-thermal emission will begin to dominate the thermal component when $U_{\rm th}'/U_{\rm inj}'<1$, where
\begin{equation}
    \frac{U_{\rm th}'}{U_{\rm inj}'} = 4\times10^{-2}\frac{L_{\Omega,52}^{7/15}\lamep_8^{14/45}}{\epsilon_e\Gamma_{\infty,3}^{7/9}r_{12}^{7/9}}\,.
\end{equation}
For the fiducial parameters chosen in Fig.~\ref{fig:time-res-spec}, the above condition is not satisfied before dissipation ceases, 
and therefore the non-thermal synchrotron component never fully dominates over the thermal component. The above estimate is strictly 
valid when the flow is optically thick for which the radiation field energy density follows the scaling $U_{\rm th}'\propto r^{-28/9}$. 
Adiabatic cooling of the radiation field stops once the flow becomes optically thin, at which point it only suffers density dilution due 
to the volume expansion but no cooling.

\subsection{Distributed Heating of Particles}\label{sec:scenario-II}
Earlier we explored the scenario where a fraction of the incoming baryonic electrons are directly accelerated into a power law energy 
distribution at magnetic reconnection sites. Here we consider an alternative, where magnetic energy dissipation in the flow, e.g. due 
to MHD instabilities, leads to distributed heating of all electrons 
\citep{Thompson-94, Ghisellini-Celotti-99, Giannios-06, Giannios-Spruit-07, Giannios-08}. 
The comoving energy dissipation rate per unit volume, $dU_{\rm diss}'/dt'$, is given in Eq.~(\ref{eq:dUdt}) out of which only a fraction 
$\epsilon_e/2$ goes into heating the electrons in the emission region, such that the volumetric heating rate is 
$dU_e'/dt' = (\epsilon_e/2)dU_{\rm diss}'/dt'$. Deeper in the flow, at larger optical depths $\tau_T\gg1$, the thermal radiation 
field is the dominant coolant (see Eq.~\ref{eq:UB_Uth}). The continuous heating and simultaneous cooling of particles drives their 
energy distribution to peak at a critical temperature at which point heating is balanced by cooling. The Compton cooling 
rate per unit volume for a thermal electron distribution is given by
\begin{equation}
    \frac{dU_c'}{dt'} = 4n_e'\fracb{k_BT_e'}{m_ec^2}\sigma_TcU_{\rm th}'\,,
\end{equation}
where again we make the simplifying assumption that approximately half of the dissipated energy goes directly towards accelerating the 
flow and the remaining half converts to the thermal radiation field with energy density $U_{\rm th}'$. 
By equating the cooling rate to that of particle heating, $dU_c'/dt' = dU_e'/dt'$, we find the critical temperature at which particles congregate
\begin{equation}\label{eq:T_e,crit}
    k_BT_{e,\rm crit}' 
    = 138\frac{\epsilon_e\Gamma_{\infty,3}^{5/3}r_{12}^{5/3}}{L_{\Omega,52}\lamep_8^{2/3}}\,{\rm keV}\,
    \approx 132\frac{\epsilon_e}{\tau_{T,e}}\,{\rm keV}\,.
\end{equation}
This temperature is smaller at smaller radii or at larger optical depths, however, it cannot become smaller than that of the thermal radiation field. 
Therefore, below an \textit{equilibrium} radius or above the optical depth \citep{Giannios-06}, 
\begin{eqnarray}
    &&r_{\rm eq} = 5\times10^{10}\frac{L_{\Omega,52}^{5/9}\lamep_8^{1/3}}{\epsilon_e^{4/9}\Gamma_{\infty,3}^{8/9}}\,{\rm cm}\\
    &&\tau_{\rm eq} = 133 \frac{\epsilon_e^{20/27}L_{\Omega,52}^{2/27}\lamep_8^{1/9}}{\Gamma_{\infty,3}^{5/27}}\,,
\end{eqnarray}
radiation and particles are in thermal equilibrium. 
Above that radius, electrons fall out of equilibrium and attain a higher \textit{effective} 
temperature (since the distribution becomes narrowly peaked and does not remain Maxwellian) as compared to the thermal radiation field. The details 
of how distributed heating is implemented in the simulation are presented in Appendix~(\ref{sec:part-inject}).

As the flow expands, the energy density of the thermal radiation field declines. This increases the timescale over which particles are cooled 
by Comptonization. Particles are also cooling due to adiabatic expansion, the timescale for which is (see Appendix~(\ref{sec:part-inject})) 
$t_{\rm ad}' = (3/7) r/\Gamma c\propto r^{2/3}$ for $r<r_s$. The Compton cooling timescale is $t_c'=3m_ec/4\sigma_T\gamma_e U_{\rm th}'$, 
for particles with LF $\gamma_e$, where $U_{\rm th}'=U_0'(r/r_{\tau0})^{-28/9}$ and $r_{\tau0}$ is the radius corresponding to $\tau_{T0}$ where 
heating of particles commences. Comparison of the two timescales yields
\begin{equation}
    \frac{t_c'}{t_{\rm ad}'} = \frac{7}{4} \frac{m_ec^2}{\sigma_T U_{\rm th}'\gamma_e}\frac{\Gamma}{r}
    \approx \frac{5\times10^{-2}}{\gamma_e} \frac{r_{12}^{22/9}\Gamma_{\infty,3}^{22/9}}{L_{\Omega,52}^{22/15}\lamep_8^{44/45}}\,,
\end{equation}
which suggest that particles cool predominantly via Comptonization.

\begin{figure}
    \centering
    \includegraphics[width=0.4\textwidth]{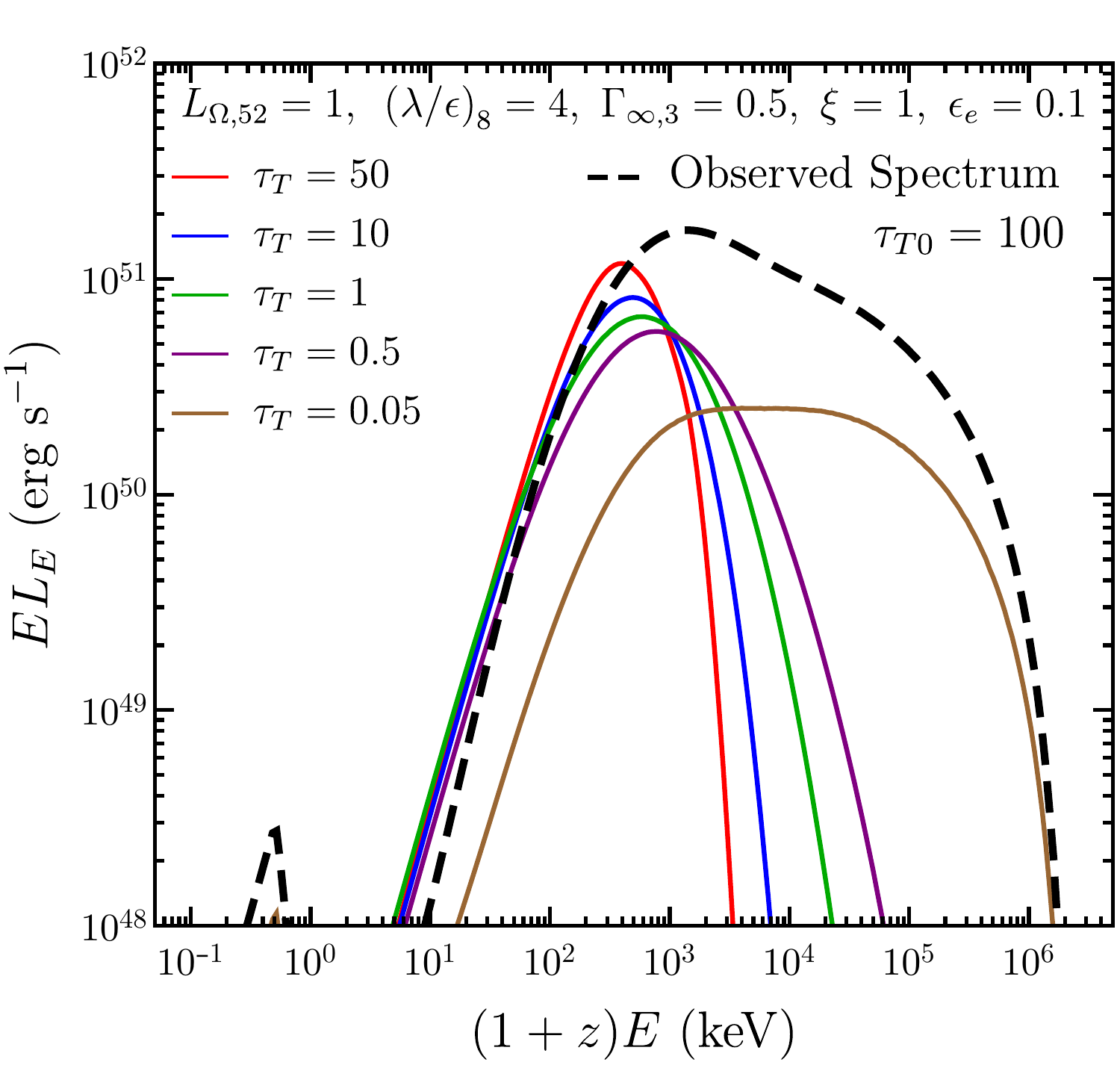}
    \includegraphics[width=0.4\textwidth]{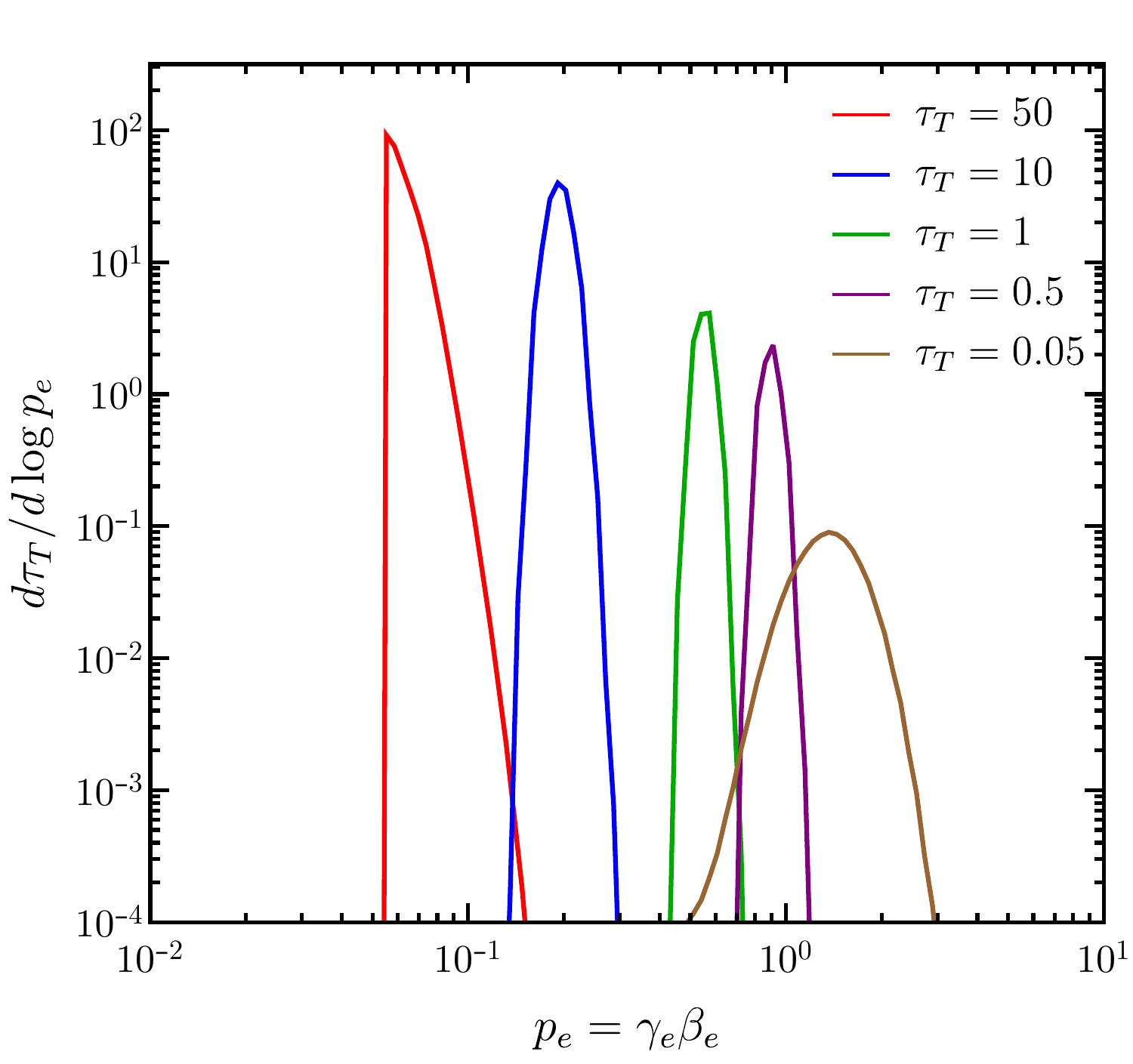}
    \includegraphics[width=0.4\textwidth]{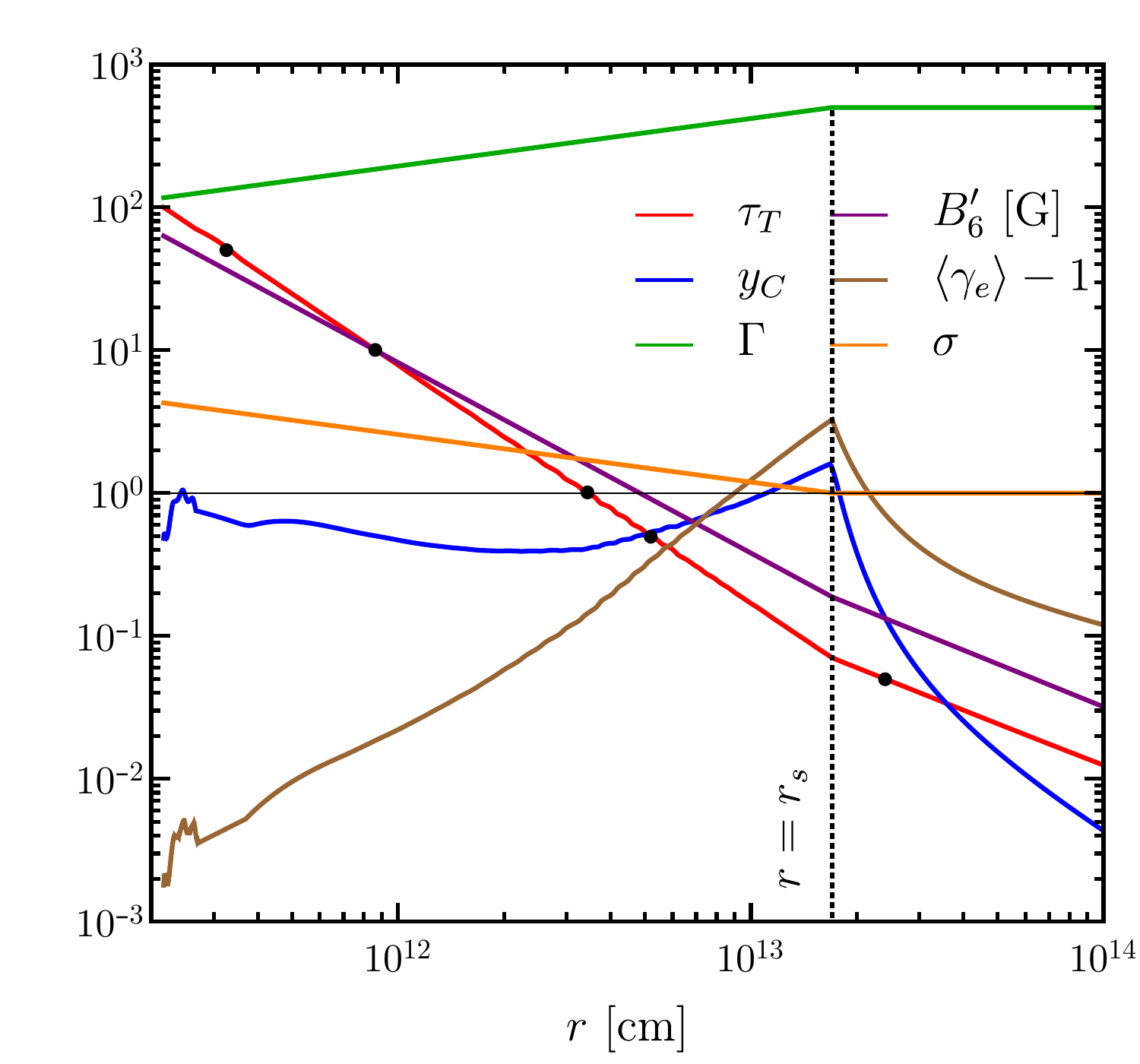}
    \caption{{\bf Top:} Evolution of the spectrum with radius for the distributed heating scenario, obtained at different optical depths $\tau_T$ 
    where the spectrum was emitted within half a dynamical time around the radius corresponding to $\tau_T$. The observed spectrum, shown using a 
    black dashed line, is effectively a sum over the optically thin spectra with emission arsing from $r>r_{\rm ph}(\tilde\theta)$. 
    {\bf Middle:} The corresponding particle distribution that remains sharply peaked at the critical momentum where particle heating and cooling are 
    balanced.  
    {\bf Bottom:} Evolution of flow parameters. Black dots mark the optical depth for which spectra and particle distribution are shown.}
    \label{fig:time-resolved-HEATED}
\end{figure}

\begin{figure*}
    \centering
    \includegraphics[width=0.4\textwidth,height=0.42\textwidth]{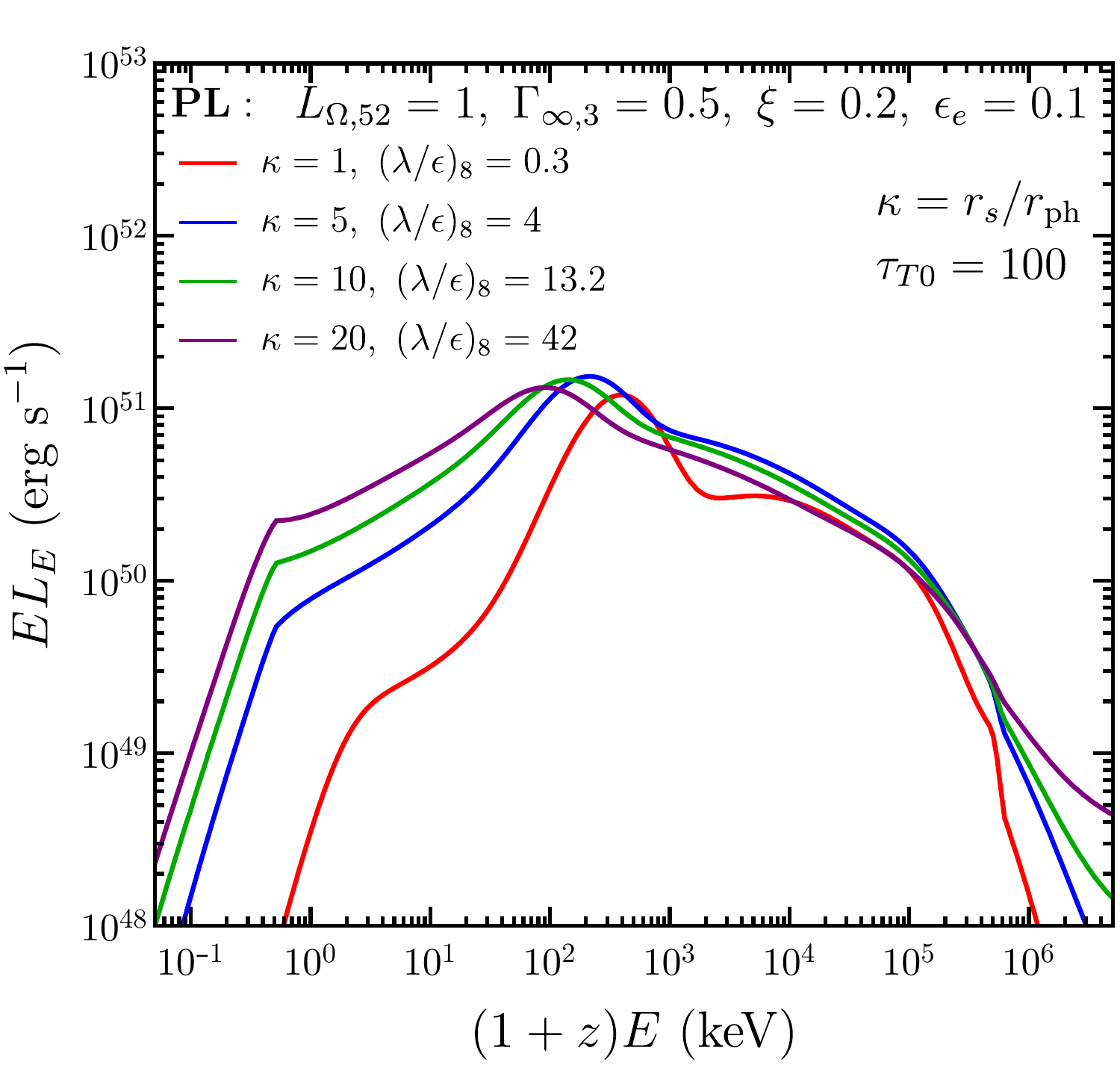}\quad\quad\quad
    \includegraphics[width=0.4\textwidth,height=0.42\textwidth]{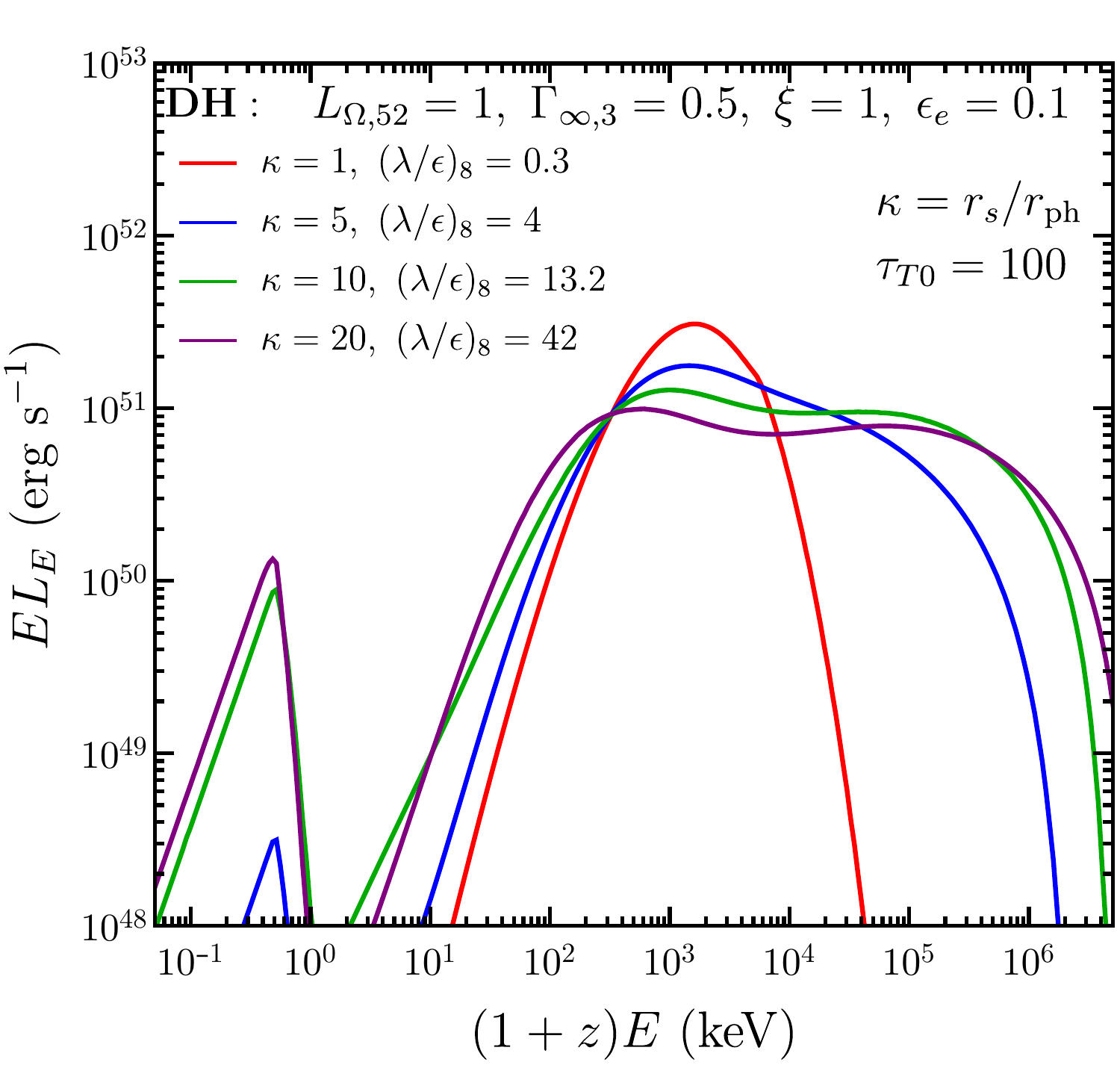}
    \includegraphics[width=0.4\textwidth,height=0.42\textwidth]{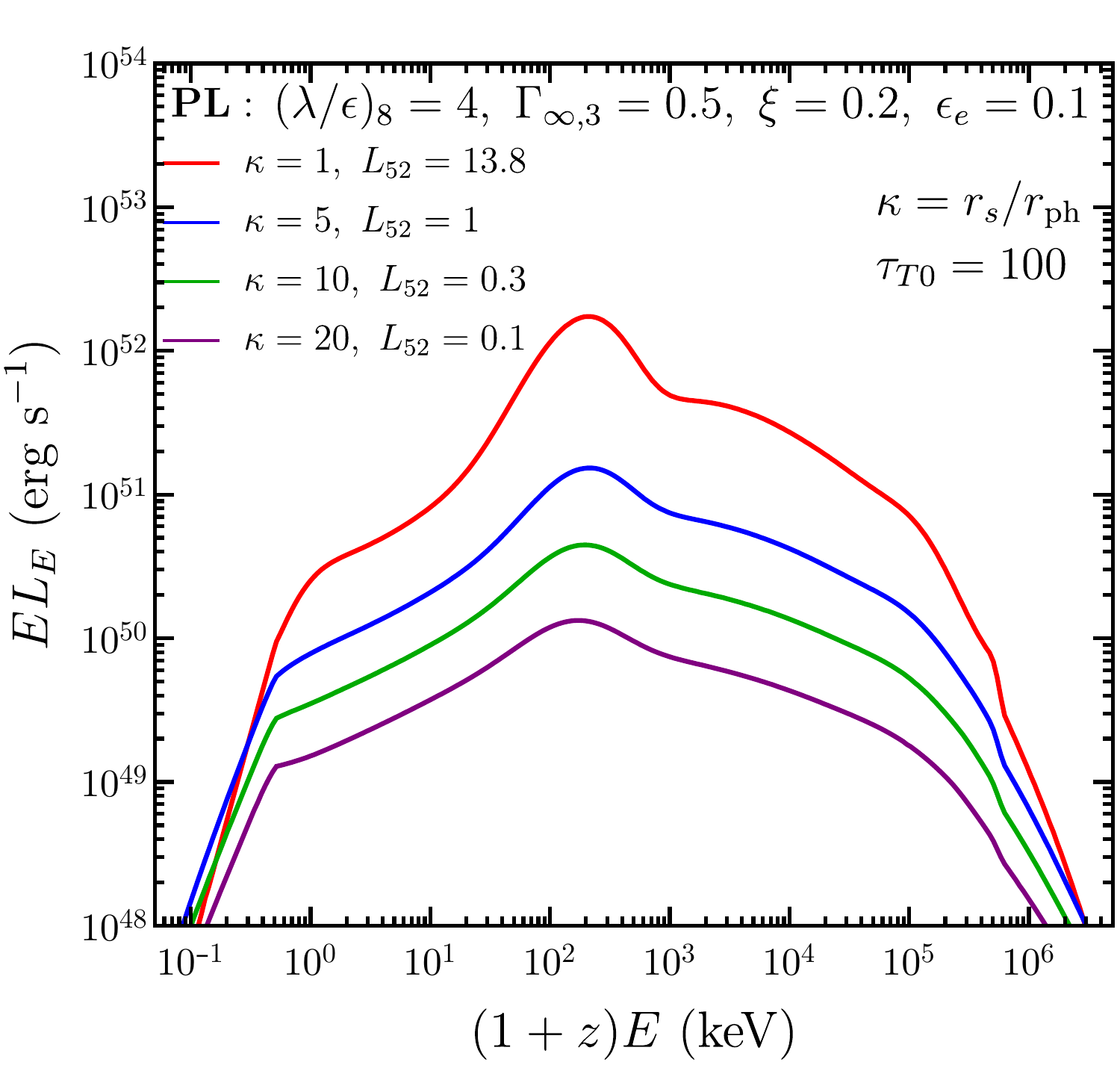}\quad\quad\quad
    \includegraphics[width=0.4\textwidth,height=0.42\textwidth]{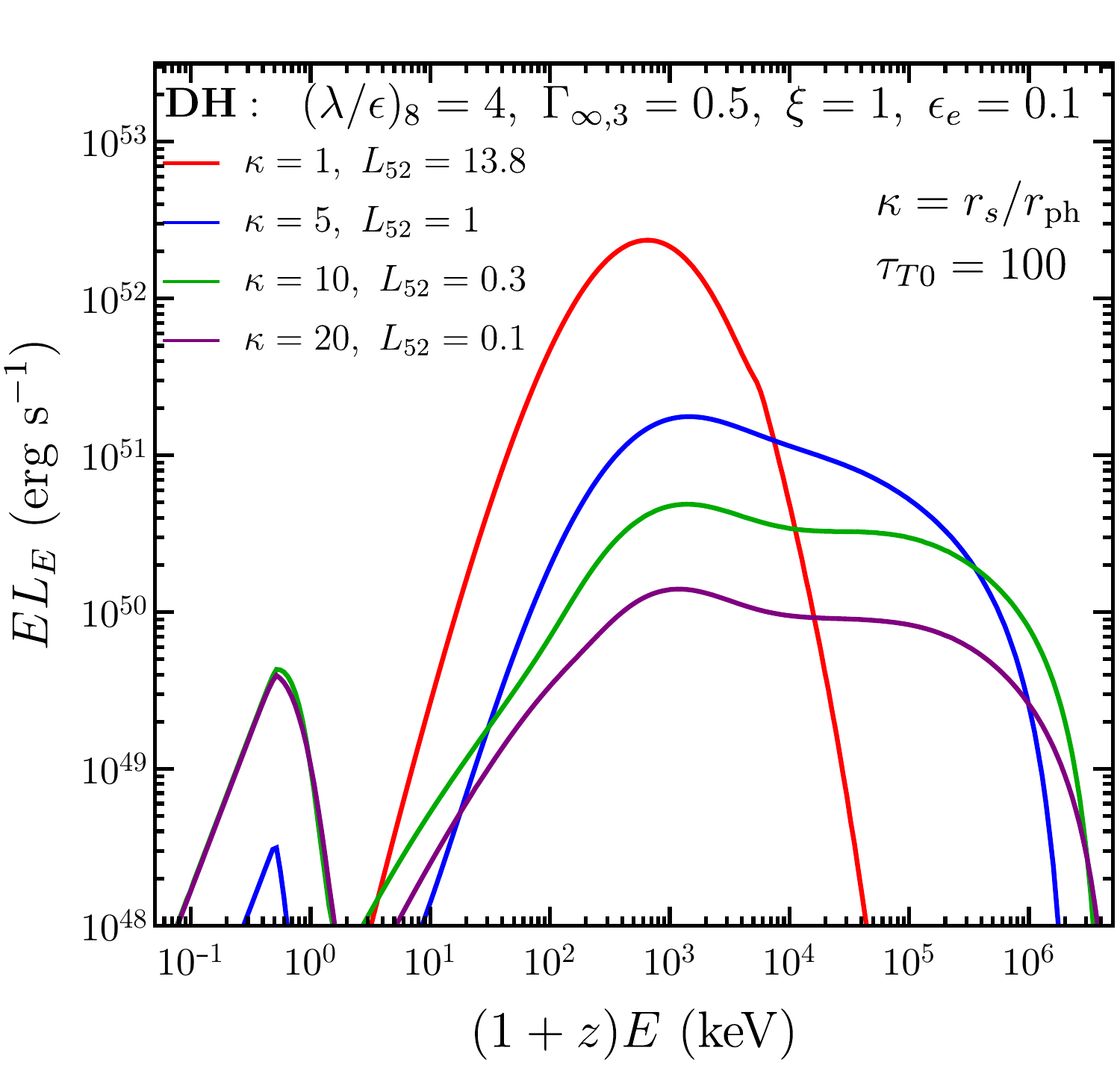}
    \includegraphics[width=0.4\textwidth,height=0.42\textwidth]{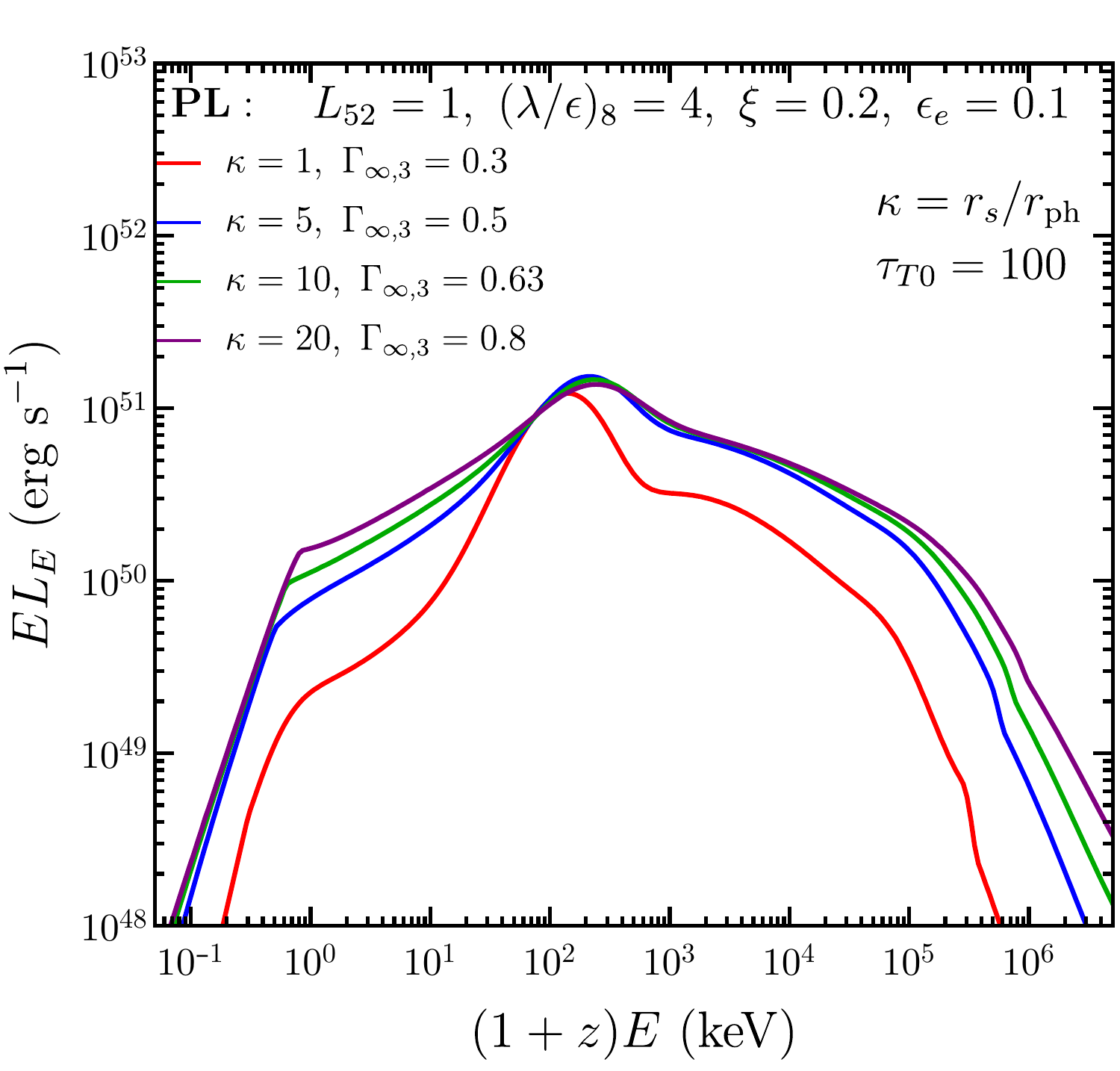}\quad\quad\quad
    \includegraphics[width=0.4\textwidth,height=0.42\textwidth]{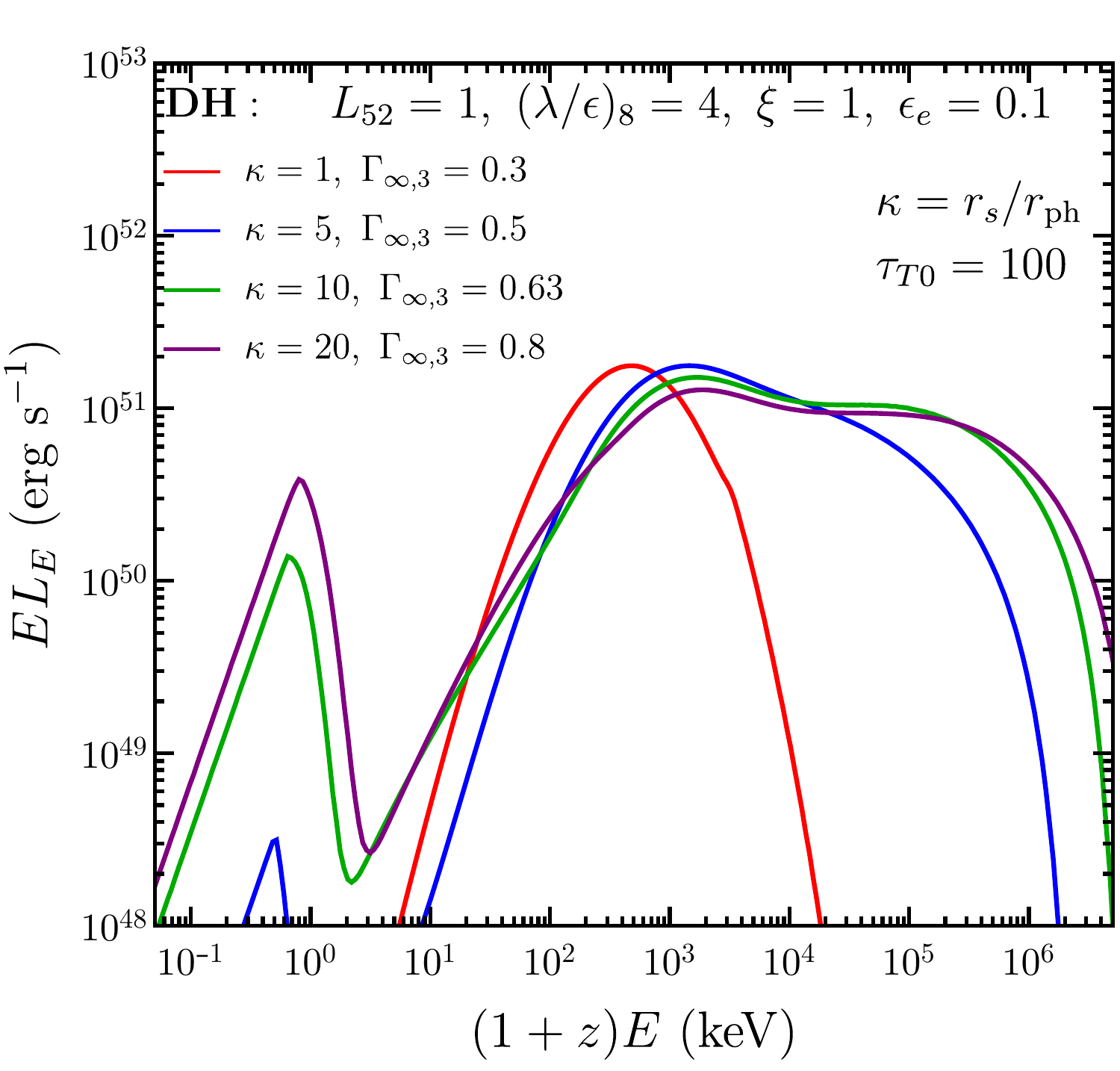}
    \caption{Spectra for the two heating scenarios, power-law particle injection (left) and distributed heating (right), 
    shown for different values of $\kappa=r_s/r_{\rm ph}$ and for correspondingly different values of $\lamep_8$ (top), $L_{52}$ (middle), 
    and $\Gamma_{\infty,3}$ (bottom) according to Eq.~(\ref{eq:kappa}) while the remaining two parameters are kept fixed.  
    }
    \label{fig:diff-kappa}
\end{figure*}

In the top-left panel of Fig.~\ref{fig:heated-spectrum-diff-epse}, we present the observed steady-state spectrum for a heated flow for different values of 
$\epsilon_e$. For the chosen fiducial parameters, $\tau_{\rm eq}\approx32$, and therefore the condition for thermal equilibrium 
holds for $\tau_{T0}=100$, the initial baryonic electron Thomson optical depth at which the simulation is initialized. 
In all cases, due to multiple Compton scattering, the spectrum extends smoothly to high energies above the adiabatically cooled thermal peak that 
appears at $(1+z)E\sim1\,$MeV. This energy is higher due to Comptonization than that expected from adiabatic cooling which freezes at the photospheric 
radius. These results are consistent with that shown in \citet{Giannios-06,Giannios-Spruit-07,Giannios-08} who used a Monte-Carlo code without 
pair cascade effects. The effect of increasing $\epsilon_e$ is to put more energy into the non-thermal Comptonized spectral component and to make 
the spectrum harder above the thermal peak. In addition, for larger $\epsilon_e$ a pronounced peak around $(1+z)E\approx0.5\,$keV develops due to 
self-absorbed synchrotron emission from mildly relativistic electrons.

The low-energy spectral index, as shown in the top-right panel of Fig.~\ref{fig:heated-spectrum-diff-epse}, becomes softer with increasing $\epsilon_e$. 
It only approaches the photon index $\alpha=-2+d\log(EL_E)/d\log E\sim-1$ typically observed in prompt GRB emission for $\epsilon_e\gtrsim0.2$, and below that 
the low-energy spectrum appears to be too hard. This can be understood by looking at the evolution of the Compton-$y$ parameter shown in the bottom-right 
panel of Fig.~\ref{fig:heated-spectrum-diff-epse}. For larger $\epsilon_e$, $y_C$ is also larger and substantially exceeds unity in the $\epsilon_e=0.2$ 
case. This results in the Comptonization of the softer synchrotron photons towards the thermal peak which softens the low-energy spectrum. For smaller 
$\epsilon_e$, $y_C$ remains below unity and the soft synchrotron photons are not efficiently Comptonized to higher energies, leading to harder low-energy 
spectral slopes.

The optical depth in all cases remains unaltered from the trend expected for baryonic electrons, which suggests that pair-production is mostly insignificant 
in the cases shown here. This can also be seen in the bottom-left panel of Fig.~\ref{fig:heated-spectrum-diff-epse} where the optical depth is dominated 
by baryonic electrons. Only for $\epsilon_e=0.2$, copious pair-production ensues at $r\sim r_s$ as the high-energy spectrum exceeds the pair-production 
threshold due to $y_C>1$. The particle distributions are sharply peaked at the momentum where heating and cooling of particles are in balance. This is in 
stark contrast with the particle distribution in the scenario with power-law electron injection.

Since $e^\pm$-pairs are subdominant and the optical depth is dominated by the baryonic electrons, the dimensionless momentum of electrons, $p_e=\gamma_e\beta_e=(\gamma_e^2-1)^{1/2}$, at which they congregate after the heating commences at 
$r>r_{\tau0}$, can be obtained by comparing their heating and Compton cooling rates. The cooling rate for a mono-energetic distribution is 
$dU_c'/dt' = (4/3)\sigma_Tcp_e^2n_e'U_{\rm th}'$, where $U_{\rm th}'=U_0'(r/r_{\tau0})^{-28/9}$ is the energy density of the adiabatically cooled 
thermal radiation. Then, heating and cooling balance yields 
\begin{equation}\label{eq:p_e-at-r_s}
    p_e(r) = 2.6\frac{\epsilon_e^{1/2}\Gamma_{\infty,3}^{11/9}r_{12}^{11/9}}{L_{\Omega,52}^{11/15}\lamep_8^{22/45}}
    = 2.6\frac{\epsilon_e^{1/2}}{\tau_{T,e}^{11/15}}\,.
\end{equation}
Note the above estimate is different from that derived in Eq.~(\ref{eq:T_e,crit}) in two respects. First, it assumes monoenergetic particles and not 
a Maxwellian distribution, where the former is relevant here. Second, for the cooling rate it assumes the adiabatically cooled energy density 
of the radiation field normalized at $r=r_{\tau0}$, whereas the scaling of $U_{\rm th}'\propto r^{-7/3}$ is assumed for Eq.~(\ref{eq:T_e,crit}) since 
the energy given to particles is assumed to be completely thermalized, which results in a shallower decay profile for the radiation field 
energy density. This estimate for $p_e$ strictly assumes that $e^\pm$-pairs are subdominant, but they may become important for some model parameters in which case 
the above estimate will not hold. For the fiducial parameters in Fig.~\ref{fig:heated-spectrum-diff-epse}, the electrons attain a maximum 
$p_{e,\rm max}\simeq5$ at $r=r_s$ when $\epsilon_e=0.1$, as shown in the bottom-left panel of the figure.

The radial evolution of the spectrum and particle distribution, along with that of the flow parameters, for the case of $\epsilon_e=0.1$ is presented 
in Fig.~\ref{fig:time-resolved-HEATED}. 
The initial spectrum at large optical depths is dominated by the thermal component which adiabatically cools and dilutes as the flow expands. 
Meanwhile, continuous dissipation in the flow heats up the baryonic electrons, as evident from the rightward shift of the narrowly peaked particle 
distribution (middle panel) as well as from the rising $\langle\gamma_e\rangle-1$ (bottom panel). This leads to gradual broadening of the Wien distribution 
as well as the shift of the thermal peak to higher energies. The high-energy spectrum only develops when the flow has become sufficiently optically thin 
\citep[also see, e.g., Fig. 1 of ][]{Giannios-08}. Heating of the flow terminates at $\tau_T=0.08$, and therefore, the particle distribution at $\tau_T=0.05$ 
lacks a sharp peak that is expected due to the balance between heating and cooling. Notice that the particle distribution shown in the middle panel of Fig.~\ref{fig:time-resolved-HEATED} is the instantaneous distribution at a given $\tau_T$, and the corresponding spectrum is integrated over 
$\Delta r/r = 1/2$ while centered at the radius corresponding to the chosen $\tau_T$.

A consequence of mildly relativistic mono-energetic electrons is the emergence of a self-absorbed cyclo-synchrotron peak at $(1+z)E\approx0.5\,$ keV. 
In the earlier scenario of power-law electron injection, the spectrum showed a break around the same energy rather than a narrow peak. This signature is a 
potential discriminant between the two scenarios, and prompt GRB observations in soft X-rays should be able to distinguish between the two particle heating 
mechanisms.  

\section{Parameter space study}\label{sec:Params-Study}

We consider here different outflow parameters to see their effect on the final spectrum. From Eq.~(\ref{eq:r_s}) and (\ref{eq:r_ph}), we see 
that a change in the model parameters is reflected in the relative position of the saturation radius with respect to the photospheric radius, 
which is parameterized by their ratio,
\begin{equation}\label{eq:kappa}
    \kappa\equiv\frac{r_s}{r_{\rm ph}} = 17\frac{\Gamma_{\infty,3}^3\lamep_8^{3/5}}{L_{52}^{3/5}}\,.
\end{equation}
Therefore, it is more intuitive to scale a given model parameter, while the other two parameters are kept fixed, using the ratio $\kappa$ with a clear 
expectation. If $\kappa$ is much larger than unity, a large fraction of the dissipated energy will be injected into the emission region when it is optically thin. 
This will result in a predominantly non-thermal spectrum. On the other hand, when $\kappa\approx1$, the spectrum should have a pronounced quasi-thermal component 
since most of the energy was dissipated below the photosphere that led to its thermalization.

In Fig.~\ref{fig:diff-kappa}, we present a survey of the model parameter space by showing the spectra for different values of $\kappa$ for the two 
particle heating scenarios. For both scenarios, the $\kappa=1$ spectrum is dominated by a quasi-thermal component, albeit with low-energy spectral slope 
much softer than a pure thermal spectrum. The spectrum in the distributed 
heating scenario is particularly interesting as it shows a more narrowly peaked spectral component. This type of spectra have been observed in a small number 
of GRBs, e.g. GRB~990123 \citep{Briggs+99}, GRB~090902B \citep{Abdo+09}, GRB~130427A \citep{Ackermann+14}, which would suggest that dissipation in 
these bursts was mostly sub-photospheric with minimal energy dissipated in the optically thin parts of the flow. 
The typical spectrum of GRBs is 
non-thermal that can be explained when $\kappa>1$ and for which significant fraction of the energy is dissipated in the optically thin parts of the flow. 
In some bursts, e.g. GRB 110731A \citep{Ackermann+13}, a time-resolved analysis finds a quasi-thermal spectrum for the initial few pulses followed by 
non-thermal spectrum for the later pulses. This kind of spectral evolution can be explained if $\kappa\approx1$ for the initial few pulses that were 
emitted from a smaller radius compared to the pulses that arrived later with $\kappa>1$ that were emitted from a relatively larger radius (or alternatively 
corresponding to a larger source variability time, $t_v\sim\lambda/c$, or a somewhat larger $\Gamma_\infty\approx\sigma_0$).

The effect of changing the model parameters can be seen by comparing the final spectrum in the different panels. An increase in $\lamep_8$, 
which characterizes the lab-frame width of the magnetic field polarity reversal in a striped wind, with $\epsilon\sim0.1$, shifts the thermal 
peak towards lower energies since $E_{\rm pk,th}(r_{\rm ph})\propto \lamep_8^{-7/20}$. Likewise, in the distributed heating scenario, the adiabatically 
cooled but also Comptonized thermal peak energy becomes softer. When the jet power 
per unit solid angle $L_\Omega$ is varied, the main effect is to increase the overall normalization of the spectrum. In the distributed heating 
case, the low-energy spectrum below the thermal peak also becomes softer as $\kappa$ is increased. Finally, as the 
saturation LF $\Gamma_\infty$ is increased for $\kappa>1$, the spectrum shows minimal changes above the thermal peak but becomes increasingly softer 
below the thermal peak for larger $\Gamma_\infty$ in both heating scenarios. In addition, the soft X-ray peak becomes more prominent with increasing $\kappa$. 
In general, larger $\kappa$ tends to produce broader Band function peaks with softer low-energy spectra simply due to the fact that most of the injected 
energy went into the non-thermal spectral component rather than being thermalized.

\begin{figure*}
    \centering
    \includegraphics[width=0.4\textwidth]{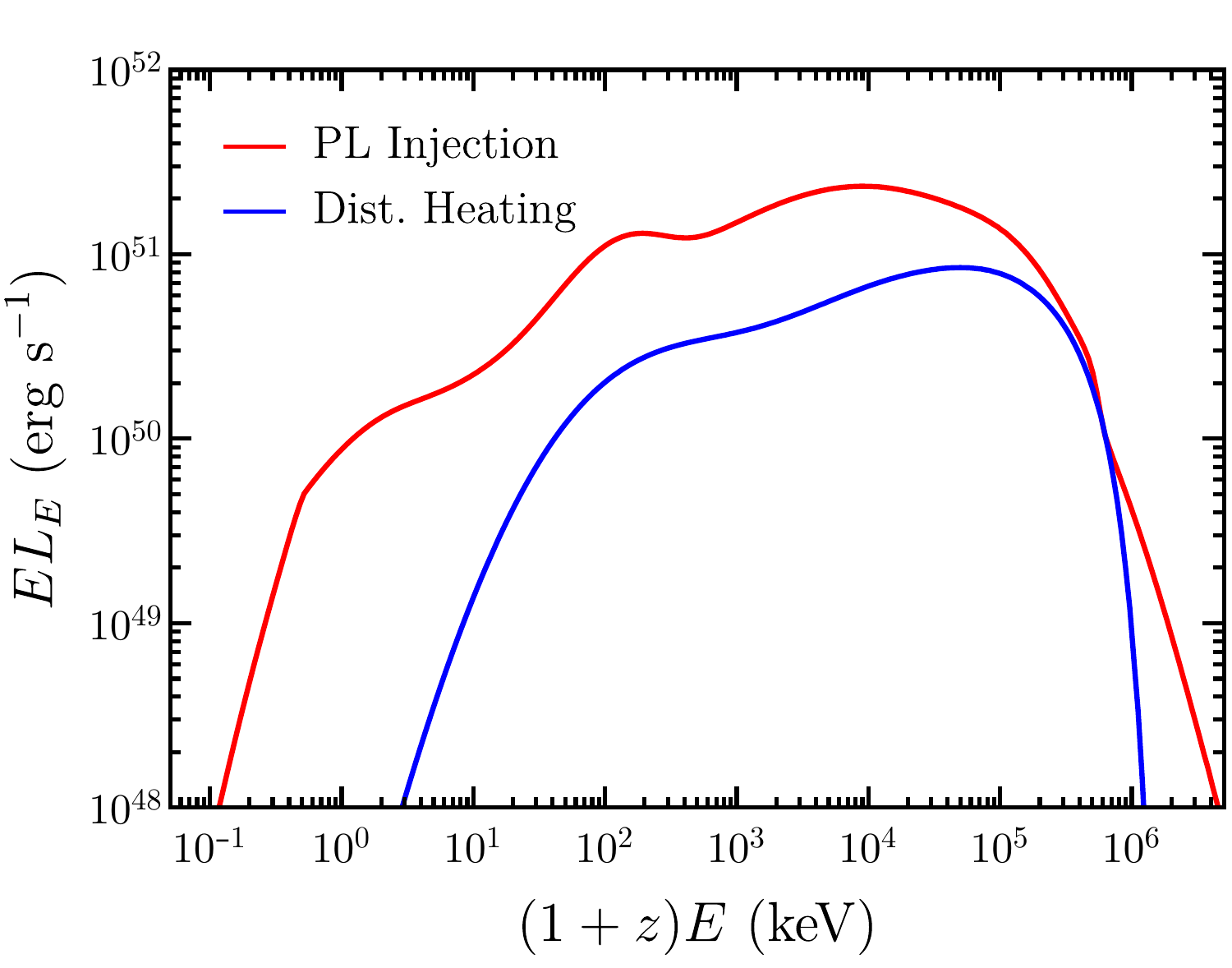}\quad\quad\quad
    \includegraphics[width=0.4\textwidth]{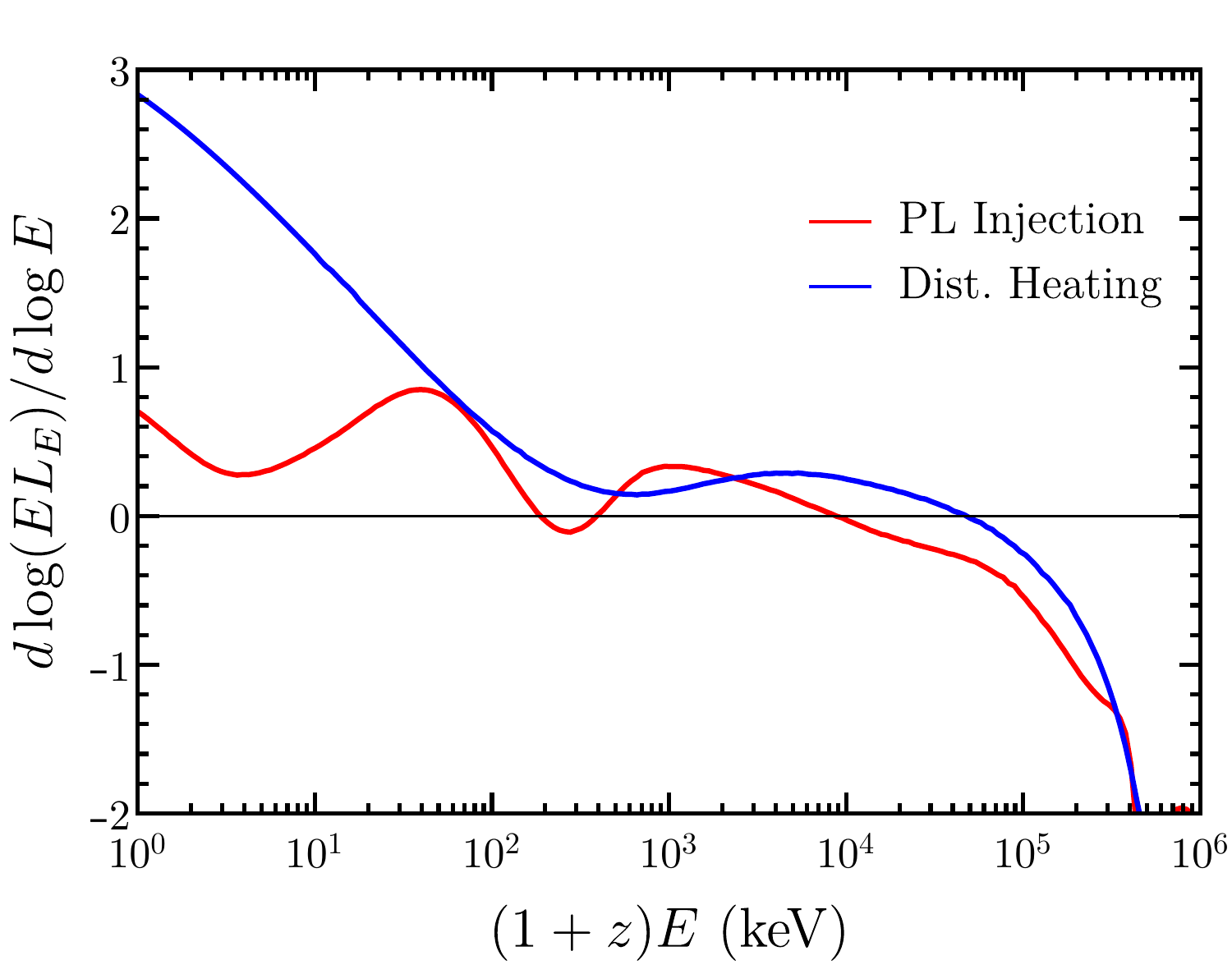}
    \caption{
    \textbf{(Left)} Spectral break below the peak energy can be obtained in both particle heating scenarios. The model parameters 
    for the two cases are: (i) PL injection -- $L_{52}=1$, $\Gamma_{\infty,3}=0.5$, $\lamep_8=4$, $\epsilon_e=0.4$, $\xi=0.2$, 
    (ii) Distributed heating -- $L_{52}=0.1$, $\Gamma_{\infty,3}=0.15$, $\lamep_8=10^3$, $\epsilon_e=0.4$, $\xi=1$. 
    \textbf{(Right)} Spectral slopes for the two scenarios.
    }
    \label{fig:spectral-break}
\end{figure*}

\section{Spectral Break Below the Peak}\label{sec:spectral-break}

A spectral break below the $EL_E$-peak ($E_{\rm pk})$ has been observed in the time-integrated as well as time-resolved spectra 
of a small number of bursts. Such cases have been modeled as having two spectral components where the low-energy component is thermal 
with its peak located at the break energy, $E_{\rm th}=E_{\rm br}\sim100\,$keV, while the high-energy non-thermal component peaks at 
$E=E_{\rm pk}\sim500\,$keV, where both components show fluctuations in the break and peak energies by a factor of $\sim2$ (or more for the latter) in 
a time-resolved analysis \citep{Guiriec+15a}. Alternatively, \citealt{Ravasio+18,Ravasio+19} find that a double smoothly broken power law (2SBPL) 
function obtains a slightly better fit over a two component thermal plus non-thermal fitting function with 
$\langle E_{\rm br}\rangle\sim100\,$keV and $\langle E_{\rm pk}\rangle\sim1\,$MeV. The two spectral profiles do differ in 
one way. A 2SBPL spectral profile features a power law at energies $E_{\rm br}<E<E_{\rm pk}$ whereas a two component model shows 
separate humps with a slight depression in the middle in some cases depending on the relative normalization of the thermal and non-thermal 
components. How well can these two spectral profiles be distinguished from each other depends on the quality of the data, but generally 
both are found to be consistent.

In the left panel of Fig.~\ref{fig:spectral-break} we show two spectra featuring a low-energy break for the two particle heating scenarios 
considered in this work.  
When power-law particles are injected the resulting spectrum features two humps, where the low-energy hump is the 
thermal component and high-energy hump is the non-thermal 
synchrotron emission. In this scenario it would be difficult to get a smoothly connecting spectrum without any depression between the two peaks, 
unless the synchrotron component completely dominates. In the case of distributed heating of particles, a hard Comptonized spectrum above the 
thermal peak can result for particle heating efficiency $\epsilon_e\gtrsim0.3$. The spectrum in this case extends more smoothly to high energies, 
although a slight depression can be present, due to its origin in multiple Compton scatterings. However, the distributed heating scenario produces 
the second hump, that would be identified as the $EL_E$-peak, at much higher energies than observed, even after correcting for the source 
redshift. The problem lies in requiring a significantly high ($y_C\gtrsim5$) Compton-$y$ parameter, which for an optically-thin flow necessarily 
demands high particle momenta $p_e$ (see Eq.~\ref{eq:p_e-at-r_s}). As a consequence, the Comptonized hump always appears at large energies.

In the right-panel of Fig.~\ref{fig:spectral-break}, we show the spectral slopes. For the power-law injection case, the low energy spectrum below 
the thermal hump has photon index $-1.8\lesssim\alpha_1\lesssim-1.2$, whereas the distributed heated scenario yields a harder spectrum with 
$-1.8\lesssim\alpha_1\lesssim1$ for the chosen fiducial model parameters. The spectrum above the thermal hump has $\alpha_2\sim-1.8$ in both cases. 
While we do not carry out an exhaustive analysis to determine these two spectral slopes for different model parameters, and the cases shown 
here do not necessarily agree with the results of \citealt{Ravasio+19}, it remains to be seen if the two particle heating scenarios can find agreement with prompt 
emission spectra that show low-energy breaks. The spectrum above the second hump shows distinct behavior for the two particle 
heating cases. When the second hump forms due to Comptonization, the spectrum shows an exponential decline above the peak whereas 
in the fast-cooling synchrotron emission the spectrum declines more slowly. This feature can be potentially used to distinguish between 
the two scenarios by fitting to observations.

\section{Summary and Discussion}
\label{sec:discussion}
In this work, we have carried out self-consistent one-zone kinetic simulations of a photon-electron-positron plasma 
in a magnetized outflow with a striped-wind magnetic field structure. The flow is launched with high magnetization, 
$\sigma_0\gg1$, which declines with radius as the flow expands and accelerates. The gradual dissipation of the magnetic field energy, 
either due to magnetic reconnection or MHD instabilities, accelerates the flow to terminal bulk Lorentz factors $\Gamma_\infty>100$. 
About half of the dissipated energy is assumed to heat the particles in the flow, out of which a fraction $\epsilon_e$ is 
given to the electrons to power the prompt GRB emission. As the flow expands, the causal volume of the emission region expands 
with it. This brings in new particles into the causal region where the energy in the outflow is dissipated, a fraction $\xi\leq1$ of 
which may be accelerated into a power-law energy distribution at magnetic reconnection sites. Alternatively, distributed heating by MHD instabilities 
in the flow may volumetrically heat all the particles ($\xi=1$) at the same rate in the dissipation region. 
Gradual energy dissipation commences in the optically thick part of the flow ($r<r_{\rm ph}<r_s$) and the flow is continuously heated across the 
photospheric radius as it becomes optically thin until $r=r_s$. The underlying model has 4 to 5 parameters: (i) the jet power per 
unit solid angle $L_\Omega$, (ii) flow terminal LF $\Gamma_\infty$, (iii) characteristic length scale over which the magnetic field changes 
polarity in the lab-frame $\lamep$, (iv) efficiency of particle heating $\epsilon_e$, and (v) fraction $\xi$ of injected electrons that are 
accelerated into a power-law energy distribution -- not needed for the distributed heated scenario for which $\xi=1$.

The two particle heating scenarios lead to different spectra and corresponding particle distributions. In both cases, 
the spectrum exhibits two main components: a {\it thermal} component peaking at $(1+z)E_{\rm th}\sim0.2-1\,$MeV and a 
{\it non-thermal} component extending to high energies from the thermal peak. The origin of the non-thermal component 
is different in the two scenarios. 

When power-law electrons are injected into the dissipation region, the non-thermal component arises due to the fast-cooling 
synchrotron emission. It dominates the spectrum below the thermal peak at energies $(1+z)E\lesssim50\,$keV, where it becomes self-absorbed at 
$(1+z)E\simeq0.5$keV, and above the thermal peak at $1\,{\rm MeV}\lesssim(1+z)E\lesssim100\,$MeV, where the emission is suppressed 
at higher energies due to $\gamma\gamma$-annihilation. 
In our model, the spectral slope of the synchrotron emission at $E>E_m$, with spectral 
power $L_E\propto E^{-p/2}$, depends on the power-law index $p = 4\sigma^{-0.3}$ of power-law electrons, which varies with 
the magnetization of the flow. This tends to produce high-energy spectra with photon indices $-2.5\lesssim\beta\lesssim-2.2$, consistent 
with prompt GRB observations. Harder power-laws above the thermal peak that have been observed in only a few cases, e.g. GRB$\,$090902B, 
can be produced for fixed values of electron power-law index with $p\sim2$. The low-energy photon index, $-1.6\lesssim\alpha\lesssim-1.2$ 
arises from the combination of the thermal component and the synchrotron emission at $E<E_m$, for which $L_E\propto E^{-1/2}$.

When the dissipated energy is distributed among all the electrons (and the produced $e^\pm$-pairs that are mostly subdominant in number in 
the cases simulated here), the non-thermal spectrum above the thermal peak arises due to Comptonization of the softer thermal peak photons. 
This also leads to softening of the spectrum below the thermal peak as the Compton-$y$ parameter grows above unity when the flow becomes optically 
thin. The thermal peak is pushed to higher energies at $(1+z)E\sim1\,$MeV and the photon index of the high-energy spectrum is 
$-2.6\lesssim\beta\lesssim-2.0$ for $0.05\lesssim\epsilon_e\lesssim0.2$, where the emission is suppressed at $(1+z)E\gtrsim100\,$MeV due 
to $\gamma\gamma$-annihilation. 
The low-energy photon index also depends on $\epsilon_e$ with $\alpha\gtrsim-1$ for $\epsilon_e\lesssim0.2$. 
A distinguishing feature of the distributed heating scenario is the appearance of soft X-ray emission at $(1+z)E\sim0.5\,$keV due to 
self-absorbed synchrotron emission from mildly relativistic mono-energetic particles.

Whether the final spectrum is Band-like or features a predominant quasi-thermal component depends on $\kappa=r_s/r_{\rm ph}$, 
the ratio of the saturation radius to the photospheric radius, and which is always larger than unity in this work. For $\kappa\sim1$, 
the final spectrum in both scenarios has a dominant quasi-thermal component since most of the energy is dissipated below the 
photosphere, which leads to suppression of the non-thermal component. On the other hand, when $\kappa>1$, the final spectrum is more 
Band-like where the non-thermal component build up as more energy is dissipated beyond the photosphere when the flow becomes optically 
thin. Since most GRBs do show a Band-like prompt spectrum, $\kappa$ is typically much larger than unity.

A broken power law spectrum with spectral indices $\alpha_1=-2/3$ and $\alpha_2=-3/2$ below the $EL_E$-peak energy are expected from 
a fast cooling optically-thin synchrotron emission for energies $E_{\rm sa}<E<E_c<E_m$ and $E_c<E<E_m$, respectively \citep{Sari+98}. A serious 
argument against this emission model is the observation of harder low-energy spectral slopes in a sizable fraction of GRBs \citep{Preece+98,Ghirlanda+03}. 
Nevertheless, low-energy spectral breaks at $E_{\rm br}\sim50-100\,$keV with spectral indices expected from optically-thin synchrotron 
emission have been found in a number of GRBs \citep[e.g.,][]{Oganesyan+17,Ravasio+18,Ravasio+19}. If the break is indeed the cooling break, 
the energy of which scales with radius as $E_c\propto r^3$ (see Eq.~(\ref{eq:Ec})), then having it at $\sim$\,100\;keV would demand that 
the emission must be at $r\gtrsim3.4\times10^{15}\,$cm. At such a large radius the comoving magnetic field, with scaling $B'\propto r^{-4/3}$, 
would reduce to merely $\lesssim81\,$G. The constraints on the emission radius and the magnetic field become even more severe if dissipation 
occurs in a coasting flow \citep{BP2014,Ravasio+18,Ravasio+19}. To circumvent these issues while keeping the same underlying emission model, 
\citet{Ghisellini+19} have recently proposed that the prompt GRB emission may be produced by power-law protons. However, knowing that protons 
are inefficient at radiating away their thermal energy as compared to electrons, the significant reduction in radiative efficiency 
must be compensated by having a much larger total energy budget, a requirement that may be too demanding. Alternatively, these conditions 
can be realized within magnetically dominated outflow models for the prompt emission, where dissipation is due to large scale reconnection events 
that are triggered far from the central engine \citep{BG2016,BBG2018}.

In this work we show that low-energy breaks can be produced in the two particle heating scenarios considered here. 
This may avoid the need to invoke protons as the main radiators and to push the 
emission region farther out to large radii. At the same time, the main downside of the smaller emission radii required 
in this scenario explored here, is that their associated variability time is significantly shorter than that observed in GRB light-curves. 
The source of variability in the model explored here is therefore the variability of the engine itself. 
For the impulsive magnetic acceleration model $\lambda\sim ct_\varv$ and indeed $t_\varv$ should be reflected in the observed lightcurve variability. 
Observationally, models invoking 
dissipation at relatively small emission radii (as considered here) and large emission radii (as discussed above) may be distinguished in 
the optical and $\gtrsim100\,$MeV bands. In both bands, models including dissipation at smaller radii should be strongly suppressed, due to either 
synchrotron self-absorption (in the optical) or the pair-production threshold (above $\sim$$100\,{\rm MeV}$). Indeed, a significant fraction of GRBs show a 
suppression of $>$$100\,$MeV flux, that is consistent with the existence of a pair production threshold \citep{Beniamini2011,Vianello+18,Gill-Granot-18a}.

The double-hump two-component spectrum has been seen in both the time-resolved and time-integrated spectra of a few 
short-hard \citep[e.g.,][]{Guiriec+10} and long-soft GRBs \citep[e.g.,][]{Guiriec+11}. In all cases, it was shown that a Band-function plus power 
law or a Band-function plus blackbody spectrum resulted in a better fit over a single component Band-function. Furthermore, it was found 
that the non-thermal power-law component dominated the spectrum below few tens of keV and above $\sim$MeV. This behavior is the characteristic 
of having power-law electrons, as seen in the top-left panel of Fig.~\ref{fig:diff-epse-spectrum}, where the non-thermal component is synchrotron 
emission. Double hump spectra were also produced in \citet{Gill-Thompson-14} who carried out one-zone kinetic simulations for a pure electron-positron 
plasma with strong magnetic fields. In that work, emission above the thermal peak was produced by inverse Compton scattering by mildly relativistic 
and mono-energetic $e^\pm$-pairs, however the flow was not continuously heated across the photosphere.

In two bright GRBs (080916C and 090926A) more than two spectral components have been shown to yield the best fit over 
a single Band-function \citep{Guiriec+15a}. In this case, the first two components are the thermal and non-thermal, as discussed above, and the third component 
is modeled as a cutoff power law. In the models presented here, this third spectral component is difficult to obtain and might 
require a subdominant but hotter particle distribution in addition to the baryonic electrons. Since the mean energy per particle is assumed to 
be the same for all particles in the emission region, a separate hotter particle distribution cannot be obtained here.

The final spectrum from both particle heating scenarios shows good agreement with observations in general. To break the degeneracy between 
the two cases, additional diagnostics are needed. One such diagnostic is the suppression of X-ray emission for $(1+z)E\lesssim1\,$keV 
when power law electrons are injected as compared to observable X-ray emission in the distributed heating case \citep[also see, e.g.,][]{Giannios-08}. 
Therefore, broadband X-ray to gamma-ray observations \citep[e.g.,][]{Page+07} during the prompt emission phase are needed to shed more light on 
this issue.

Another important diagnostic that can probe the underlying emission mechanism for the high-energy emission above the thermal peak is the 
detection of prompt linear polarization \citep[e.g.,][]{Granot-03,Gill+20}. Synchrotron emission is partially linearly polarized, 
and depending on the spectral 
index $s=-1-\alpha$ when $L_E\propto E^{-s}$, the maximum polarization from an ordered magnetic field in the dissipation region is 
$0.5\leq\Pi_{\max}=(s+1)/(s+5/3)\lesssim0.75$. Therefore, when synchrotron dominates the high-energy emission, as in the power-law electron 
injection case, high levels of polarization are expected. On the other hand, negligible polarization is expected when the high-energy emission 
is attributed to Comptonization. In this case, even though a singly Compton scattered photon is polarized, multiple such scatterings 
washes out the polarization when averaged over the entire GRB image on the plane of the sky. Polarization with $\Pi\lesssim0.2$ is expected 
if the outflow has angular structure. Polarization is also expected from synchrotron emission below the thermal peak but above the 
self-absorption break in the scenario with power-law electron injection. Therefore, in this case, energy resolved broadband polarimetry should 
reveal a high level polarization at energies both below and above the peak but not near the peak.

Prompt GRB polarization in the range $0.1\lesssim\Pi\lesssim0.9$ has been possibly detected, albeit with only $\sim3\sigma$ significance in most cases, for 
a number of GRBs \citep[see Table 1 of][]{Gill+20}. However, a conclusive picture has not emerged yet. More sensitive upcoming/proposed 
X-ray and gamma-ray polarimetry missions, e.g. POLAR-II, eXTP, will be instrumental in furthering our understanding of the underlying 
prompt GRB emission mechanism.

\section*{Acknowledgements}
RG thanks Sylvain Guiriec for useful discussions. 
RG and JG's research was supported by the ISF-NSFC joint research program (grant No. 3296/19). 
PB’s research was funded in part by the Gordon and Betty Moore Foundation through Grant GBMF5076. 

\begin{appendix}

\section{Particle Injection and Heating}\label{sec:part-inject}
We briefly describe here how the two particle heating scenarios are implemented in this work.

The comoving number density of baryonic electrons in the flow at any given radius $r$ can be expressed as
\begin{equation}\label{eq:n-prime}
    n'(r) = \frac{L_\Omega}{\Gamma_\infty m_pc^3}\frac{1}{r^2\Gamma}\propto r^{-7/3}\,,
\end{equation}
where the expression after the proportionality assumes $r<r_s$ for which $\Gamma\propto r^{1/3}$. 
The number density of particles in the comoving causal volume $\tilde V'=4\pi r^3/\Gamma$ is the 
same as above, such that $\tilde n'(r)=n'(r)$. The number of particles in the causal volume $\tilde N(r)$ 
grows with radius as the volume expands,
\begin{equation}\label{eq:causal-N}
    \tilde N(r) = n'(r)\tilde V' = \tilde n'(r)\tilde V' = \frac{4\pi L_\Omega}{\Gamma_\infty m_pc^3}\frac{r}{\Gamma^2}\propto r^{1/3}\,,
\end{equation}
where this represents a fraction $(r/r_s)^{1/3}$ of the total particle number $N$ in the lab-frame volume of a 
spherical shell whose characteristic width $\lambda\approx r_s/\Gamma_\infty^2$ remains constant. Since no external particles 
are added to this shell as it expands, $N$ remains constant. Therefore, when $r=r_s$ all the particles in this shell 
are found within the causal volume.

The one-zone kinetic code used in this work \citep[see][for more details]{Gill-Thompson-14} evolves particle number density 
instead of particle number in the causal volume. The evolution of electron number density distribution as energy is added in 
the form of power-law particles can be obtained from the evolution of the particle number distribution
\begin{equation}\label{eq:number-evol}
    \frac{\partial\tilde N(p_e)}{\partial t'} 
    = \frac{\partial}{\partial p_e}\left[\frac{\gamma_e}{p_e}A_{\rm ad}\tilde N(p_e)\right] + S'(p_e)\,,
\end{equation}
where $\tilde N(p_e)=\partial\tilde N/\partial p_e$ is the momentum space particle number distribution. The first term on the 
RHS describes the movement of particles in momentum space, using a Fokker-Planck advection term, due to adiabatic cooling 
($A_{\rm ad}$), and in general other energy exchange processes, namely Compton scattering, synchrotron cooling, and Coulomb 
cooling, that are not discussed here but included in the numerical code. The second term is the source term, 
$S'(p_e)=d\tilde{N}/dt'dp_e$, which describes the injection of new power-law particles into the causal volume. In general, $S'(p_e)$ 
receives contributions from other processes, such as Compton scattering, pair-production and annihilation.

The advection coefficient for the adiabatic cooling of particles is given by
\begin{equation}
    A_{\rm ad} = -\frac{1}{m_ec^2}\frac{dE_e'}{dt'} 
    = \frac{(\hat\gamma-1)E_e'}{m_ec^2}\frac{d\ln V'}{dt'}
    = \frac{(\hat\gamma-1)(\gamma_e-1)}{t_{\rm ad}'}\,.
\end{equation}
Here we have used the scaling of particle energy with comoving volume due to adiabatic expansion, such that 
$E_e' = (\gamma_e-1)m_ec^2\propto V'^{1-\hat\gamma}$ where $\hat\gamma=4/3~(5/3)$ is the adiabatic index for 
a relativistic (non-relativistic) particle distribution. In general, the advection coefficient includes contributions 
from other processes that change the energy of particles, e.g. Compton scattering, synchrotron cooling, Coulomb interactions, 
and particle heating (as described below). The adiabatic cooling timescale is obtained from the rate of volume expansion, where
\begin{equation}\label{eq:t_ad}
    \frac{1}{t_{\rm ad}'} = \frac{d\ln V'}{dt'} 
    = \frac{1}{r^2\Gamma}\frac{d(r^2\Gamma)}{dt'} 
    = \frac{c}{r^2}\frac{d(r^2\Gamma)}{dr}
    = \zeta\frac{\Gamma c}{r}\,,
\end{equation}
where $\zeta = 7/3$ for $r<r_s$ when $\Gamma\propto r^{1/3}$, and $\zeta=2$ for $r\geq r_s$ when $\Gamma=\Gamma_\infty$.

By expressing the LHS of (\ref{eq:number-evol}) using the number density, we find
\begin{equation}\label{eq:dNdt-dndt}
\frac{\partial\tilde{n}'(p_e)}{\partial t'} =
\frac{\partial}{\partial t'}\fracb{\tilde{N}(p_e)}{\tilde{V}'}=
\frac{1}{\tilde V'}\frac{\partial\tilde N(p_e)}{\partial t'} - 
\frac{\tilde{n}'(p_e)}{t'_{\tilde{V}'}}\;,
\end{equation}
where we have conveniently defined the comoving growth time of the causal volume, $t'_{\tilde{V}'}$, through
\begin{equation}
\frac{1}{t'_{\tilde{V}'}}= \frac{d\ln\tilde V'}{dt'}\;.   
\end{equation}
Next, we use (\ref{eq:causal-N}) to express the rate of change of the comoving number density with comoving time in terms 
of $\tilde N$ and the causal volume $\tilde V'$, so that
\begin{equation}
\frac{d\ln \tilde{n}'}{dt'} =
    \frac{d\ln n'}{dt'} = -\frac{d\ln V'}{dt'} = \frac{d\ln\tilde N}{dt'} - \frac{d\ln\tilde V'}{dt'}=-\frac{1}{t'_{\rm ad}}=\frac{\dot{\tilde{n}}'}{\tilde{n}'}\,,
\end{equation}
which in turn implies
\begin{equation}
\frac{1}{t'_{\tilde{V}'}}= \frac{1}{t'_{\rm ad}}+\frac{1}{t'_{\rm inj}}\;,
\end{equation}
where the comoving particle injection time into the causal volume is defined by
\begin{equation}
    \frac{1}{t_{\rm inj}'} = \frac{d\ln\tilde N}{dt'} = \frac{d\ln\tilde{V}'}{dt'}+\frac{d\ln \tilde{n}'}{dt'}
    = \frac{1}{t'_{\tilde{V}'}}+\frac{\dot{\tilde{n}}'}{\tilde{n}'}=\frac{1}{t'_{\tilde{V}'}}-\frac{1}{t'_{\rm ad}}\;,
\end{equation}
and $\dot n_{\rm inj}'$ is the particle injection rate per unit volume. Note that the causal comoving volume $\tilde{V}'$ grows 
faster than the total comoving volume $V'$ since it occupies an increasing fraction of it, $f=\tilde{V}'/V'=\tilde{N}/N=\min[1,(r/r_s)^{1/3}]$, 
where $1/t'_{\rm inj}=d\ln f/dt'$.

Now, using (\ref{eq:number-evol}), 
and (\ref{eq:dNdt-dndt}), leads to the equation for the number density evolution
\begin{eqnarray}\label{eq:part-evol-eq}
    \frac{\partial\tilde{n}'(p_e)}{\partial t'} 
    &=& \frac{\partial}{\partial p_e}\left[\frac{\gamma_e}{p_e}A_{\rm ad}\frac{\tilde N(p_e)}{\tilde V'}\right] 
    + \frac{S'(p_e)}{\tilde V'} 
    -\frac{\tilde{n}'(p_e)}{t'_{\tilde{V}'}}
    \nonumber\\
    &=& \frac{\partial}{\partial p_e}\left[\frac{\gamma_e}{p_e}A_{\rm ad} n'(p_e)\right] + Q'(p_e) 
    - \frac{\tilde{n}'(p_e)}{t'_{\tilde{V}'}}\,,
\end{eqnarray}
where $Q'(p_e)=S'(p_e)/\tilde{V}'=d\tilde{N}/d\tilde{V}'dt'dp_e$.
By integrating the above equation over $p_e$, it can be shown that the adiabatic cooling term vanishes and 
$Q'=\int Q'(p_e)dp_e =S'/\tilde{V}'=\tilde{n}'/t_{\rm inj}'$ where $S'=\int S'(p_e)dp_e=d\tilde{N}/dt'$
so that the equation reads $d\tilde{n}'/dt'=(1/t_{\rm inj}'-1/t'_{\tilde{V}'})\tilde{n}'=-\tilde{n}'/t'_{\rm ad}=\tilde{n}'(d\ln\tilde{n}'/dt')=\dot{\tilde{n}}'$.
The comoving rate of particle injection per unit volume per unit dimensionless momentum comprises of two terms, where a fraction $\xi$ of the total is accelerated 
into a power-law (non-thermal) distribution and the remaining fraction $(1-\xi)$ forms a thermal distribution,
\begin{equation}
    Q'(p_e) = (1-\xi)Q'(p_e) + \xi Q'(p_e)
    =
    (1-\xi)Q'\hat{Q}'_{\rm th}(p_e) + \xi Q' \hat{Q}'_{\rm nth}(p_e)\,,
\end{equation}
where $\hat Q'(p_e) = Q'(p_e)/Q'$ represents the normalized momentum distribution. The total energy density 
per unit rest mass energy of the injected distribution can be expressed as
\begin{equation}
    \int(\gamma_e-1)Q'(p_e)dp = Q'[(1-\xi)(\langle\gamma_e\rangle_{\rm th}-1) 
    + \xi(\langle\gamma_e\rangle_{\rm nth}-1)]\,,
\end{equation}
where $\langle\gamma_e\rangle_{\rm th}$ is the mean energy per rest mass energy of the thermal distribution and for the power-law 
distribution $\langle\gamma_e\rangle_{\rm nth}$ is given in Eq.(\ref{eq:gamma_nth}). Here we make the assumption that the injected 
thermal particles are cold, so that $\langle\gamma_e\rangle_{\rm th}-1\ll\frac{\xi}{1-\xi}(\langle\gamma_e\rangle_{\rm nth}-1)$. The momentum distribution of power-law electrons is 
given by
\begin{equation}
    Q'_{\rm nth}(p_e) = \frac{d\gamma_e}{dp_e}Q'_{\rm nth}(\gamma_e)=\frac{p_e}{\gamma_e}Q'_{\rm nth}(\gamma_e) = Q_0'p_e\gamma_e^{-p-1}\,.
\end{equation}
The normalization $Q_0'$ is obtained by equating the rate of energy injection per unit volume to that given to the electrons 
due to dissipation, such that
\begin{equation}
    \frac{1}{m_ec^2}\frac{dU_e'}{dt'} = \int(\gamma_e-1)Q'_{\rm nth}(p_e)dp_e = \int(\gamma_e-1)Q'_{\rm nth}(\gamma_e)d\gamma_e\,, 
\end{equation}
which for $\gamma_M>\gamma_m\gg1$ gives
\begin{equation}
    Q_0' = \frac{(p-2)}{(\gamma_m^{2-p}-\gamma_M^{2-p})}\frac{\epsilon_e}{2m_ec^2}\frac{dU_{\rm diss}'}{dt'}\,,
\end{equation}
where we take $\gamma_M=\gamma_{\rm max}$, the maximum LF corresponding to the outer boundary of the particle momentum grid in the simulation. 
The efficiency of acceleration is controlled by the mean energy of the power-law electrons, which depends on $\xi$, 
and ultimately by the minimal energy of power-law electrons $\gamma_m = \left[(p-2)/(p-1)\right]\langle\gamma_e\rangle_{\rm nth}$.



In the distributed heating scenario, all particles (including produced $e^\pm$-pairs) in the causal volume are heated at the same rate. 
This volumetric heating can again be described using a Fokker-Planck equation with only the advective term
\begin{equation}
    \frac{\partial \tilde n_\pm'(p_e)}{\partial t'} 
    = \frac{\partial}{\partial p_e}\left[\frac{\gamma_e}{p_e}(A_{\rm ad}+A_{\pm,\rm heat}) \tilde n_\pm'(p_e)\right]
    -\frac{\tilde n'_\pm(p_e)}{t_{\rm ad}'}\,,
\end{equation}
where $A_{\pm,\rm heat}$ is the advection coefficient. The rate of change of particle energy due to heating 
is obtained from the volumetric rate of energy injection that yields 
\begin{equation}
    A_{\pm,\rm heat} = \mathcal{A}\frac{\tilde n_\pm'}{(\tilde n_+'+\tilde n_-')}\frac{\epsilon_e}{2m_ec^2}\frac{dU_{\rm diss}'}{dt'}\,,
\end{equation}
where the normalization is given by
\begin{equation}
    \mathcal{A} = \left[\left\{\frac{\gamma_e^2}{p_e}\tilde n_\pm'(p_e)\right\}_{p_{e,\min}}^{p_{e,\max}}-\tilde n_\pm'\right]^{-1}
\end{equation}
with the term in braces evaluated at the particle momentum grid boundaries. The electron and positron distributions are evolved separately 
as described by the above equations with `-' and `+' subscripts, respectively.

\end{appendix}

\section*{Data Availability}
No new data were generated or analysed in support of this research.

\bibliographystyle{mnras}
\bibliography{refs.bib}

\begin{thebibliography}{}
\makeatletter
\relax
\def\mn@urlcharsother{\let\do\@makeother \do\$\do\&\do\#\do\^\do\_\do\%\do\~}
\def\mn@doi{\begingroup\mn@urlcharsother \@ifnextchar [ {\mn@doi@}
  {\mn@doi@[]}}
\def\mn@doi@[#1]#2{\def\@tempa{#1}\ifx\@tempa\@empty \href
  {http://dx.doi.org/#2} {doi:#2}\else \href {http://dx.doi.org/#2} {#1}\fi
  \endgroup}
\def\mn@eprint#1#2{\mn@eprint@#1:#2::\@nil}
\def\mn@eprint@arXiv#1{\href {http://arxiv.org/abs/#1} {{\tt arXiv:#1}}}
\def\mn@eprint@dblp#1{\href {http://dblp.uni-trier.de/rec/bibtex/#1.xml}
  {dblp:#1}}
\def\mn@eprint@#1:#2:#3:#4\@nil{\def\@tempa {#1}\def\@tempb {#2}\def\@tempc
  {#3}\ifx \@tempc \@empty \let \@tempc \@tempb \let \@tempb \@tempa \fi \ifx
  \@tempb \@empty \def\@tempb {arXiv}\fi \@ifundefined
  {mn@eprint@\@tempb}{\@tempb:\@tempc}{\expandafter \expandafter \csname
  mn@eprint@\@tempb\endcsname \expandafter{\@tempc}}}

\bibitem[\protect\citeauthoryear{{Abdo} et~al.,}{{Abdo} et~al.}{2009}]{Abdo+09}
{Abdo} A.~A.,  et~al., 2009, \mn@doi [\apjl] {10.1088/0004-637X/706/1/L138},
  \href {https://ui.adsabs.harvard.edu/abs/2009ApJ...706L.138A} {706, L138}

\bibitem[\protect\citeauthoryear{{Abramowicz}, {Novikov}  \&
  {Paczynski}}{{Abramowicz} et~al.}{1991}]{Abramowicz+91}
{Abramowicz} M.~A.,  {Novikov} I.~D.,   {Paczynski} B.,  1991, \mn@doi [\apj]
  {10.1086/169748}, \href
  {https://ui.adsabs.harvard.edu/abs/1991ApJ...369..175A} {369, 175}

\bibitem[\protect\citeauthoryear{{Ackermann} et~al.,}{{Ackermann}
  et~al.}{2013}]{Ackermann+13}
{Ackermann} M.,  et~al., 2013, \mn@doi [\apj] {10.1088/0004-637X/763/2/71},
  \href {https://ui.adsabs.harvard.edu/abs/2013ApJ...763...71A} {763, 71}

\bibitem[\protect\citeauthoryear{{Ackermann} et~al.,}{{Ackermann}
  et~al.}{2014}]{Ackermann+14}
{Ackermann} M.,  et~al., 2014, \mn@doi [Science] {10.1126/science.1242353},
  \href {https://ui.adsabs.harvard.edu/abs/2014Sci...343...42A} {343, 42}

\bibitem[\protect\citeauthoryear{{Band} et~al.,}{{Band} et~al.}{1993}]{Band+93}
{Band} D.,  et~al., 1993, \mn@doi [\apj] {10.1086/172995}, \href
  {https://ui.adsabs.harvard.edu/abs/1993ApJ...413..281B} {413, 281}

\bibitem[\protect\citeauthoryear{{B{\'e}gu{\'e}} \& {Pe'er}}{{B{\'e}gu{\'e}} \&
  {Pe'er}}{2015}]{Begue-Peer-15}
{B{\'e}gu{\'e}} D.,  {Pe'er} A.,  2015, \mn@doi [\apj]
  {10.1088/0004-637X/802/2/134}, \href
  {https://ui.adsabs.harvard.edu/abs/2015ApJ...802..134B} {802, 134}

\bibitem[\protect\citeauthoryear{{B{\'e}gu{\'e}}, {Pe'er}  \&
  {Lyubarsky}}{{B{\'e}gu{\'e}} et~al.}{2017}]{Begue+17}
{B{\'e}gu{\'e}} D.,  {Pe'er} A.,   {Lyubarsky} Y.,  2017, \mn@doi [\mnras]
  {10.1093/mnras/stx237}, \href
  {https://ui.adsabs.harvard.edu/abs/2017MNRAS.467.2594B} {467, 2594}

\bibitem[\protect\citeauthoryear{{Beloborodov}}{{Beloborodov}}{2010}]{Beloborodov-10}
{Beloborodov} A.~M.,  2010, \mn@doi [\mnras]
  {10.1111/j.1365-2966.2010.16770.x}, \href
  {https://ui.adsabs.harvard.edu/abs/2010MNRAS.407.1033B} {407, 1033}

\bibitem[\protect\citeauthoryear{{Beloborodov}}{{Beloborodov}}{2011}]{Beloborodov-11}
{Beloborodov} A.~M.,  2011, \mn@doi [\apj] {10.1088/0004-637X/737/2/68}, \href
  {https://ui.adsabs.harvard.edu/abs/2011ApJ...737...68B} {737, 68}

\bibitem[\protect\citeauthoryear{{Beloborodov}}{{Beloborodov}}{2013}]{Beloborodov-13}
{Beloborodov} A.~M.,  2013, \mn@doi [\apj] {10.1088/0004-637X/764/2/157}, \href
  {https://ui.adsabs.harvard.edu/abs/2013ApJ...764..157B} {764, 157}

\bibitem[\protect\citeauthoryear{{Beloborodov} \&
  {M{\'e}sz{\'a}ros}}{{Beloborodov} \&
  {M{\'e}sz{\'a}ros}}{2017}]{Beloborodov-Meszaros-17}
{Beloborodov} A.~M.,  {M{\'e}sz{\'a}ros} P.,  2017, \mn@doi [\ssr]
  {10.1007/s11214-017-0348-6}, \href
  {https://ui.adsabs.harvard.edu/abs/2017SSRv..207...87B} {207, 87}

\bibitem[\protect\citeauthoryear{{Beniamini} \& {Giannios}}{{Beniamini} \&
  {Giannios}}{2017}]{Beniamini-Giannios-17}
{Beniamini} P.,  {Giannios} D.,  2017, \mn@doi [\mnras] {10.1093/mnras/stx717},
  \href {https://ui.adsabs.harvard.edu/abs/2017MNRAS.468.3202B} {468, 3202}

\bibitem[\protect\citeauthoryear{{Beniamini} \& {Granot}}{{Beniamini} \&
  {Granot}}{2016}]{BG2016}
{Beniamini} P.,  {Granot} J.,  2016, \mn@doi [\mnras] {10.1093/mnras/stw895},
  \href {https://ui.adsabs.harvard.edu/abs/2016MNRAS.459.3635B} {459, 3635}

\bibitem[\protect\citeauthoryear{{Beniamini} \& {Piran}}{{Beniamini} \&
  {Piran}}{2013}]{Beniamini2013}
{Beniamini} P.,  {Piran} T.,  2013, \mn@doi [\apj]
  {10.1088/0004-637X/769/1/69}, \href
  {https://ui.adsabs.harvard.edu/abs/2013ApJ...769...69B} {769, 69}

\bibitem[\protect\citeauthoryear{{Beniamini} \& {Piran}}{{Beniamini} \&
  {Piran}}{2014}]{BP2014}
{Beniamini} P.,  {Piran} T.,  2014, \mn@doi [\mnras] {10.1093/mnras/stu2032},
  \href {https://ui.adsabs.harvard.edu/abs/2014MNRAS.445.3892B} {445, 3892}

\bibitem[\protect\citeauthoryear{{Beniamini}, {Guetta}, {Nakar}  \&
  {Piran}}{{Beniamini} et~al.}{2011}]{Beniamini2011}
{Beniamini} P.,  {Guetta} D.,  {Nakar} E.,   {Piran} T.,  2011, \mn@doi
  [\mnras] {10.1111/j.1365-2966.2011.19259.x}, \href
  {https://ui.adsabs.harvard.edu/abs/2011MNRAS.416.3089B} {416, 3089}

\bibitem[\protect\citeauthoryear{{Beniamini}, {Barniol Duran}  \&
  {Giannios}}{{Beniamini} et~al.}{2018}]{BBG2018}
{Beniamini} P.,  {Barniol Duran} R.,   {Giannios} D.,  2018, \mn@doi [\mnras]
  {10.1093/mnras/sty340}, \href
  {https://ui.adsabs.harvard.edu/abs/2018MNRAS.476.1785B} {476, 1785}

\bibitem[\protect\citeauthoryear{{Bhattacharya} \& {Kumar}}{{Bhattacharya} \&
  {Kumar}}{2020}]{Bhattacharya-Kumar-20}
{Bhattacharya} M.,  {Kumar} P.,  2020, \mn@doi [\mnras]
  {10.1093/mnras/stz3182}, \href
  {https://ui.adsabs.harvard.edu/abs/2020MNRAS.491.4656B} {491, 4656}

\bibitem[\protect\citeauthoryear{{Briggs} et~al.,}{{Briggs}
  et~al.}{1999}]{Briggs+99}
{Briggs} M.~S.,  et~al., 1999, \mn@doi [\apj] {10.1086/307808}, \href
  {https://ui.adsabs.harvard.edu/abs/1999ApJ...524...82B} {524, 82}

\bibitem[\protect\citeauthoryear{{Daigne}, {Bo{\v{s}}njak}  \&
  {Dubus}}{{Daigne} et~al.}{2011}]{Daigne2011}
{Daigne} F.,  {Bo{\v{s}}njak} {\v{Z}}.,   {Dubus} G.,  2011, \mn@doi [\aap]
  {10.1051/0004-6361/201015457}, \href
  {https://ui.adsabs.harvard.edu/abs/2011A&A...526A.110D} {526, A110}

\bibitem[\protect\citeauthoryear{{Drenkhahn}}{{Drenkhahn}}{2002}]{Drenkhahn-02}
{Drenkhahn} G.,  2002, \mn@doi [\aap] {10.1051/0004-6361:20020390}, \href
  {https://ui.adsabs.harvard.edu/abs/2002A&A...387..714D} {387, 714}

\bibitem[\protect\citeauthoryear{{Drenkhahn} \& {Spruit}}{{Drenkhahn} \&
  {Spruit}}{2002}]{Drenkhahn-Spruit-02}
{Drenkhahn} G.,  {Spruit} H.~C.,  2002, \mn@doi [\aap]
  {10.1051/0004-6361:20020839}, \href
  {http://cdsads.u-strasbg.fr/abs/2002A%26A...391.1141D} {391, 1141}

\bibitem[\protect\citeauthoryear{{Eichler} \& {Levinson}}{{Eichler} \&
  {Levinson}}{2000}]{Eichler-Levinson-00}
{Eichler} D.,  {Levinson} A.,  2000, \mn@doi [\apj] {10.1086/308245}, \href
  {https://ui.adsabs.harvard.edu/abs/2000ApJ...529..146E} {529, 146}

\bibitem[\protect\citeauthoryear{{Ghirlanda}, {Celotti}  \&
  {Ghisellini}}{{Ghirlanda} et~al.}{2003}]{Ghirlanda+03}
{Ghirlanda} G.,  {Celotti} A.,   {Ghisellini} G.,  2003, \mn@doi [\aap]
  {10.1051/0004-6361:20030803}, \href
  {https://ui.adsabs.harvard.edu/abs/2003A&A...406..879G} {406, 879}

\bibitem[\protect\citeauthoryear{{Ghisellini} \& {Celotti}}{{Ghisellini} \&
  {Celotti}}{1999}]{Ghisellini-Celotti-99}
{Ghisellini} G.,  {Celotti} A.,  1999, \mn@doi [\apjl] {10.1086/311845}, \href
  {https://ui.adsabs.harvard.edu/abs/1999ApJ...511L..93G} {511, L93}

\bibitem[\protect\citeauthoryear{{Ghisellini} et~al.,}{{Ghisellini}
  et~al.}{2019}]{Ghisellini+19}
{Ghisellini} G.,  et~al., 2019, arXiv e-prints, \href
  {https://ui.adsabs.harvard.edu/abs/2019arXiv191202185G} {p. arXiv:1912.02185}

\bibitem[\protect\citeauthoryear{{Giannios}}{{Giannios}}{2006}]{Giannios-06}
{Giannios} D.,  2006, \mn@doi [\aap] {10.1051/0004-6361:20065000}, \href
  {https://ui.adsabs.harvard.edu/abs/2006A&A...457..763G} {457, 763}

\bibitem[\protect\citeauthoryear{{Giannios}}{{Giannios}}{2008}]{Giannios-08}
{Giannios} D.,  2008, \mn@doi [\aap] {10.1051/0004-6361:20079085}, \href
  {https://ui.adsabs.harvard.edu/abs/2008A&A...480..305G} {480, 305}

\bibitem[\protect\citeauthoryear{{Giannios} \& {Spruit}}{{Giannios} \&
  {Spruit}}{2005}]{Giannios-Spruit-05}
{Giannios} D.,  {Spruit} H.~C.,  2005, \mn@doi [\aap]
  {10.1051/0004-6361:20047033}, \href
  {https://ui.adsabs.harvard.edu/abs/2005A&A...430....1G} {430, 1}

\bibitem[\protect\citeauthoryear{{Giannios} \& {Spruit}}{{Giannios} \&
  {Spruit}}{2007}]{Giannios-Spruit-07}
{Giannios} D.,  {Spruit} H.~C.,  2007, \mn@doi [\aap]
  {10.1051/0004-6361:20066739}, \href
  {https://ui.adsabs.harvard.edu/abs/2007A&A...469....1G} {469, 1}

\bibitem[\protect\citeauthoryear{{Gill} \& {Granot}}{{Gill} \&
  {Granot}}{2018}]{Gill-Granot-18a}
{Gill} R.,  {Granot} J.,  2018, \mn@doi [\mnras] {10.1093/mnrasl/slx199}, \href
  {http://adsabs.harvard.edu/abs/2018MNRAS.475L...1G} {475, L1}

\bibitem[\protect\citeauthoryear{{Gill} \& {Thompson}}{{Gill} \&
  {Thompson}}{2014}]{Gill-Thompson-14}
{Gill} R.,  {Thompson} C.,  2014, \mn@doi [\apj] {10.1088/0004-637X/796/2/81},
  \href {https://ui.adsabs.harvard.edu/abs/2014ApJ...796...81G} {796, 81}

\bibitem[\protect\citeauthoryear{{Gill}, {Granot}  \& {Lyubarsky}}{{Gill}
  et~al.}{2018}]{Gill+18}
{Gill} R.,  {Granot} J.,   {Lyubarsky} Y.,  2018, \mn@doi [\mnras]
  {10.1093/mnras/stx3000}, \href
  {https://ui.adsabs.harvard.edu/abs/2018MNRAS.474.3535G} {474, 3535}

\bibitem[\protect\citeauthoryear{{Gill}, {Granot}  \& {Kumar}}{{Gill}
  et~al.}{2020}]{Gill+20}
{Gill} R.,  {Granot} J.,   {Kumar} P.,  2020, \mn@doi [\mnras]
  {10.1093/mnras/stz2976}, \href
  {https://ui.adsabs.harvard.edu/abs/2020MNRAS.491.3343G} {491, 3343}

\bibitem[\protect\citeauthoryear{{Granot}}{{Granot}}{2003}]{Granot-03}
{Granot} J.,  2003, \mn@doi [\apjl] {10.1086/379110}, \href
  {https://ui.adsabs.harvard.edu/abs/2003ApJ...596L..17G} {596, L17}

\bibitem[\protect\citeauthoryear{{Granot}}{{Granot}}{2012}]{Granot12b}
{Granot} J.,  2012, \mn@doi [\mnras] {10.1111/j.1365-2966.2012.20474.x}, \href
  {https://ui.adsabs.harvard.edu/abs/2012MNRAS.421.2467G} {421, 2467}

\bibitem[\protect\citeauthoryear{{Granot} \& {Sari}}{{Granot} \&
  {Sari}}{2002}]{Granot-Sari-02}
{Granot} J.,  {Sari} R.,  2002, \mn@doi [\apj] {10.1086/338966}, \href
  {http://adsabs.harvard.edu/abs/2002ApJ...568..820G} {568, 820}

\bibitem[\protect\citeauthoryear{{Granot}, {Piran}  \& {Sari}}{{Granot}
  et~al.}{1999}]{Granot+99}
{Granot} J.,  {Piran} T.,   {Sari} R.,  1999, \mn@doi [\apj] {10.1086/306884},
  \href {http://adsabs.harvard.edu/abs/1999ApJ...513..679G} {513, 679}

\bibitem[\protect\citeauthoryear{{Granot}, {Piran}  \& {Sari}}{{Granot}
  et~al.}{2000}]{Granot+00}
{Granot} J.,  {Piran} T.,   {Sari} R.,  2000, \mn@doi [\apjl] {10.1086/312661},
  \href {https://ui.adsabs.harvard.edu/abs/2000ApJ...534L.163G} {534, L163}

\bibitem[\protect\citeauthoryear{{Granot}, {Cohen-Tanugi}  \& {Silva}}{{Granot}
  et~al.}{2008}]{Granot+08}
{Granot} J.,  {Cohen-Tanugi} J.,   {Silva} E. d. C.~e.,  2008, \mn@doi [\apj]
  {10.1086/526414}, \href
  {https://ui.adsabs.harvard.edu/abs/2008ApJ...677...92G} {677, 92}

\bibitem[\protect\citeauthoryear{{Granot}, {Komissarov}  \&
  {Spitkovsky}}{{Granot} et~al.}{2011}]{Granot+11}
{Granot} J.,  {Komissarov} S.~S.,   {Spitkovsky} A.,  2011, \mn@doi [\mnras]
  {10.1111/j.1365-2966.2010.17770.x}, \href
  {https://ui.adsabs.harvard.edu/abs/2011MNRAS.411.1323G} {411, 1323}

\bibitem[\protect\citeauthoryear{{Guiriec} et~al.,}{{Guiriec}
  et~al.}{2010}]{Guiriec+10}
{Guiriec} S.,  et~al., 2010, \mn@doi [\apj] {10.1088/0004-637X/725/1/225},
  \href {https://ui.adsabs.harvard.edu/abs/2010ApJ...725..225G} {725, 225}

\bibitem[\protect\citeauthoryear{{Guiriec} et~al.,}{{Guiriec}
  et~al.}{2011}]{Guiriec+11}
{Guiriec} S.,  et~al., 2011, \mn@doi [\apjl] {10.1088/2041-8205/727/2/L33},
  \href {https://ui.adsabs.harvard.edu/abs/2011ApJ...727L..33G} {727, L33}

\bibitem[\protect\citeauthoryear{{Guiriec} et~al.,}{{Guiriec}
  et~al.}{2015a}]{Guiriec+15a}
{Guiriec} S.,  et~al., 2015a, \mn@doi [\apj] {10.1088/0004-637X/807/2/148},
  \href {https://ui.adsabs.harvard.edu/abs/2015ApJ...807..148G} {807, 148}

\bibitem[\protect\citeauthoryear{{Guiriec}, {Mochkovitch}, {Piran}, {Daigne},
  {Kouveliotou}, {Racusin}, {Gehrels}  \& {McEnery}}{{Guiriec}
  et~al.}{2015b}]{Guiriec+15b}
{Guiriec} S.,  {Mochkovitch} R.,  {Piran} T.,  {Daigne} F.,  {Kouveliotou} C.,
  {Racusin} J.,  {Gehrels} N.,   {McEnery} J.,  2015b, \mn@doi [\apj]
  {10.1088/0004-637X/814/1/10}, \href
  {https://ui.adsabs.harvard.edu/abs/2015ApJ...814...10G} {814, 10}

\bibitem[\protect\citeauthoryear{{Guiriec}, {Gonzalez}, {Sacahui},
  {Kouveliotou}, {Gehrels}  \& {McEnery}}{{Guiriec}
  et~al.}{2016a}]{Guiriec+16a}
{Guiriec} S.,  {Gonzalez} M.~M.,  {Sacahui} J.~R.,  {Kouveliotou} C.,
  {Gehrels} N.,   {McEnery} J.,  2016a, \mn@doi [\apj]
  {10.3847/0004-637X/819/1/79}, \href
  {https://ui.adsabs.harvard.edu/abs/2016ApJ...819...79G} {819, 79}

\bibitem[\protect\citeauthoryear{{Guiriec} et~al.,}{{Guiriec}
  et~al.}{2016b}]{Guiriec+16b}
{Guiriec} S.,  et~al., 2016b, \mn@doi [\apjl] {10.3847/2041-8205/831/1/L8},
  \href {https://ui.adsabs.harvard.edu/abs/2016ApJ...831L...8G} {831, L8}

\bibitem[\protect\citeauthoryear{{Guiriec}, {Gehrels}, {McEnery}, {Kouveliotou}
   \& {Hartmann}}{{Guiriec} et~al.}{2017}]{Guiriec+17}
{Guiriec} S.,  {Gehrels} N.,  {McEnery} J.,  {Kouveliotou} C.,   {Hartmann}
  D.~H.,  2017, \mn@doi [\apj] {10.3847/1538-4357/aa81c2}, \href
  {https://ui.adsabs.harvard.edu/abs/2017ApJ...846..138G} {846, 138}

\bibitem[\protect\citeauthoryear{{Guo}, {Liu}, {Daughton}  \& {Li}}{{Guo}
  et~al.}{2015}]{Guo+15}
{Guo} F.,  {Liu} Y.-H.,  {Daughton} W.,   {Li} H.,  2015, \mn@doi [\apj]
  {10.1088/0004-637X/806/2/167}, \href
  {https://ui.adsabs.harvard.edu/abs/2015ApJ...806..167G} {806, 167}

\bibitem[\protect\citeauthoryear{{Kagan}, {Sironi}, {Cerutti}  \&
  {Giannios}}{{Kagan} et~al.}{2015}]{Kagan+15}
{Kagan} D.,  {Sironi} L.,  {Cerutti} B.,   {Giannios} D.,  2015, \mn@doi [\ssr]
  {10.1007/s11214-014-0132-9}, \href
  {https://ui.adsabs.harvard.edu/abs/2015SSRv..191..545K} {191, 545}

\bibitem[\protect\citeauthoryear{{Kaneko}, {Preece}, {Briggs}, {Paciesas},
  {Meegan}  \& {Band}}{{Kaneko} et~al.}{2006}]{Kaneko+06}
{Kaneko} Y.,  {Preece} R.~D.,  {Briggs} M.~S.,  {Paciesas} W.~S.,  {Meegan}
  C.~A.,   {Band} D.~L.,  2006, \mn@doi [\apjs] {10.1086/505911}, \href
  {https://ui.adsabs.harvard.edu/abs/2006ApJS..166..298K} {166, 298}

\bibitem[\protect\citeauthoryear{{Katz}}{{Katz}}{1994}]{Katz-94}
{Katz} J.~I.,  1994, \mn@doi [\apjl] {10.1086/187523}, \href
  {https://ui.adsabs.harvard.edu/abs/1994ApJ...432L.107K} {432, L107}

\bibitem[\protect\citeauthoryear{{Komissarov}}{{Komissarov}}{2012}]{Komissarov12}
{Komissarov} S.~S.,  2012, \mn@doi [\mnras] {10.1111/j.1365-2966.2012.20609.x},
  \href {https://ui.adsabs.harvard.edu/abs/2012MNRAS.422..326K} {422, 326}

\bibitem[\protect\citeauthoryear{{Kouveliotou}, {Meegan}, {Fishman}, {Bhat},
  {Briggs}, {Koshut}, {Paciesas}  \& {Pendleton}}{{Kouveliotou}
  et~al.}{1993}]{Kouveliotou+93}
{Kouveliotou} C.,  {Meegan} C.~A.,  {Fishman} G.~J.,  {Bhat} N.~P.,  {Briggs}
  M.~S.,  {Koshut} T.~M.,  {Paciesas} W.~S.,   {Pendleton} G.~N.,  1993,
  \mn@doi [\apjl] {10.1086/186969}, \href
  {https://ui.adsabs.harvard.edu/abs/1993ApJ...413L.101K} {413, L101}

\bibitem[\protect\citeauthoryear{{Kumar} \& {McMahon}}{{Kumar} \&
  {McMahon}}{2008}]{KM2008}
{Kumar} P.,  {McMahon} E.,  2008, \mn@doi [\mnras]
  {10.1111/j.1365-2966.2007.12621.x}, \href
  {https://ui.adsabs.harvard.edu/abs/2008MNRAS.384...33K} {384, 33}

\bibitem[\protect\citeauthoryear{{Kumar} \& {Zhang}}{{Kumar} \&
  {Zhang}}{2015}]{Kumar-Zhang-15}
{Kumar} P.,  {Zhang} B.,  2015, \mn@doi [\physrep]
  {10.1016/j.physrep.2014.09.008}, \href
  {http://adsabs.harvard.edu/abs/2015PhR...561....1K} {561, 1}

\bibitem[\protect\citeauthoryear{{Lightman} \& {Zdziarski}}{{Lightman} \&
  {Zdziarski}}{1987}]{Lightman-Zdziarski-87}
{Lightman} A.~P.,  {Zdziarski} A.~A.,  1987, \mn@doi [\apj] {10.1086/165485},
  \href {https://ui.adsabs.harvard.edu/abs/1987ApJ...319..643L} {319, 643}

\bibitem[\protect\citeauthoryear{{Lyubarsky}}{{Lyubarsky}}{2010}]{Lyubarsky-10}
{Lyubarsky} Y.,  2010, \mn@doi [\apjl] {10.1088/2041-8205/725/2/L234}, \href
  {https://ui.adsabs.harvard.edu/abs/2010ApJ...725L.234L} {725, L234}

\bibitem[\protect\citeauthoryear{{Lyubarsky} \& {Kirk}}{{Lyubarsky} \&
  {Kirk}}{2001}]{Lyubarsky-Kirk-01}
{Lyubarsky} Y.,  {Kirk} J.~G.,  2001, \mn@doi [\apj] {10.1086/318354}, \href
  {https://ui.adsabs.harvard.edu/abs/2001ApJ...547..437L} {547, 437}

\bibitem[\protect\citeauthoryear{{Lyutikov} \& {Blandford}}{{Lyutikov} \&
  {Blandford}}{2003}]{Lyutikov-Blandford-03}
{Lyutikov} M.,  {Blandford} R.,  2003, ArXiv Astrophysics e-prints, \href
  {http://cdsads.u-strasbg.fr/abs/2003astro.ph.12347L} {}

\bibitem[\protect\citeauthoryear{{McKinney} \& {Uzdensky}}{{McKinney} \&
  {Uzdensky}}{2012}]{Mckinney-Uzdensky-12}
{McKinney} J.~C.,  {Uzdensky} D.~A.,  2012, \mn@doi [\mnras]
  {10.1111/j.1365-2966.2011.19721.x}, \href
  {https://ui.adsabs.harvard.edu/abs/2012MNRAS.419..573M} {419, 573}

\bibitem[\protect\citeauthoryear{{M{\'e}sz{\'a}ros} \&
  {Rees}}{{M{\'e}sz{\'a}ros} \& {Rees}}{2000}]{Meszaros-Rees-00}
{M{\'e}sz{\'a}ros} P.,  {Rees} M.~J.,  2000, \mn@doi [\apj] {10.1086/308371},
  \href {https://ui.adsabs.harvard.edu/abs/2000ApJ...530..292M} {530, 292}

\bibitem[\protect\citeauthoryear{{Metzger}, {Giannios}, {Thompson},
  {Bucciantini}  \& {Quataert}}{{Metzger} et~al.}{2011}]{Metzger+11}
{Metzger} B.~D.,  {Giannios} D.,  {Thompson} T.~A.,  {Bucciantini} N.,
  {Quataert} E.,  2011, \mn@doi [\mnras] {10.1111/j.1365-2966.2011.18280.x},
  \href {http://adsabs.harvard.edu/abs/2011MNRAS.413.2031M} {413, 2031}

\bibitem[\protect\citeauthoryear{{Oganesyan}, {Nava}, {Ghirlanda}  \&
  {Celotti}}{{Oganesyan} et~al.}{2017}]{Oganesyan+17}
{Oganesyan} G.,  {Nava} L.,  {Ghirlanda} G.,   {Celotti} A.,  2017, \mn@doi
  [\apj] {10.3847/1538-4357/aa831e}, \href
  {https://ui.adsabs.harvard.edu/abs/2017ApJ...846..137O} {846, 137}

\bibitem[\protect\citeauthoryear{{Page} et~al.,}{{Page} et~al.}{2007}]{Page+07}
{Page} K.~L.,  et~al., 2007, \mn@doi [\apj] {10.1086/518821}, \href
  {https://ui.adsabs.harvard.edu/abs/2007ApJ...663.1125P} {663, 1125}

\bibitem[\protect\citeauthoryear{{Parfrey}, {Giannios}  \&
  {Beloborodov}}{{Parfrey} et~al.}{2015}]{Parfrey+15}
{Parfrey} K.,  {Giannios} D.,   {Beloborodov} A.~M.,  2015, \mn@doi [\mnras]
  {10.1093/mnrasl/slu162}, \href
  {https://ui.adsabs.harvard.edu/abs/2015MNRAS.446L..61P} {446, L61}

\bibitem[\protect\citeauthoryear{{Pe'er}}{{Pe'er}}{2008}]{Peer-08}
{Pe'er} A.,  2008, \mn@doi [\apj] {10.1086/588136}, \href
  {https://ui.adsabs.harvard.edu/abs/2008ApJ...682..463P} {682, 463}

\bibitem[\protect\citeauthoryear{{Pe'er}}{{Pe'er}}{2017}]{Peer-17}
{Pe'er} A.,  2017, \mn@doi [\apj] {10.3847/1538-4357/aa974e}, \href
  {https://ui.adsabs.harvard.edu/abs/2017ApJ...850..200P} {850, 200}

\bibitem[\protect\citeauthoryear{{Pe'er} \& {Waxman}}{{Pe'er} \&
  {Waxman}}{2004}]{Peer-Waxman-04}
{Pe'er} A.,  {Waxman} E.,  2004, \mn@doi [\apj] {10.1086/422989}, \href
  {https://ui.adsabs.harvard.edu/abs/2004ApJ...613..448P} {613, 448}

\bibitem[\protect\citeauthoryear{{Pe'er} \& {Waxman}}{{Pe'er} \&
  {Waxman}}{2005}]{Peer-Waxman-05}
{Pe'er} A.,  {Waxman} E.,  2005, \mn@doi [\apj] {10.1086/431139}, \href
  {https://ui.adsabs.harvard.edu/abs/2005ApJ...628..857P} {628, 857}

\bibitem[\protect\citeauthoryear{{Pe'er}, {M{\'e}sz{\'a}ros}  \&
  {Rees}}{{Pe'er} et~al.}{2006}]{Peer+06}
{Pe'er} A.,  {M{\'e}sz{\'a}ros} P.,   {Rees} M.~J.,  2006, \mn@doi [\apj]
  {10.1086/501424}, \href
  {https://ui.adsabs.harvard.edu/abs/2006ApJ...642..995P} {642, 995}

\bibitem[\protect\citeauthoryear{{Piran}}{{Piran}}{2004}]{Piran-04}
{Piran} T.,  2004, \mn@doi [Reviews of Modern Physics]
  {10.1103/RevModPhys.76.1143}, \href
  {http://cdsads.u-strasbg.fr/abs/2004RvMP...76.1143P} {76, 1143}

\bibitem[\protect\citeauthoryear{{Preece}, {Briggs}, {Mallozzi}, {Pendleton},
  {Paciesas}  \& {Band}}{{Preece} et~al.}{1998}]{Preece+98}
{Preece} R.~D.,  {Briggs} M.~S.,  {Mallozzi} R.~S.,  {Pendleton} G.~N.,
  {Paciesas} W.~S.,   {Band} D.~L.,  1998, \mn@doi [\apjl] {10.1086/311644},
  \href {https://ui.adsabs.harvard.edu/abs/1998ApJ...506L..23P} {506, L23}

\bibitem[\protect\citeauthoryear{{Preece}, {Briggs}, {Mallozzi}, {Pendleton},
  {Paciesas}  \& {Band}}{{Preece} et~al.}{2000}]{Preece+00}
{Preece} R.~D.,  {Briggs} M.~S.,  {Mallozzi} R.~S.,  {Pendleton} G.~N.,
  {Paciesas} W.~S.,   {Band} D.~L.,  2000, \mn@doi [\apjs] {10.1086/313289},
  \href {https://ui.adsabs.harvard.edu/abs/2000ApJS..126...19P} {126, 19}

\bibitem[\protect\citeauthoryear{{Ravasio}, {Oganesyan}, {Ghirlanda}, {Nava},
  {Ghisellini}, {Pescalli}  \& {Celotti}}{{Ravasio} et~al.}{2018}]{Ravasio+18}
{Ravasio} M.~E.,  {Oganesyan} G.,  {Ghirlanda} G.,  {Nava} L.,  {Ghisellini}
  G.,  {Pescalli} A.,   {Celotti} A.,  2018, \mn@doi [\aap]
  {10.1051/0004-6361/201732245}, \href
  {https://ui.adsabs.harvard.edu/abs/2018A&A...613A..16R} {613, A16}

\bibitem[\protect\citeauthoryear{{Ravasio}, {Ghirlanda}, {Nava}  \&
  {Ghisellini}}{{Ravasio} et~al.}{2019}]{Ravasio+19}
{Ravasio} M.~E.,  {Ghirlanda} G.,  {Nava} L.,   {Ghisellini} G.,  2019, \mn@doi
  [\aap] {10.1051/0004-6361/201834987}, \href
  {https://ui.adsabs.harvard.edu/abs/2019A&A...625A..60R} {625, A60}

\bibitem[\protect\citeauthoryear{{Rees} \& {Meszaros}}{{Rees} \&
  {Meszaros}}{1994}]{Rees-Meszaros-94}
{Rees} M.~J.,  {Meszaros} P.,  1994, \mn@doi [\apjl] {10.1086/187446}, \href
  {https://ui.adsabs.harvard.edu/abs/1994ApJ...430L..93R} {430, L93}

\bibitem[\protect\citeauthoryear{{Rees} \& {M{\'e}sz{\'a}ros}}{{Rees} \&
  {M{\'e}sz{\'a}ros}}{2005}]{Rees-Meszaros-05}
{Rees} M.~J.,  {M{\'e}sz{\'a}ros} P.,  2005, \mn@doi [\apj] {10.1086/430818},
  \href {https://ui.adsabs.harvard.edu/abs/2005ApJ...628..847R} {628, 847}

\bibitem[\protect\citeauthoryear{{Ryde}}{{Ryde}}{2004}]{Ryde-04}
{Ryde} F.,  2004, \mn@doi [\apj] {10.1086/423782}, \href
  {https://ui.adsabs.harvard.edu/abs/2004ApJ...614..827R} {614, 827}

\bibitem[\protect\citeauthoryear{{Ryde}}{{Ryde}}{2005}]{Ryde-05}
{Ryde} F.,  2005, \mn@doi [\apjl] {10.1086/431239}, \href
  {https://ui.adsabs.harvard.edu/abs/2005ApJ...625L..95R} {625, L95}

\bibitem[\protect\citeauthoryear{{Sari}, {Piran}  \& {Narayan}}{{Sari}
  et~al.}{1998}]{Sari+98}
{Sari} R.,  {Piran} T.,   {Narayan} R.,  1998, \mn@doi [\apjl]
  {10.1086/311269}, \href {http://adsabs.harvard.edu/abs/1998ApJ...497L..17S}
  {497, L17}

\bibitem[\protect\citeauthoryear{{Sironi} \& {Spitkovsky}}{{Sironi} \&
  {Spitkovsky}}{2014}]{Sironi-Spitkovsky-14}
{Sironi} L.,  {Spitkovsky} A.,  2014, \mn@doi [\apjl]
  {10.1088/2041-8205/783/1/L21}, \href
  {https://ui.adsabs.harvard.edu/abs/2014ApJ...783L..21S} {783, L21}

\bibitem[\protect\citeauthoryear{{Stern} \& {Poutanen}}{{Stern} \&
  {Poutanen}}{2004}]{Stern-Poutanen-04}
{Stern} B.~E.,  {Poutanen} J.,  2004, \mn@doi [\mnras]
  {10.1111/j.1365-2966.2004.08163.x}, \href
  {https://ui.adsabs.harvard.edu/abs/2004MNRAS.352L..35S} {352, L35}

\bibitem[\protect\citeauthoryear{{Tavani}}{{Tavani}}{1996}]{Tavani-96}
{Tavani} M.,  1996, \mn@doi [\apj] {10.1086/177551}, \href
  {https://ui.adsabs.harvard.edu/abs/1996ApJ...466..768T} {466, 768}

\bibitem[\protect\citeauthoryear{{Thompson}}{{Thompson}}{1994}]{Thompson-94}
{Thompson} C.,  1994, \mn@doi [\mnras] {10.1093/mnras/270.3.480}, \href
  {http://cdsads.u-strasbg.fr/abs/1994MNRAS.270..480T} {270, 480}

\bibitem[\protect\citeauthoryear{{Thompson} \& {Gill}}{{Thompson} \&
  {Gill}}{2014}]{Thompson-Gill-14}
{Thompson} C.,  {Gill} R.,  2014, \mn@doi [\apj] {10.1088/0004-637X/791/1/46},
  \href {https://ui.adsabs.harvard.edu/abs/2014ApJ...791...46T} {791, 46}

\bibitem[\protect\citeauthoryear{{Vianello}, {Gill}, {Granot}, {Omodei},
  {Cohen-Tanugi}  \& {Longo}}{{Vianello} et~al.}{2018}]{Vianello+18}
{Vianello} G.,  {Gill} R.,  {Granot} J.,  {Omodei} N.,  {Cohen-Tanugi} J.,
  {Longo} F.,  2018, \mn@doi [\apj] {10.3847/1538-4357/aad6ea}, \href
  {https://ui.adsabs.harvard.edu/abs/2018ApJ...864..163V} {864, 163}

\bibitem[\protect\citeauthoryear{{Vurm} \& {Beloborodov}}{{Vurm} \&
  {Beloborodov}}{2016}]{Vurm-Beloborodov-16}
{Vurm} I.,  {Beloborodov} A.~M.,  2016, \mn@doi [\apj]
  {10.3847/0004-637X/831/2/175}, \href
  {https://ui.adsabs.harvard.edu/abs/2016ApJ...831..175V} {831, 175}

\bibitem[\protect\citeauthoryear{{Vurm}, {Beloborodov}  \& {Poutanen}}{{Vurm}
  et~al.}{2011}]{Vurm+11}
{Vurm} I.,  {Beloborodov} A.~M.,   {Poutanen} J.,  2011, \mn@doi [\apj]
  {10.1088/0004-637X/738/1/77}, \href
  {https://ui.adsabs.harvard.edu/abs/2011ApJ...738...77V} {738, 77}

\bibitem[\protect\citeauthoryear{{Vurm}, {Lyubarsky}  \& {Piran}}{{Vurm}
  et~al.}{2013}]{Vurm+13}
{Vurm} I.,  {Lyubarsky} Y.,   {Piran} T.,  2013, \mn@doi [\apj]
  {10.1088/0004-637X/764/2/143}, \href
  {https://ui.adsabs.harvard.edu/abs/2013ApJ...764..143V} {764, 143}

\bibitem[\protect\citeauthoryear{{Werner}, {Uzdensky}, {Cerutti}, {Nalewajko}
  \& {Begelman}}{{Werner} et~al.}{2016}]{Werner+16}
{Werner} G.~R.,  {Uzdensky} D.~A.,  {Cerutti} B.,  {Nalewajko} K.,   {Begelman}
  M.~C.,  2016, \mn@doi [\apjl] {10.3847/2041-8205/816/1/L8}, \href
  {https://ui.adsabs.harvard.edu/abs/2016ApJ...816L...8W} {816, L8}

\makeatother
\end{thebibliography}

\bsp	
\label{lastpage}
\end{document}